\documentclass[twocolumn,showpacs,preprintnumbers,amsmath,amssymb,superscriptaddress,prc]{revtex4-2}

\usepackage{graphicx}
\usepackage{dcolumn}
\usepackage{bm}

\begin{document}

\preprint{PRC preprint}

\title{Emerging collectivity and phase transition in mass A$\approx$150 region.\\
                New information for Nd isotopes.}

\author{W.~Urban}
\author{T. Rz\c{a}ca-Urban}
\author{J. Wi\'sniewski}
\affiliation{Faculty of Physics, University of Warsaw, ulica Pasteura 5, PL-02-093 Warsaw, Poland}

\author{A.G. Smith}
\affiliation{Department of Physics and Astronomy, The University of Manchester, 
             M13 9PL Manchester, UK}

\author{J.P. Greene}
\affiliation{Argonne National Laboratory, Argonne, IL 60439, USA}
\date{\today}

\begin{abstract}
Low and medium spin excitations in $^{146,148,150,152}$Nd isotopes, populated in $\beta^-$ decay
of corresponding Pr isotopes or in prompt-$\gamma$ fission of $^{252}$Cf have been studied using
Gammasphere array of Ge spectrometers. 159 new levels, including two new isomers, 305 new $\gamma$
transitions and 83 new spin-parity assignments were added in the four studied nuclei. The structure
of excited levels in the studied Nd isotopes is discussed using phenomenological classifications
and systematics and compared to calculations reported in other works. Particular attention is paid
to $0^+$ and $2^+$ excitations related to the emerging quadrupole collectivity and to the role of
the 11/2$^-$[505] neutron extruder in the process.
\end{abstract}


\maketitle

\section{Introduction}

Coexisting nuclear structures often manifest their presence via low-lying 0$^+$ excitations
\cite{HW2011}. The nature of these levels is not yet fully understood. Their interpretation
as $\beta$ vibrations has been questioned \cite{Gar01,Gar10,Sha11,Gar16}, though the existence
of such type of excitations in well deformed nuclei is still considered possible
\cite{Cla03,Pie04,Gup17,Apr18,Urb19,Gup19}.

Recent works on 0$^+$ excitations in the A$\approx$100 region \cite{Urb19,Urb21} have shown that
many of them, especially those with low energies, are due to excitations of nucleon pairs at
crossings of up sloping and down sloping Nilsson orbitals. Special role plays here the steeply
up sloping 9/2$^+$[404] neutron extruder active around neutron number N=59 \cite{Urb03,Urb04},
where pronounced shape change and coexistence effects are observed \cite{Urb01,HW2011}. We have
shown how the 0$^+_2$ level in $^{98}$Sr is formed with the involvement of the $\nu$9/2$^+$[404]
extruder \cite{Urb19} and proposed that any up sloping orbital, both, neutron and proton, can act
analogously to the extruder \cite{Urb20,Wis23}.

\begin{figure}
\centering
\scalebox{.24}{\includegraphics{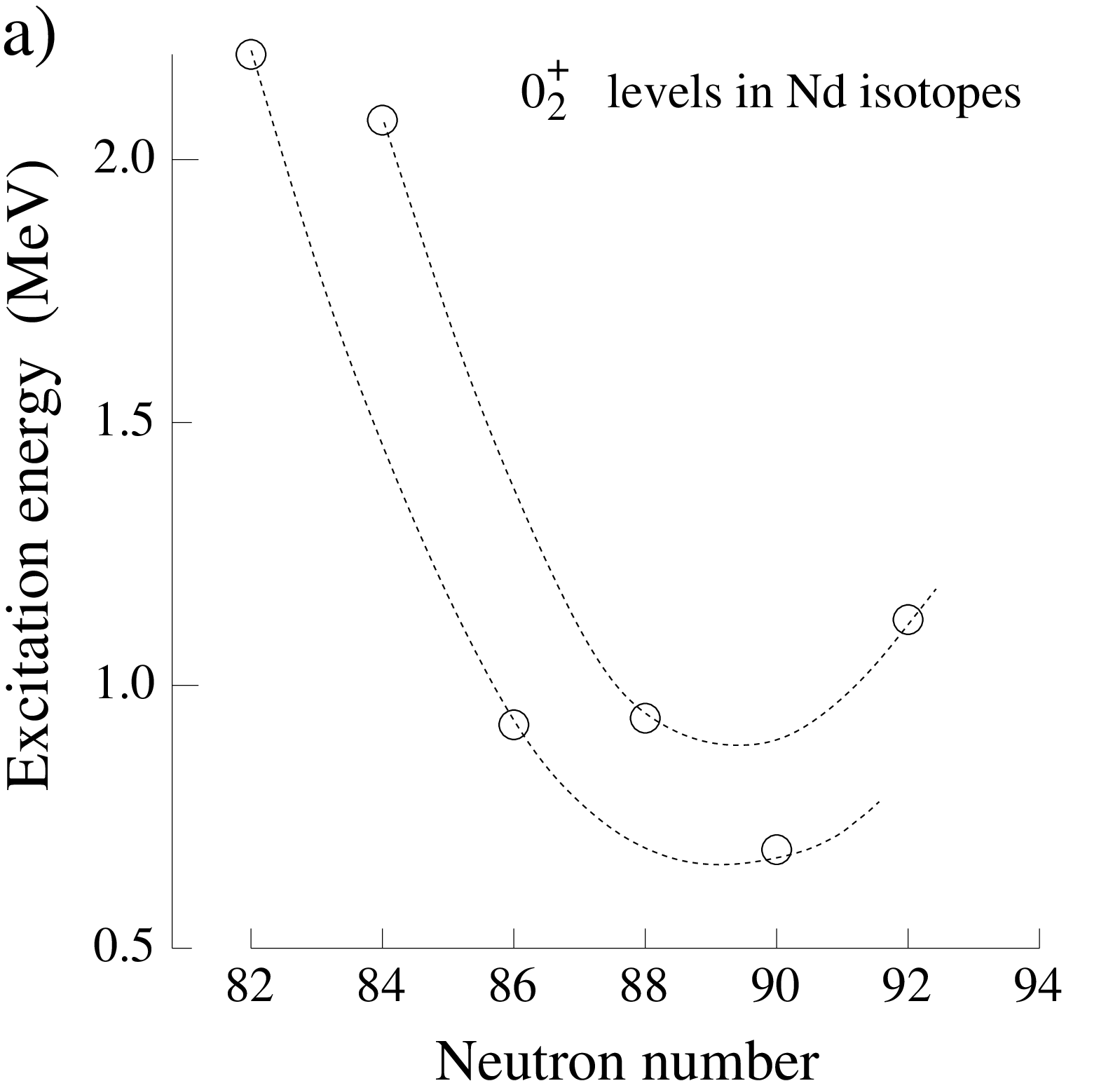}}
\scalebox{.27}{\includegraphics{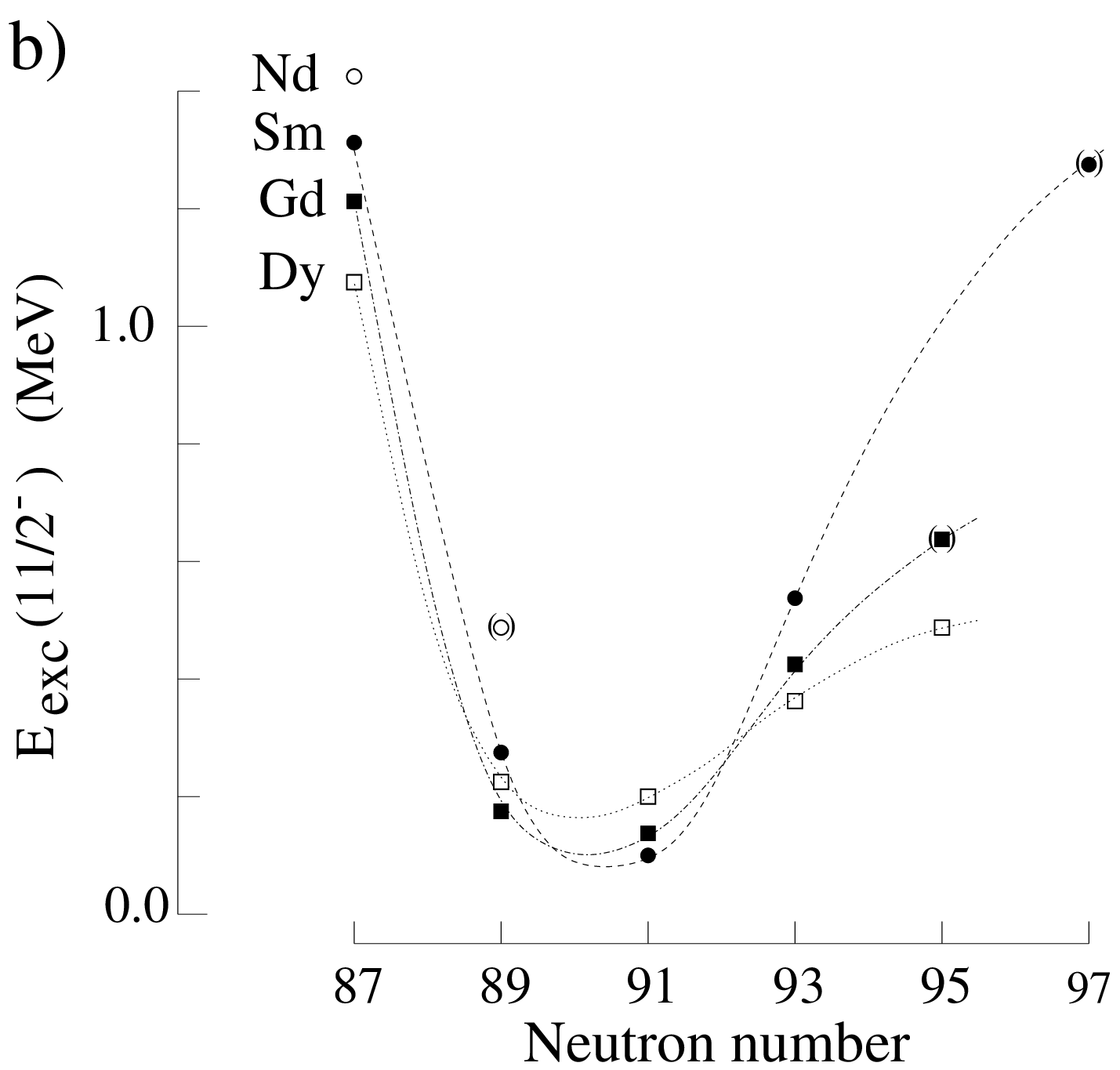}}
\scalebox{.24}{\includegraphics{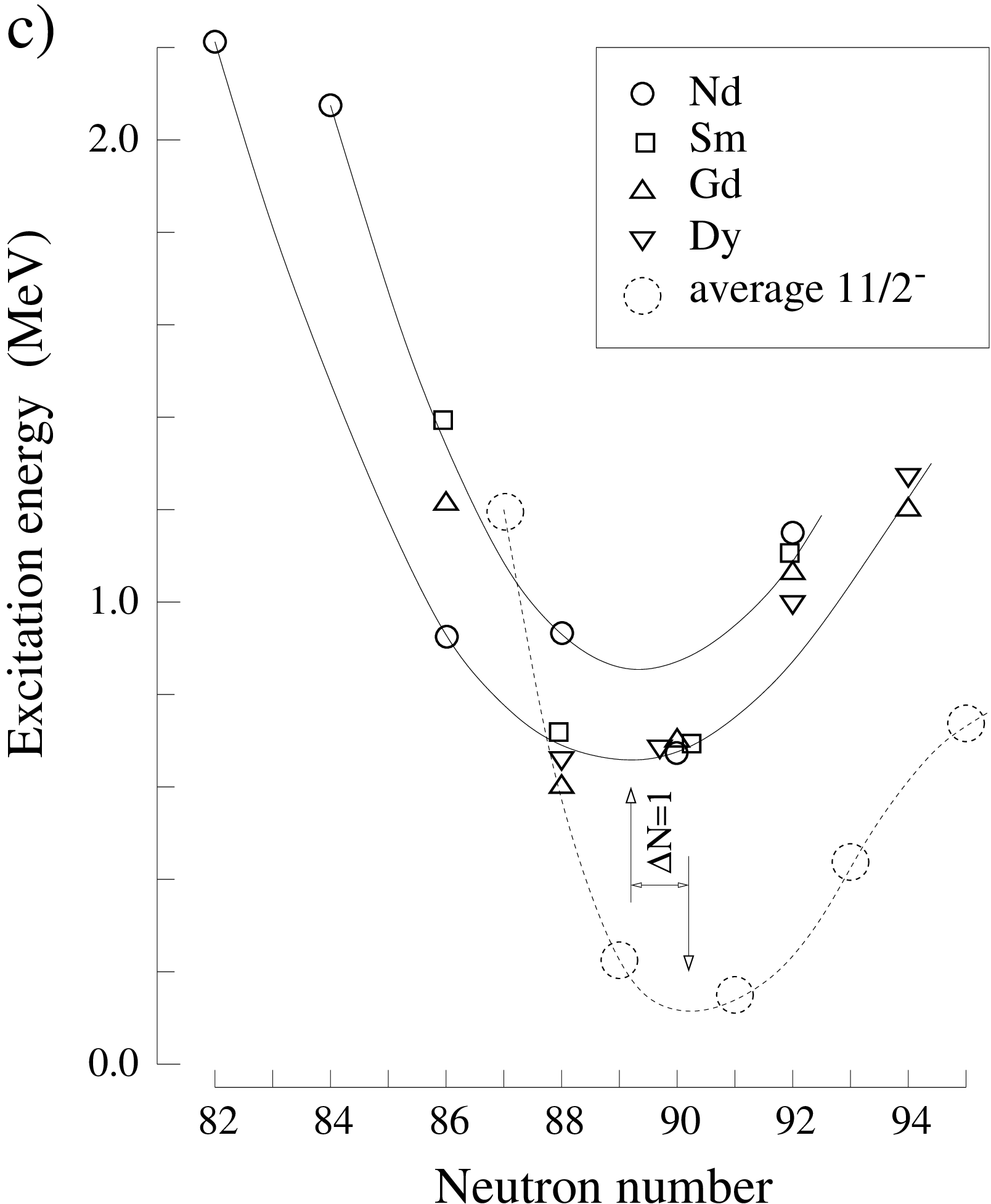}}
\caption{a) Excitation energies of $0^+_2$ levels in even-even Nd isotopes.
         b) Excitation energies of 11/2$^-$ levels in the A$\approx$150 region. Points in
         parentheses are tentative.
         c) Excitation energies of $0^+_2$ levels compared on a common scale to average energies
         of 11/2$^-$ levels in the A$\approx$150 region.
         The data are taken from Ref. \cite{Urb09} and the compilation \cite{ENSDF}. Dashed
         lines are drawn to guide the eye.}
\label{A150_Nd_0plus_N}
\end{figure}

It was proposed that in the A$\approx$150 region the 11/2$^-$[505] neutron extruder is involved in
creating low-lying, 0$^+$ excitations \cite{Sha11,Sha11b,Sha19}. Figure \ref{A150_Nd_0plus_N} (a)
shows 0$^+_2$ excitation energies in even-even Nd isotopes whereas Fig. \ref{A150_Nd_0plus_N} (b)
(an upgraded Fig. 6 from Ref. \cite{Urb09}) shows excitation energies of 11/2$^-$ levels in odd-A
isotopic chains, which are due to the $\nu$11/2$^-$[505] extruder, as first shown for N=87 isotones
in Refs. \cite{Kle74,Kle77}. Figure \ref{A150_Nd_0plus_N} (c) compares, on a common scale, more
0$^+_2$ levels in even-even nuclei in the region to average energies of 11/2$^-$ levels from Fig.
\ref{A150_Nd_0plus_N} (b). The observed correlation suggests the contribution of the
$\nu$11/2$^-$[505] orbital to the structure of 0$^+_2$ excitations  in the region. The minimum of
the average 11/2$^-$ energy is one neutron higher than the minimum of 0$^+$ levels. This may
tell how the $\nu$11/2$^-$[505] orbital is positioned relative to the low-$\Omega$, down-slopping
orbitals in the corresponding even-even cores, to which it passes its pair of neutrons in a process
of producing 0$^+$ levels.

An intriguing observation in Fig. \ref{A150_Nd_0plus_N} is that the 0$^+_2$ energies follow two
different ``parabolas'', which, in addition, are wider than the ``parabola'' corresponding to the
$\nu$11/2$^-$[505] extruder. One may ask whether these wide parabolas represent some physical
effect or are a kind of an envelope encompassing fine, underlying structures, yet to be uncovered.

Disentangling of this intricate picture requires good experimental knowledge of excited levels in
the region. Valuable information on neutron orbitals were obtained from studies of two-quasi-particle
(2-qp) isomers in $^{152,154,156}$Nd and $^{156,158,160}$Sm \cite{Gau98,Sim09}. The 2-qp isomer in
$^{152}$Nd was studied further in Ref. \cite{Yeo10} revealing some inconsistencies with the
expectations. In lighter Nd isotopes, crucial for studying the build up of collectivity in the
A$\approx$150 region, such isomers are not yet known.

The aim of the present work is to verify and extend the experimental information on Nd isotopes
reported previously. Using this information we will then discuss properties of low-spin excitations
to learn more about the emergence and the evolution of various collective modes and the, so called,
quantum phase transition in A$\approx$150 nuclei, in particular in even Nd isotopes (see Refs. \cite{Nik07,Li009} and references therein).

The paper contains the Introduction, the Measurements and Results sections and section discussing  excitations in Nd isotopes in a wider context of low-energy excitations in the A$\approx$150
region. The work is concluded by the Summary and Outlook section.

\section{Measurements and results}

New experimental results on Nd isotopes have been obtained from measurements of $\gamma$ rays
following spontaneous fission of $^{252}$Cf using Gammasphere array of Compton-suppressed Ge
spectrometers \cite{Lee90}. The experiment was described in our previous works \cite{Urb09b,Nai17}.

\begin{figure}
\centering
\scalebox{.36}{\includegraphics{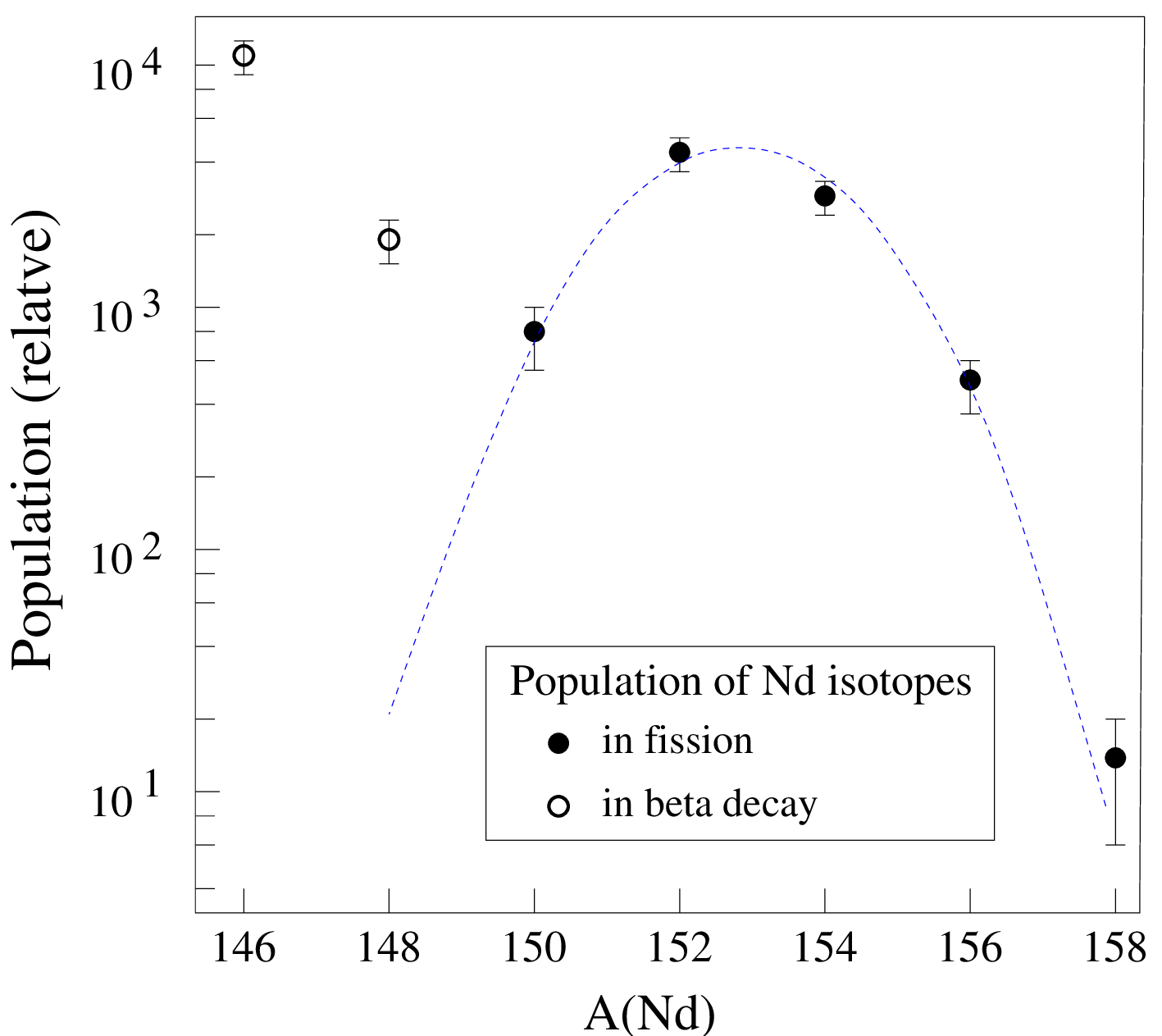}}
\caption{Relative population (in arbitrary units) of even-even Nd isotopes following spontaneous
         fission of $^{252}$Cf. Full circles represent prompt-$\gamma$ intensities of triple
         coincidences in the $8_1^+ - 6_1^+ - 4_1^+ - 2_1^+$ cascade in $^{150-158}$Nd. Empty
         circles show $\gamma$ intensities of triple coincidences in the
         $2_6^+ - 3_1^- - 2_1^+ - 0_1^+$ cascade in $^{146}$Nd and of triple coincidences in the
         $2_4^+ - 3_1^- - 2_1^+ - 0_1^+$ cascade in $^{148}$Nd, populated in $\beta^-$ decay of
         $^{146}$Pr $^{148}$Pr, respectively. The dashed line shows Gaussian fit to the
         prompt-$\gamma$ data points, as described in Refs. \cite{Urb97,Urb20}.}
\label{A150_Nd_population}
\end{figure}

Figure \ref{A150_Nd_population} shows relative intensities of triple-$\gamma$ coincidences in
$\gamma$ cascades of Nd isotopes in the measurement. In $^{146}$Nd and $^{148}$Nd the population
is significantly increased due to $\beta^-$ decay of $^{146}$Pr and $^{148}$Pr nuclei, respectively,
which collect high cumulative yield produced in fission at A=146 and A=148 isobaric chains.

The data point for $^{158}$Nd in Fig. \ref{A150_Nd_population} represents intensity in the
$8_1^+ - 6_1^+ - 4_1^+ - 2_1^+$ ground state cascade, identified up to spin 6$^+$ in Ref.
\cite{Ide16} with the $8_1^+ - 6_1^+$ transition of 310.5(3) keV newly observed in the present
work.

\subsection{Results for $^{146}$Nd}

Excited states in $^{146}$Nd were studied at medium spins in Ref. \cite{Iac96}. The present study
adds new levels with low spins. Figure \ref{A150_Nd_population} indicates that $^{146}$Nd is
populated exclusively in $\beta^-$ decay of $^{146}$Pr. Previous $\beta^-$ decay results for
$^{146}$Pr \cite{Ike78,NDS146} are extended in the present work by 15 new excited levels and 67
new $\gamma$ transitions.

\begin{figure*}
\centering
\scalebox{.67}{\includegraphics{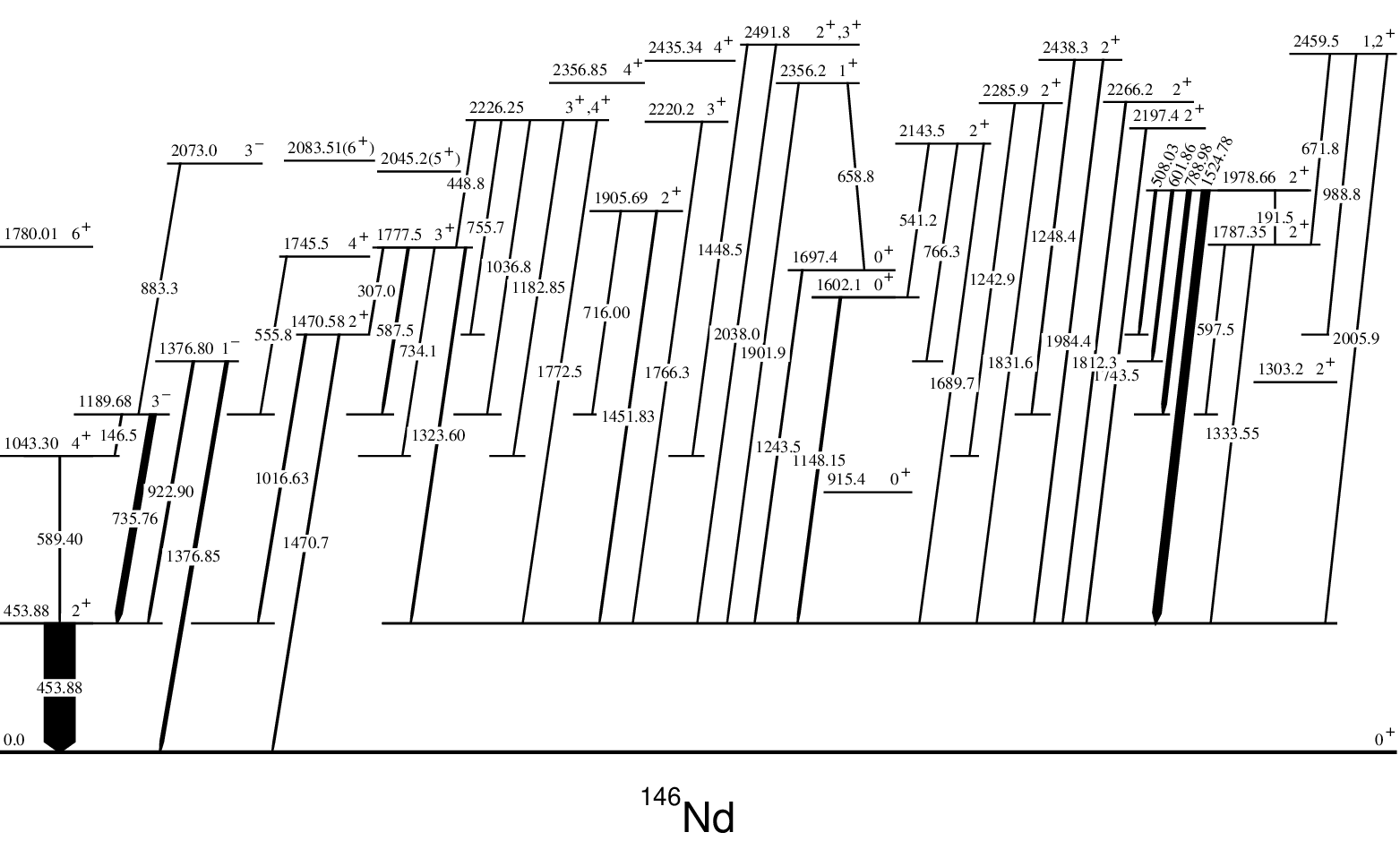}}
\caption{Partial level scheme of $^{146}$Nd populated in $\beta^-$ decay of $^{146}$Pr, as observed
         in the present work. The 915.5- and 1303.2-keV levels are drawn after Ref. \cite{Ike78}.
         The 1780.01-, 2045.2- and 2083.51-keV levels, not populated in $\beta^-$ decay, are drawn
         after Ref. \cite{NDS146} to assist the discussion. See Tables \ref{table_Nd146_levels_I} - \ref{table_Nd146_levels_III} for all excited levels in $^{146}$Nd observed in the present
         work.}
\label{A150_Nd_146_scheme}
\end{figure*}

Partial scheme of excited states in $^{146}$Nd observed in the present work is shown in Fig.
\ref{A150_Nd_146_scheme}. To assist further discussions we included in the scheme the 915.4-,
1303.2-, 1780.01-, 2045.2-keV, 2083.51- and 2435.34-keV levels reported previously in $\beta^-$
decay \cite{Ike78,NDS146}, which are not observed in our work.

All levels and their $\gamma$ decays in $^{146}$Nd observed in the present work are listed in
Tables \ref{table_Nd146_levels_I}, \ref{table_Nd146_levels_II} and \ref{table_Nd146_levels_III}.
Levels and transitions in Tables \ref{table_Nd146_levels_I}-\ref{table_Nd146_levels_III} have
been determined using double- and triple-$\gamma$ coincidences. Transitions reported in Ref.
\cite{NDS146}, which feed the ground state (g.s.) and are not in cascade with other transitions,
were not analysed.

Intensities of the 453.88-, 735.76-, 788.98- and 1524.78-keV transitions are contaminated in our
data by 453.70-737.25-keV and 737.25-1523.5-keV cascades in $^{144}$Ce, strongly populated in
$\beta^-$ decay of $^{144}$La \cite{Nai17}. In Table \ref{table_Nd146_levels_I} intensities of
these and a few other strong transitions are taken from the compilation \cite{NDS146}. Intensities
of other transitions have been normalized to them. For several transitions they differ from
intensities reported previously \cite{Ike78,NDS146}.

We confirm 0$^+$ levels at 1602.1 and 1697.4 keV reported previously \cite{Ike78,NDS146}. The
1148.15-keV decay of the 1602.1-keV level is new in $\beta^-$ decay scheme.

The 466.4-keV decay from 2143.5-keV, 2$^+$ to the 1697.4-keV, 0$^+$ level,
reported in previous $\beta^-$ decay work \cite{Ide16}, is not seen in the present work. The
low-energy, 1303.2-keV level reported in the (p,t) study \cite{Pon96} with spin-parity 0$^+$
was assigned later spin-parity 2$^+$ \cite{NDS146}.

An important finding of the present work is the 307.0-keV decay from the 3$^+$, 1777.5-keV level to
the 2$^+$ level at 1470.58 keV expected to be the head of a $\gamma$ band in $^{146}$Nd. This confirms
$\gamma$ collectivity already at N=86 as suggested by Fig. 8 of Ref. \cite{Nai17}. The band includes
the 1745.5-, 1777.5-, 2045.2- and 2083.51-keV levels shown in Fig. \ref{A150_Nd_146_scheme}.
The 2045.2-keV level reported in the compilation \cite{NDS146} with spin-parity 4$^-$ or spin 5
is assigned tentative spin-parity (5$^+$) because of no decay to the 3$^-$ level at 1189.68 keV.
It has energy expected for the 5$^+$ member of the $\gamma$ band (see Fig. 8 of Ref. \cite{Nai17}).

The 3$^+$ level at 2220.2 keV and further candidates for 3$^+$ excitations in $^{146}$Nd at 2226.25
and 2491.8 keV, reported in the compilation \cite{NDS146} but newly observed in $\beta^-$ decay,
suggests the presence other structures related to $\gamma$ collectivity.

\begin{table}[]
\caption{Excited levels and their $\gamma$ decays in $^{146}$Nd populated in $\beta^-$ decay of
         $^{146}$Pr as observed in the present work in coincidence data. New data are marked by
         the star symbol and those marked with superscript {\it ``a''} are taken from Refs.
         \cite{NDS146,Ike78}. See text for more comments.}
\begin{center}
\begin{tabular}{l c r c r }
\hline
~~Initial      &level    &~~~${\gamma}$~~~~-&decay& Final level~~~                           \\
E$_{exc}$ (keV)&I$^{\pi}$&~~~E$_{\gamma}$(keV)& I$_{\gamma}$(rel)&E$_{exc}$ (keV), I$^{\pi}$ \\
\hline
\\
   453.88(3)   &   2$^+$ &   453.88(3)        &  1000$^a$        &     0.00, 0$^+$           \\
  1043.30(8)   &   4$^+$ &   589.40(7)        &     7.2(4)$^a$   &   453.88, 2$^+$           \\
  1189.68(5)   &   3$^-$ &   146.5(1)         &     3(1)         &  1043.30, 4$^+$           \\
               &         &   735.76(3)        &   156(8)$^a$     &   453.88, 2$^+$           \\
  1376.80(5)   &   1$^-$ &   922.90(3)        &    48.5(24)$^a$  &   453.88, 2$^+$           \\
               &         &  1376.85(8)        &    91(5)$^a$     &     0.00, 0$^+$           \\
  1470.58(7)   &   2$^+$ &  1016.63(5)        &    28(3)         &   453.88, 2$^+$           \\
               &         &  1470.7(1)         &    24.8(13)$^a$  &     0.00, 0$^+$           \\
  1602.1(1)    &   0$^+$ &  1148.15(5)*       &     4(1)         &   453.88, 2$^+$           \\
  1697.4(2)    &   0$^+$ &  1243.5(1)         &     12(3)        &   453.88, 2$^+$           \\
  1745.5(3)*   &   4$^+$*&   555.8(2)*        &     1.1(3)       &  1189.68, 3$^-$           \\
  1777.5(1)    &   3$^+$ &   307.0(2)*        &     0.2(1)       &  1470.58, 2$^+$           \\
               &         &   587.5(3)         &     0.6(2)       &  1189.68, 3$^-$           \\
               &         &   734.1(1)*        &     0.8(2)       &  1043.30, 4$^+$           \\
               &         &  1323.60(5)        &    22(4)         &   453.88, 2$^+$           \\
  1787.35(7)   &   2$^+$ &   597.5(1)         &     1.8(4)       &  1189.68, 3$^-$           \\
               &         &  1333.55(5)        &    19(3)         &   453.88, 2$^+$           \\
  1905.69(6)   &   2$^+$ &   716.00(6)        &     2.0(4)       &  1189.68, 3$^-$           \\
               &         &  1451.83(5)        &    30(5)         &   453.88, 2$^+$           \\
  1978.66(5)   &   2$^+$ &   191.5(2)         &     3(1)         &  1787.35, 2$^+$           \\
               &         &   201.3(1)*        &     4(1)         &  1777.5,  3$^+$           \\
               &         &   508.03(5)        &    10(1)         &  1470.58, 2$^+$           \\
               &         &   601.86(3)        &    70(10)        &  1376.80, 1$^-$           \\
               &         &   788.98(5)        &   131(7)$^a$     &  1189.68, 3$^-$           \\
               &         &  1524.78(3)        &   260(10)        &   453.88, 2$^+$           \\
  2073.0(2)*   &   3$^-$*&   883.3(1)*        &     1.0(2)       &  1189.68, 3$^-$           \\
  2143.5(1)    &   2$^+$ &   541.2(2)*        &     3(1)         &  1602.1 , 0$^+$           \\
               &         &   766.3(2)         &     2(1)         &  1376.80, 1$^-$           \\
               &         &  1689.7(1)         &    11(1)         &   453.88, 2$^+$           \\
  2197.4(3)    &   2$^+$ &  1743.5(2)         &     3(1)         &   453.88, 2$^+$           \\
  2220.2(2)    &   3$^+$ &  1766.3(1)         &    12(2)         &   453.88, 2$^+$           \\
  2226.25(7)*  &3$^+$,4$^+$*& 448.8(2)*       &     0.5(2)       &  1777.5 , 3$^+$           \\
               &         &   755.7(2)*        &     1.5(3)       &  1470.58, 2$^+$           \\
               &         &  1036.8(1)*        &     2.1(3)       &  1189.68, 3$^-$           \\
               &         &  1182.85(5)*       &     2.0(3)       &  1043.30, 4$^+$           \\
               &         &  1772.5(2)*        &     2.5(5)       &   453.88, 2$^+$           \\
  2266.2(2)    &  2$^+$  &  1812.3(2)         &     3.4(6)       &   453.88, 2$^+$           \\
  2285.9(3)*   &  2$^+$  &  1242.9(2)*        &     0.3(1)       &  1043.30, 4$^+$           \\
               &         &  1831.6(2)*        &     3.0(6)       &   453.88, 2$^+$           \\
  2336.3(2)    &  3$^-$  &  1292.7(2)         &     0.6(2)       &  1043.30, 4$^+$           \\
               &         &  1882.4(2)         &     2.5(6)       &   453.88, 2$^+$           \\
  2356.2(2)    &  1$^+$  &   658.8(2)         &     3(1)         &  1697.4 , 0$^+$           \\
               &         &  1901.9(2)         &     6(2)         &   453.88, 2$^+$           \\
  2438.3(2)    &  2$^+$  &  1248.4(3)         &     0.4(2)       &  1189.68, 3$^-$           \\
               &         &  1984.4(2)         &     4.9(6)       &   453.88, 2$^+$           \\
  2459.5(3)    & 1,2$^+$ &   671.8(3)*        &     3(1)         &  1787.35, 2$^+$           \\
               &         &   988.8(2)*        &     1.6(3)       &  1470.58, 2$^+$           \\
               &         &  2005.9(3)         &     5(1)         &   453.88, 2$^+$           \\
  2491.8(3)*   &2$^+$,3$^+$*&1448.5(2)*       &     0.8(2)       &  1043.30, 4$^+$           \\
               &         &  2038.0(3)*        &     5(2)         &   453.88, 2$^+$           \\
\hline
\end{tabular}
\end{center}
\label{table_Nd146_levels_I}
\end{table}

\begin{table}[]
\caption{ continuation of Table \ref{table_Nd146_levels_I}.}
\begin{center}
\begin{tabular}{l c r c r }
\hline
~~Initial      &level    &~~~${\gamma}$~~~~-&decay& Final level~~~                           \\
E$_{exc}$ (keV)&I$^{\pi}$&~~~E$_{\gamma}$(keV)& I$_{\gamma}$(rel)&E$_{exc}$ (keV), I$^{\pi}$ \\
\hline
  2551.9(1)    &  2$^+$  &   646.1(1)*        &     1.2(3)       &  1905.69, 2$^+$           \\
               &         &   774.50(5)        &     6(1)         &  1777.5,  3$^+$           \\
               &         &  1081.30(5)        &     17(1)        &  1470.58, 2$^+$           \\
               &         &  1362.5(2)*        &     1.1(2)       &  1189.68, 3$^-$           \\
  2681.1(2)    &  1$^-$  &   983.8(1)         &     2.5(5)       &  1697.4,  0$^+$           \\
               &         &  2227.1(2)         &     3.9(6)       &   453.88, 2$^+$           \\
  2705.8(1)    &2$^{(+)}$*&  349.5(2)*        &     0.6(2)       &  2356.2,  1$^+$           \\
               &         &   562.20(3)        &     15(3)        &  2143.5,  2$^+$           \\
               &         &   727.3(1)         &     11.8(6)$^a$  &  1978.66, 2$^+$           \\
               &         &   800.2(2)*        &     0.7(2)       &  1905.69, 2$^+$           \\
               &         &   928.25(5)        &     5(1)         &  1777.5,  3$^+$           \\
               &         &  1235.16(5)        &     9(1)         &  1470.58, 2$^+$           \\
               &         &  1328.90(5)        &     13(1)        &  1376.80, 1$^-$           \\
               &         &  1516.02(7)        &     2.5(3)       &  1189.68, 3$^-$           \\
 2775.25(15)   & 1,2$^+$ &  1077.9(1)*        &     2.0(4)       &  1697.4,  0$^+$           \\
               &         &  1304.2(2)*        &     1.6(3)       &  1470.58, 2$^+$           \\
               &         &  2321.5(2)*        &     6(2)         &   453.88, 2$^+$           \\
 2855.6(3)*    &  2$^+$  &  1665.9(2)*        &     0.4(2)       &  1189.68, 3$^-$           \\
 2930.9(2)*    &  4$^+$  &  1741.25(15)*      &     1.3(2)       &  1189.68, 3$^-$           \\
 2970.85(15)   &  2$^+$  &  1499.9(2)         &     1.4(3)       &  1470.58, 2$^+$           \\
               &         &  1594.05(9)        &     4(1)         &  1376.80, 1$^-$           \\
               &         &  1781.3(1)         &     2.2(3)       &  1189.68, 3$^-$           \\
 3026.9(1)*    &         &  1649.9(1)*        &     2(1)         &  1376.80, 1$^-$           \\
               &         &  1837.5(2)*        &     0.8(2)       &  1189.68, 3$^-$           \\
 3171.5(2)*    &         &  1794.6(2)*        &     3(1)         &  1376.80, 1$^-$           \\
 3231.4(3)*    &         &  1760.8(2)*        &     0.8(2)       &  1470.58, 2$^+$           \\
 3291.8(2)     &  1      &  1313.2(2)*        &     0.6(2)       &  1978.66, 2$^+$           \\
               &         &  1914.9(2)         &     3(1)         &  1376.80, 1$^-$           \\
 3316.8(1)*    & (2)$^-$*&  1172.8(1)*        &     3(1)         &  2143.5,  2$^+$           \\
               &         &  1338.0(1)*        &     2.4(4)       &  1978.66, 2$^+$           \\
               &         &  1411.0(1)*        &     2.5(5)       &  1905.69, 2$^+$           \\
               &         &  1529.3(1)*        &     6(1)         &  1787.35, 2$^+$           \\
               &         &  1539.2(2)*        &     0.9(2)       &  1777.5,  3$^+$           \\
               &         &  1846.0(1)*        &     1.1(2)       &  1470.58, 2$^+$           \\
               &         &  1940.2(1)*        &     1.9(3)       &  1376.80, 1$^-$           \\
               &         &  2127.1(1)*        &     0.7(2)       &  1189.68, 3$^-$           \\
 3335.3(2)     & (2$^-$)*&  1429.9(1)*        &     1.0(2)       &  1905.69, 2$^+$           \\
               &         &  1864.4(2)*        &     1.1(2)       &  1470.58, 2$^+$           \\
               &         &  1958.3(1)         &     5(1)         &  1376.80, 1$^-$           \\
               &         &  2145.5(2)*        &     0.9(2)       &  1189.68, 3$^-$           \\
 3368.7(1)     & 1$^-$,2 &   816.8(3)*        &     0.3(1)       &  2551.9,  2$^+$           \\
               &         &  1148.0(1)*        &     3.2(6)       &  2220.2,  3$^+$           \\
               &         &  1170.9(1)*        &     0.7(2)       &  2197.4,  2$^+$           \\
               &         &  1463.1(2)         &     1.1(2)       &  1905.69, 2$^+$           \\
               &         &  1766.5(1)*        &     5(1)         &  1602.1,  0$^+$           \\
               &         &  1898.1(1)*        &     1.7(3)       &  1470.58, 2$^+$           \\
               &         &  1991.7(1)         &     2.0(4)       &  1376.80, 1$^-$           \\
               &         &  2179.06(5)        &     1.6(4)       &  1189.68, 3$^-$           \\
 3391.3(2)     & (2,3)*  &   839.4(2)         &     3(1)         &  2551.9,  2$^+$           \\
               &         &  1165.0(2)*        &     0.3(1)       &  2226.25, 3$^+$,4$^+$     \\
               &         &  1170.8(2)*        &     0.7(2)       &  2220.2,  3$^+$           \\
               &         &  1248.0(3)*        &     2(1)         &  2143.5,  2$^+$           \\
               &         &  1613.8(1)         &     1.7(3)       &  1777.5,  3$^+$           \\
               &         &  1920.6(1)         &     1.9(3)       &  1470.58, 2$^+$           \\
               &         &  2202.1(3)         &     0.4(2)       &  1189.68, 3$^-$           \\
 3507.3(3)*    &         &  2317.6(2)*        &     0.5(2)       &  1189.68, 3$^-$           \\
\hline
\end{tabular}
\end{center}
\label{table_Nd146_levels_II}
\end{table}

\begin{table}[]
\caption{ continuation of Table \ref{table_Nd146_levels_I}.}
\begin{center}
\begin{tabular}{l c r c r }
\hline
~~Initial      &level    &~~~${\gamma}$~~~~-&decay& Final level~~~                           \\
E$_{exc}$ (keV)&I$^{\pi}$&~~~E$_{\gamma}$(keV)& I$_{\gamma}$(rel)&E$_{exc}$ (keV), I$^{\pi}$ \\
\hline
 3534.1(2)     & (2$^-$)*&  1042.2(2)*        &     0.7(2)       &  2491.8,  2$^+$,3$^+$     \\
               &         &  1313.7(2)*        &     0.5(2)       &  2220.2,  3$^+$           \\
               &         &  1390.5(3)*        &     1.5(5)       &  2143.5,  2$^+$           \\
               &         &  1555.6(3)         &     1.6(3)       &  1978.66, 2$^+$           \\
               &         &  1746.9(3)*        &     6(1)         &  1787.35, 2$^+$           \\
               &         &  2063.1(2)*        &     0.4(1)       &  1470.58, 2$^+$           \\
               &         &  2157.4(2)         &     1.2(2)       &  1376.80, 1$^-$           \\
               &         &  2344.3(2)*        &     0.5(2)       &  1189.68, 3$^-$           \\
               &         &  3080.1(4)         &     1.0(4)       &   453.88, 2$^+$           \\
 3585.1(3)*    & (2$^+$) &  1679.3(2)*        &     0.5(2)       &  1905.69, 2$^+$           \\
 3586.3(2)*    & (2$^+$) &  1608.0(3)*        &     0.6(2)       &  1978.66, 2$^+$           \\
               &         &  2396.4(2)*        &     0.7(2)       &  1189.68, 3$^-$           \\
 3594.2(2)     & (2$^+$) &  1374.0(2)*        &     0.6(2)       &  2220.2,  3$^+$           \\
               &         &  1816.7(2)*        &     0.7(2)       &  1777.5,  3$^+$           \\
               &         &  2217.5(2)         &     1.8(4)       &  1376.80, 1$^-$           \\
               &         &  2404.2(2)         &     0.6(2)       &  1189.68, 3$^-$           \\
 3618.6(4)     &         &  2148.3(3)         &     0.8(1)       &  1470.58, 2$^+$           \\
               &         &  2241.8(3)*        &     1.3(3)       &  1376.80, 1$^-$           \\
 3639.8(2)*    &   (2)   &  1661.5(3)*        &     0.7(2)       &  1978.66, 2$^+$           \\
               &         &  1852.1(2)*        &     7(1)         &  1787.35, 2$^+$           \\
               &         &  2449.9(3)*        &     0.4(2)       &  1189.68, 3$^-$           \\
\hline
\end{tabular}
\end{center}
\label{table_Nd146_levels_III}
\end{table}

\begin{table}[h]
\caption{Angular correlation results for $\gamma \gamma$ cascades in $^{146}$Nd populated
in $\beta^-$ decay of $^{146}$Pr, as observed in the
present work. Superscript ``m'' indicates mixed transition for which $\delta_{exp}$ value
(with corresponding $\chi^2$) is obtained.}
\begin{center}
\begin{tabular}{ c c c c }
\hline
E${\gamma_1}$ - E${\gamma_2}$& $A_2/A_0$  & $A_4/A_0$  &     Spins in cascade         \\
      (keV)                  &   exp.     &    exp.    &and $\delta_{exp}$; $\chi^2$  \\
\hline
  922.90$^m$-453.88          &-0.222(33) &  0.038(48)  & 1 - 2 - 0                    \\
                             &           &             & -0.023(30); 0.7              \\
 1016.63$^m$-453.88          &-0.206(63) &  0.327(82)  & 2 - 2 - 0                    \\
                             &           &             & +6.0(+94,-26); 0.1           \\
 1243.5 - 453.88             & 0.19(18)  &  1.02(28)   & 0 - 2 - 0;  $\chi^2$=1.1     \\
 1323.60$^m$-453.88          &-0.226(99) &  0.041(129) & 3 - 2 - 0                    \\
                             &           &             & -0.20(14); 0.1 or            \\
                             &           &             & -33(+27,-Inf); 0.9           \\
 1333.55$^m$-453.88          & 0.45(12)  &  0.31(17)   & 2 - 2 - 0                    \\
                             &           &             & -1.2(6); 0.6                 \\
 1451.83$^m$-453.88          & 0.375(59) & -0.061(85)  & 2 - 2 - 0                    \\
                             &           &             & -0.18(10); 0.7               \\
 1524.78$^m$-453.88          & 0.232(21) & -0.023(32)  & 2 - 2 - 0                    \\
                             &           &             & +0.024(28); 0.6              \\
 1689.7$^m$-453.88           & 0.28(12)  & -0.13(15)   & 2 - 2 - 0                    \\
                             &           &             & +0.69(27); 3.0               \\
 2227.1$^m$-453.88           & 0.35(18)  & -0.31(24)   & 1 - 2 - 0                    \\
                             &           &             & -0.64(-38,+22); 0.3          \\
\hline
\end{tabular}
\end{center}
\label{table_Nd146_ang}
\end{table}

Spins and parities of levels in $^{146}$Nd in Fig. \ref{A150_Nd_146_scheme} and in Tables
\ref{table_Nd146_levels_I} - \ref{table_Nd146_levels_III} are drawn after the compilation
\cite{NDS146} when confirmed by our work. The assignments marked as new in the tables are
proposed based on angular correlations, determined using the technique described in Ref.
\cite{Nai17} and the decay branchings. We confirm spins of levels at 1376.80, 1470.58, 1602.1,
1697.4, 1777.5, 1787.35, 1905.69 and 1978.66 keV and provide $\delta$ mixing ratios for several
transitions. One notes the large values of $A_2/A_0$ and and $A_4/A_0$ for the 0-2-0 cascade
from the 1697.4-keV level, confirming spin I=0 of this level.

The 349.5-keV new decay of the 2705.8-keV level excludes 3$^{(-)}$ spin-parity for this level
proposed in \cite{NDS146} and suggest spin-parity 2$^{(+)}$, considering the observed decay
branchings.

\subsection{Results for $^{148}$Nd}

\begin{table}[]
\caption{Excited levels and their $\gamma$ decays in $^{148}$Nd populated in $\beta^-$ decay of
         $^{148}$Pr produced in fission of $^{252}$Cf, as observed in the present work in
         coincidence data. New data are indicated by asterisks. Values with superscript
         {\it ``a''} are taken from Ref. \cite{NDS148}. See text for more comments.}
\begin{center}
\begin{tabular}{l c r c r }
\hline
~~Initial      &level    &~~~${\gamma}$~~~~-&decay& Final level~~~                           \\
E$_{exc}$ (keV)&I$^{\pi}$&~~~E$_{\gamma}$(keV)& I$_{\gamma}$(rel)&E$_{exc}$ (keV), I$^{\pi}$ \\
\hline
   301.75(3)   &   2$^+$ &   301.75(3)        &  1000$^a$        &       0.00,  0$^+$        \\
   752.35(5)   &   4$^+$ &   450.60(3)        &    95(9)         &     301.75,  2$^+$        \\
   916.80(5)   &   0$^+$ &   615.05(3)        &    40(10)        &     301.75,  2$^+$        \\
   999.35(5)   &   3$^-$ &   247.05(5)        &     8(2)         &     752.35,  4$^+$        \\
               &         &   697.55(3)        &   105(15)        &     301.75,  2$^+$        \\
  1023.00(5)   &   1$^-$ &   721.25(3)        &    64(6)         &     301.75,  2$^+$        \\
               &         &   1022.9(2)        &    70(20)        &       0.00,  0$^+$        \\
  1170.95(5)   &   2$^+$ &   171.8(1)*        &     1.6(4)       &     999.35,  3$^-$        \\
               &         &   254.3(2)*        &     0.3(1)       &     916.80,  0$^+$        \\
               &         &   418.7(1)         &     2(1)         &     752.35,  4$^+$        \\
               &         &   869.16(3)        &    66(8)$^a$     &     301.75,  2$^+$        \\
  1242.15(7)   &   5$^-$ &   489.80(3)        &    12(2)         &     752.35,  4$^+$        \\
  1248.73(7)   &   2$^+$ &   496.5(1)         &    2.5(5)        &     752.35,  4$^+$        \\
               &         &   946.93(5)        &    23(3)         &     301.75,  2$^+$        \\
               &         &  1248.6(2)         &    20(8)         &       0.00,  0$^+$        \\
  1279.80(7)   &   6$^+$ &   527.45(3)        &     5(1)         &     752.35,  4$^+$        \\
  1511.45(5)   &   3$^+$ &   759.10(5)        &     8(2)         &     752.35,  4$^+$        \\
               &         &  1209.70(4)        &    39(4)         &     301.75,  2$^+$        \\
  1515.80(7)   & (2$^-$)*&   492.80(5)        &    11(2)         &    1023.00,  1$^-$        \\
  1604.3(1)*   &  4$^+$* &   433.3(2)*        &     0.4(2)       &    1170.95,  2$^+$        \\
               &         &   605.0(1)*        &     1.3(3)       &     999.35,  3$^-$        \\
               &         &   851.9(1)*        &     2.5(5)       &     752.35,  4$^+$        \\
  1644.6(2)    &   7$^-$ &   364.8(1)         &     1.5(5)       &    1279.80,  6$^+$        \\
  1645.3(2)    & (1$^-$)*&  621.9(2)         &     2.1(4)       &    1023.00,  1$^-$        \\
               &         &  1343.7(2)         &     5(1)         &     301.75,  2$^+$        \\
  1659.50(5)   &   2$^+$ &   636.41(5)        &    18(3)$^a$     &    1023.00,  1$^-$        \\
               &         &   660.15(5)        &    28(3)$^a$     &     999.35,  3$^-$        \\
               &         &  1357.80(6)        &    35(3)         &     301.75,  2$^+$        \\
  1683.37(7)   &   2$^+$*&   512.2(2)         &     2.4(4)       &    1170.95,  2$^+$        \\
               &         &   660.4(1)*        &     1.6(3)       &    1023.00,  1$^-$        \\
               &         &   766.40(5)*       &     5(1)         &     916.80,  0$^+$        \\
               &         &  1381.62(5)        &    24(3)         &     301.75,  2$^+$        \\
  1687.85(8)   &   4$^+$*&   176.4(2)*        &     0.2(1)       &    1511.45,  3$^+$        \\
               &         &   438.95(15)*      &     2.3(3)       &    1248.73,  2$^+$        \\
               &         &   935.50(5)        &     3.2(6)       &     752.35,  4$^+$        \\
  1728.9(1)*   &   3$^+$*&   976.55(5)*       &     6(1)         &     752.35,  4$^+$        \\
  1823.2(3)*   &         &   800.2(2)*        &     1.9(4)       &    1023.00,  1$^-$        \\
  1824.5(1)*   & (4$^-$)*&   824.9(1)*        &     6(1)         &     999.35,  3$^-$        \\
               &         &  1072.4(2)*        &     1.5(4)       &     752.35,  4$^+$        \\
  1839.3(2)*   &         &   839.9(1)*        &     4(1)         &     999.35,  3$^-$        \\
  1856.0(2)    &   8$^+$ &   576.2(2)         &     1.7(4)       &    1279.80,  6$^+$        \\
  1858.6(1)    & 2$^+$,3 &   347.1(1)*        &     0.6(2)       &    1511.45,  3$^+$        \\
               &         &   859.1(2)*        &     0.7(2)       &     999.35,  3$^-$        \\
               &         &  1106.20(5)        &     7(1)         &     752.35,  4$^+$        \\
               &         &  1557.0(1)         &     7(2)         &     301.75,  2$^+$        \\
  1881.65(9)*  & (5$^+$)*&  1129.30(5)*       &     5(1)         &     752.35,  4$^+$        \\
  1935.35(7)*  &2$^-$,3$^-$*& 912.4(1)*       &     2.9(5)       &    1023.00,  1$^-$        \\
               &         &   935.95(5)*       &     2.5(5)       &     999.35,  3$^-$        \\
  1979.7(3)*   &         &   956.7(2)         &     1.3(2)       &    1023.00,  1$^-$        \\
  2004.9(2)*   &         &  1252.5(2)*        &     1.0(2)       &     752.35,  4$^+$        \\
  2038.5(1)*   &         &   526.9(2)*        &     0.7(2)       &    1511.45,  3$^+$        \\
               &         &   789.6(2)*        &     2.4(3)       &    1248.73,  2$^+$        \\
               &         &  1286.4(2)*        &     0.7(2)       &     752.35,  4$^+$        \\
\hline
\end{tabular}
\end{center}
\label{table_Nd148_levels_I}
\end{table}

\begin{table}[]
\caption{ continuation of Table \ref{table_Nd148_levels_I}.}
\begin{center}
\begin{tabular}{l c r c r }
\hline
~~Initial      &level    &~~~${\gamma}$~~~~-&decay& Final level~~~                           \\
E$_{exc}$ (keV)&I$^{\pi}$&~~~E$_{\gamma}$(keV)& I$_{\gamma}$(rel)&E$_{exc}$ (keV), I$^{\pi}$ \\
\hline
  2041.5(1)*   & (2$^+$)*&  1042.2(1)*        &     1.3(3)       &     999.35,  3$^-$        \\
               &         &  1124.5(1)*        &     3(1)         &     916.80,  0$^+$        \\
               &         &  1289.4(2)*        &     1.3(3)       &     752.35,  4$^+$        \\
  2048.8(3)*   &         &  1025.8(2)*        &     0.6(2)       &    1023.00,  1$^-$        \\
  2070.70(8)*  & (2$^+$)*&   899.6(2)*        &     1.3(2)       &    1170.95,  2$^+$        \\
               &         &  1153.92(5)*       &     15(3)        &     916.80,  0$^+$        \\
  2073.77(6)   &2$^{(+)}$&   562.40(5)        &     5(1)         &    1511.45,  3$^+$        \\
               &         &   825.06(3)        &    27(2)$^a$     &    1248.73,  2$^+$        \\
               &         &   902.66(3)        &    21(3)$^a$     &    1170.95,  2$^+$        \\
               &         &  1050.74(5)        &     6(1)         &    1023.00,  1$^-$        \\
  2080.2(2)*   &         &  1081.0(1)*        &     2.6(5)       &     999.35,  3$^-$        \\
               &         &  1327.6(2)*        &     1.0(2)       &     752.35,  4$^+$        \\
  2082.4(2)*   &         &  1330.0(2)*        &     1.9(4)       &     752.35,  4$^+$        \\
  2098.4(2)    &   6$^+$ &   410.2(3)*        &     0.4(2)       &    1687.85,  4$^+$        \\
               &         &   818.5(2)         &     3(1)         &    1279.80,  6$^+$        \\
               &         &   856.1(2)*        &     4(1)         &    1242.15,  5$^-$        \\
               &         &  1346.2(2)         &     0.5(2)       &     752.35,  4$^+$        \\
  2101.5(2)*   &         &  1079.4(1)*        &     2.1(4)       &    1023.00,  1$^-$        \\
               &         &  1102.4(1)*        &     2.5(5)       &     999.35,  3$^-$        \\
  2149.9(3)*   &(4$^+$,5$^+$)*& 638.4(2)*     &     0.5(2)       &    1511.45,  3$^+$        \\
  2155.7(2)*   &         &  1156.5(1)*        &     0.9(2)       &     999.35,  3$^-$        \\
               &         &  1403.1(1)*        &     1.0(2)       &     752.35,  4$^+$        \\
  2160.3(1)*   &(3$^+$,4$^+$)*& 649.0(2)*     &     0.5(2)       &    1511.45,  3$^+$        \\
               &         &   989.33(8)*       &     1.9(3)       &    1170.95,  2$^+$        \\
               &         &  1407.88(5)*       &     3.0(6)       &     752.35,  4$^+$        \\
  2192.7(2)*   &         &   509.7(2)*        &     0.7(2)       &    1683.37,  2$^+$        \\
               &         &  1169.4(2)*        &     1.1(2)       &    1023.00,  1$^-$        \\
  2198.1(1)*   &         &  1175.2(1)*        &     2.2(4)       &    1023.00,  1$^-$        \\
               &         &  1198.75(5)*       &     3.2(5)       &     999.35,  3$^-$        \\
  2223.7(1)*   &         &  1306.86(8)*       &     5(1)         &     916.80,  0$^+$        \\
  2237.86(8)   &         &  1066.91(5)        &     4.1(7)       &    1170.95,  2$^+$        \\
  2238.62(10)* &         &  1239.27(8)*       &     2.2(4)       &     999.35,  3$^-$        \\
  2270.8(2)*   &         &  1271.4(1)*        &     1.2(3)       &     999.35,  3$^-$        \\
  2344.0(2)*   &         &  1591.6(2)*        &     0.8(2)       &     752.35,  4$^+$        \\
  2348.6(1)*   &         &  1177.9(2)*        &     0.8(2)       &    1170.95,  2$^+$        \\
               &         &  1596.15(5)*       &     2.5(6)       &     752.35,  4$^+$        \\
  2357.3(2)*   &         &   697.8(1)*        &     0.9(2)       &    1659.50,  2$^+$        \\
  2359.8(3)*   &         &  1360.5(2)*        &     1.2(3)       &     999.35,  3$^-$        \\
  2381.05(7)*  &(3$^+$,4$^+$)*& 522.42(5)*    &     4(1)         &    1858.6,   2$^+$,3      \\
               &         &   869.35(5)*       &     3.8(8)       &    1511.45,  3$^+$        \\
               &         &  1132.7(2)*        &     6(1)         &    1248.73,  2$^+$        \\
               &         &  1210.17(5)*       &     4.4(7)       &    1170.95,  2$^+$        \\
  2397.6(2)*   &         &   752.1(2)*        &     2(1)         &    1645.3,  (1$^-$)       \\
               &         &  1227.0(2)*        &     1.4(2)       &    1170.95,  2$^+$        \\
  2399.2(4)*   &         &  1399.9(3)*        &     3.2(6)       &     999.35,  3$^-$        \\
  2404.8(2)*   &         &   721.4(1)*        &     1.3(3)       &    1683.37,  2$^+$        \\
  2406.06(5)   & (2$^+$)*&   547.6(1)*        &     1.0(3)       &    1858.6,   2$^+$,3      \\
               &         &   894.53(4)        &     8(1)$^a$     &    1511.45,  3$^+$        \\
               &         &  1157.40(3)        &    17(2)         &    1248.73,  2$^+$        \\
               &         &  1235.2(1)*        &     1.0(2)       &    1170.95,  2$^+$        \\
               &         &  1383.1(1)*        &     2.3(4)       &    1023.00,  1$^-$        \\
               &         &  1406.3(2)*        &     0.8(2)       &     999.35,  3$^-$        \\
  2431.6(1)    &   2$^+$ &   772.1(1)*        &     1.6(4)       &    1659.50,  2$^+$        \\
               &         &   919.9(1)         &     1.4(3)       &    1511.45,  3$^+$        \\
               &         &  1182.6(1)         &     3.7(4)       &    1248.73,  2$^+$        \\
               &         &  1260.60(5)        &     8(1)         &    1170.95,  2$^+$        \\
               &         &  2130.0(2)         &    11(2)         &     301.75,  2$^+$        \\
  2443.1(2)*   &         &  1443.8(1)*        &     1.7(3)       &     999.35,  3$^-$        \\
\hline
\end{tabular}
\end{center}
\label{table_Nd148_levels_II}
\end{table}

\begin{table}[]
\caption{ continuation of Table \ref{table_Nd148_levels_I}.}
\begin{center}
\begin{tabular}{l c r c r }

\hline
~~Initial      &level    &~~~${\gamma}$~~~~-&decay& Final level~~~                           \\
E$_{exc}$ (keV)&I$^{\pi}$&~~~E$_{\gamma}$(keV)& I$_{\gamma}$(rel)&E$_{exc}$ (keV), I$^{\pi}$ \\
\hline
  2444.8(3)*   &         &  1692.5(2)*        &     0.9(2)       &     752.35,  4$^+$        \\
  2454.3(2)*   &         &   942.8(1)*        &     0.9(2)       &    1511.45,  3$^+$        \\
  2478.9(2)*   &         &  1726.5(1)*        &     1.5(3)       &     752.35,  4$^+$        \\
  2480.4(2)*   &   1     &   835.3(2)*        &     1.5(5)       &    1645.3,  (1$^-$)       \\
               &         &  1563.4(2)*        &     1.5(3)       &     916.80,  0$^+$        \\
  2494.3(2)*   &         &  1251.1(1)*        &     3(1)         &    1242.15,  5$^-$        \\
  2495.35(15)* &         &  1324.4(1)*        &     0.8(2)       &    1170.95,  2$^+$        \\
  2499.2(3)*   &         &   987.7(2)*        &     0.5(2)       &    1511.45,  3$^+$        \\
  2534.8(3)*   &         &  1535.5(2)*        &     0.8(2)       &     999.35,  3$^-$        \\
  2542.2(1)*   &  (1)*   &   896.7(1)*        &     3(1)         &    1645.3,  (1$^-$)       \\
               &         &  1371.2(1)*        &     1.6(3)       &    1170.95,  2$^+$        \\
               &         &  1519.3(1)*        &     0.6(2)       &    1023.00,  1$^-$        \\
               &         &  1625.5(2)*        &     1.8(4)       &     916.80,  0$^+$        \\
  2552.7(1)*   &         &  1381.8(1)*        &     3.3(5)       &    1170.95,  2$^+$        \\
  2555.5(1)*   &         &   872.19(4)*       &    10(2)         &    1683.37,  2$^+$        \\
               &         &  1532.5(1)*        &     4.7(7)       &    1023.00,  1$^-$        \\
  2565.3(3)*   &         &  1812.9(2)*        &     0.5(2)       &     752.35,  4$^+$        \\
  2569.1(3)*   &         &  1816.7(2)*        &     1.0(2)       &     752.35,  4$^+$        \\
  2589.5(3)    &  (4)    &  1078.1(2)*        &     0.4(1)       &    1511.45,  3$^+$        \\
  2592.7(3)*   &         &  1593.4(2)*        &     1.0(2)       &     999.35,  3$^-$        \\
  2598.3(3)*   &         &  1681.5(3)*        &     1.3(3)       &     916.80,  0$^+$        \\
  2602.4(3)*   &         &  1850.3(2)*        &     0.9(2)       &     752.35,  4$^+$        \\
  2632.0(2)*   &         &  1632.6(1)*        &     0.7(2)       &     999.35,  3$^-$        \\
  2692.47(11)* &         &  1033.0(1)*        &     1.4(3)       &    1659.50,  2$^+$        \\
               &         &  1521.5(1)*        &     2.2(3)       &    1170.95,  2$^+$        \\
  2714.2(3)*   &         &  1961.8(2)*        &     0.9(2)       &     752.35,  4$^+$        \\
  2727.7(2)*   &  (1,2)* &   869.3(2)*        &     0.4(2)       &    1858.6,   2$^+$ ,3     \\
               &         &  1810.5(3)*        &     1.2(3)       &     916.80,  0$^+$        \\
  2731.1(3)    &         &  1978.8(3)*        &     1.5(3)       &     752.35,  4$^+$        \\
  2791.2(3)*   &         &  2038.8(3)*        &     2.0(5)       &     752.35,  4$^+$        \\
  2803.8(3)*   &         &  1780.8(2)*        &     0.8(2)       &    1023.00,  1$^-$        \\
  2855.6(3)*   &         &  1832.6(2)*        &     1.0()2       &    1023.00,  1$^-$        \\
  2864.7(2)*   &         &  1693.7(2)*        &     0.8(2)       &    1170.95,  2$^+$        \\
  2908.8(2)*   &  (1,2)* &  1992.0(2)*        &     1.6(3)       &     916.80,  0$^+$        \\
  2930.70(6)   & (2$^-$) &   498.9(1)*        &     1.6(5)       &    2431.6,   2$^+$        \\
               &         &   660.0(1)*        &     1.5(5)       &    2270.8                 \\
               &         &   770.5(2)*        &     0.6(2)       &    2160.3, (3$^+$,4$^+$)  \\
               &         &  1247.3(1)*        &     1.9(5)       &    1683.37,  2$^+$        \\
               &         &  1271.25(5)        &     8(2)$^a$     &    1659.50,  2$^+$        \\
               &         &  1419.20(5)        &     7(2)         &    1511.45,  3$^+$        \\
               &         &  1759.9(1)*        &     1.4(2)       &    1170.95,  2$^+$        \\
               &         &  1907.65(5)        &    15(2)         &    1023.00,  1$^-$        \\
               &         &  1931.35(5)        &    14(2)         &     999.35,  3$^-$        \\
               &         &  2629.0(1)         &     4(1)         &     301.75,  2$^+$        \\
  2979.0(2)*   &         &  1956.0(1)*        &     1.3(2)       &    1023.00,  1$^-$        \\
  2982.5(2)*   &  (1,2)* &  1299.3(1)*        &     0.9(3)       &    1683.37,  2$^+$        \\
               &         &  2065.4(2)*        &     1.9(4)       &     916.80,  0$^+$        \\
  2996.4(3)*   &         &  1997.1(3)*        &     0.5(2)       &     999.35,  3$^-$        \\
  3013.3(3)*   &         &  2014.0(2)*        &     0.8(2)       &     999.35,  3$^-$        \\
  3037.0(1)    & (2$^-$)*&  1377.8(1)*        &     1.9(4)       &    1659.50,  2$^+$        \\
               &         &  1787.9(2)*        &     1.5(3)       &    1248.73,  2$^+$        \\
               &         &  2014.2(1)*        &     4.6(7)       &    1023.00,  1$^-$        \\
               &         &  2037.7(1)*        &     0.2(1)       &     999.35,  3$^-$        \\
               &         &  2735.1(2)*        &     2.7(8)       &     301.75,  2$^+$        \\
  3042.6(2)*   &         &  2019.6(1)*        &     2.0(3)       &    1023.00,  1$^-$        \\
  3142.8(1)*   & (2$^+$)*&  1631.3(2)*        &     1.0(2)       &    1511.45,  3$^+$        \\
               &         &  2143.50(5)*       &     8(1)         &     999.35,  3$^-$        \\
  3192.4(2)*   &  (1)    &  2169.4(1)*        &     1.5(3)       &    1023.00,  1$^-$        \\
\hline
\end{tabular}
\end{center}
\label{table_Nd148_levels_III}
\end{table}

\begin{table}[]
\caption{ continuation of Table \ref{table_Nd148_levels_I}. All data listed in this Table are
new (not marked with asterisks).}
\begin{center}
\begin{tabular}{l c r c r }

\hline
~~Initial      &level    &~~~${\gamma}$~~~~-&decay& Final level~~~                           \\
E$_{exc}$ (keV)&I$^{\pi}$&~~~E$_{\gamma}$(keV)& I$_{\gamma}$(rel)&E$_{exc}$ (keV), I$^{\pi}$ \\
\hline
  3305.2(3)    &         &  2282.2(2)         &     0.7(2)       &    1023.00,  1$^-$        \\
  3374.4(2)    &         &  2374.3(1)         &     1.3(3)       &     999.35,  3$^-$        \\
  3386.8(2)    &         &  1703.5(2)         &     0.7(2)       &    1683.37,  2$^+$        \\
               &         &  1727.6(2)         &     0.8(2)       &    1659.50,  2$^+$        \\
               &         &  2387.5(2)         &     0.8(2)       &     999.35,  3$^-$        \\
  3326.9(3)    &         &  2303.9(2)         &     0.9(2)       &    1023.00,  1$^-$        \\
  3440.3(3)    &         &  1928.8(3)         &     0.3(1)       &    1511.45,  3$^+$        \\
  3473.4(2)    &         &  2474.4(1)         &     2.5(5)       &     999.35,  3$^-$        \\
  3454.3(1)    & (2$^-$) &  1771.0(1)         &     0.8(2)       &    1683.37,  2$^+$        \\
               &         &  2205.5(2)         &     1.8(3)       &    1248.73,  2$^+$        \\
               &         &  2431.2(1)         &     2.3(4)       &    1023.00,  1$^-$        \\
  3472.3(3)    &         &  1960.8(2)         &     0.5(2)       &    1511.45,  3$^+$        \\
  3474.7(3)    &         &  2450.7(2)         &     0.5(2)       &    1023.00,  1$^-$        \\
  3490.6(3)    &         &  1807.2(2)         &     0.4(1)       &    1683.37,  2$^+$        \\
  3498.8(3)    &         &  2475.8(2)         &     0.4(2)       &    1023.00,  1$^-$        \\
  3549.9(1)    & (2$^-$) &  1866.2(3)         &     0.5(2)       &    1683.37,  2$^+$        \\
               &         &  2379.0(2)         &     0.7(2)       &    1170.95,  2$^+$        \\
               &         &  2527.2(1)         &     1.1(2)       &    1023.00,  1$^-$        \\
  3568.6(2)    &         &  1885.2(1)         &     1.2(3)       &    1683.37,  2$^+$        \\
  3583.7(2)    &         &  2413.0(2)         &     0.8(2)       &    1170.95,  2$^+$        \\
               &         &  2584.1(2)         &     0.5(2)       &     999.35,  3$^-$        \\
  3751.8(2)    &         &  1680.9(1)         &     1.2(3)       &    2070.70, (2$^+$)       \\
               &         &  2068.6(1)         &     2.4(5)       &    1683.37,  2$^+$        \\
  3854.5(3)    &         &  2343.0(2)         &     0.4(1)       &    1511.45,  3$^+$        \\
  3900.5(3)    &         &  2901.2(2)         &     0.6(2)       &     999.35,  3$^-$        \\
  3928.9(3)    &         &  2905.9(2)         &     0.4(2)       &    1023.00,  1$^-$        \\
  3957.4(2)    & (2$^-$) &  1886.6(2)         &     0.8(3)       &    2070.70, (2$^+$)       \\
               &         &  2273.8(2)         &     0.4(1)       &    1683.37,  2$^+$        \\
               &         &  2786.6(2)         &     0.5(2)       &    1170.95,  2$^+$        \\
  3999.6(1)*   & (2$^-$) &  1928.6(3)         &     0.8(3)       &    2070.70, (2$^+$)       \\
               &         &  2316.1(1)         &     3.3(6)       &    1683.37,  2$^+$        \\
               &         &  2828.8(2)         &     1.6(3)       &    1170.95,  2$^+$        \\
  4004.5(3)    &         &  3005.2(2)         &     1.2(3)       &     999.35,  3$^-$        \\
  4018.6(3)    &         &  2335.2(2)         &     0.6(2)       &    1683.37,  2$^+$        \\
  4046.2(3)    &         &  2386.7(2)         &     1.0(3)       &    1659.50,  2$^+$        \\
  4063.3(1)    & (2$^-$) &  1992.5(1)         &     1.1(3)       &    2070.70, (2$^+$)       \\
               &         &  2380.0(1)         &     2.4(5)       &    1683.37,  2$^+$        \\
               &         &  2552.0(2)         &     0.7(2)       &    1511.45,  3$^+$        \\
               &         &  3040.1(2)         &     0.8(2)       &    1023.00,  1$^-$        \\
  4074.7(1)    & (2$^-$) &  2003.9(1)         &     0.4(2)       &    2070.70, (2$^+$)       \\
               &         &  2391.40(4)        &     6(1)         &    1683.37,  2$^+$        \\
               &         &  3051.8(1)         &     2.1(3)       &    1023.00,  1$^-$        \\
  4099.5(2)    & (2$^-$) &  2241.0(2)         &     0.4(2)       &    1858.6,   2$^+$,3      \\
               &         &  2416.2(1)         &     1.7(4)       &    1683.37,  2$^+$        \\
               &         &  2439.8(2)         &     0.9(3)       &    1659.50,  2$^+$        \\
               &         &  3100.3(2)         &     0.7(2)       &     999.35,  3$^-$        \\
  4100.8(2)    &         &  3077.8(1)         &     1.1(2)       &    1023.00,  1$^-$        \\
  4121.2(2)    &         &  3098.4(2)         &     0.8(2)       &    1023.00,  1$^-$        \\
               &         &  3121.6(2)         &     0.3(1)       &     999.35,  3$^-$        \\
  4138.6(3)    &         &  3139.3(2)         &     0.2(1)       &     999.35,  3$^-$        \\
  4204.3(2)    &         &  2133.6(1)         &     0.5(2)       &    2070.70, (2$^+$)       \\
  4237.5(3)    &         &  2578.0(2)         &     0.3(1)       &    1659.50,  2$^+$        \\
  4242.5(3)    &         &  2559.1(2)         &     0.4(1)       &    1683.37,  2$^+$        \\
  4250.5(3)    &         &  2567.1(2)         &     0.9(2)       &    1683.37,  2$^+$        \\
\hline
\end{tabular}
\end{center}
\label{table_Nd148_levels_IV}
\end{table}

\begin{table}[]
\caption{Angular correlation results for $\gamma \gamma$ cascades in $^{148}$Nd populated
in $\beta^-$ decay of $^{148}$Pr, as observed in the present work. Superscript ``m'' indicates
mixed transition for which $\delta_{exp}$ value, shown in the last column, is determined with
the corresponding $\chi^2$ of the fit.}
\begin{center}
\begin{tabular}{ c c c c }
\hline
E${\gamma_1}$ - E${\gamma_2}$& $A_2/A_0$  & $A_4/A_0$  &     Spins in cascade        \\
      (keV)                  &   exp.     &    exp.    & and $\delta_{exp}$; $\chi^2$\\
\hline
  247.05$^m$-450.60          & 0.037(27) & -0.053(41)  & 3 - 4 - 2;                  \\
                             &           &             &+0.21(5); 1.2                \\
  450.60-301.75              & 0.106(18) &  0.007(29)  & 4 - 2 - 0;                  \\
                             &           &             & $\chi^2$=0.1                \\
  489.80$^m$-450.60          &-0.189(73) & -0.091(109) & 5 - 4 - 2;                  \\
                             &           &             &+0.17(12); 0.7               \\
  527.45-450.60              & 0.101(30) & -0.011(43)  & 6 - 4 - 2;                  \\
                             &           &             & $\chi^2$=0.2                \\
  697.55$^m$-301.75          &-0.054(18) &  0.038(48)  & 3 - 2 - 0                   \\
                             &           &             &+0.023(24); 0.7              \\
  721.25$^m$-301.75          &-0.193(24) & -0.026(34)  & 1 - 2 - 0                   \\
                             &           &             &-0.05(2); 0.5                \\
  872.19-1381.62             & 0.42(14)  & -0.03(18)   &                             \\
  869.16$^m$-301.75          &-0.144(34) &  0.160(52)  & 2 - 2 - 0                   \\
                             &           &             &+0.60(7); 2.1                \\
  902.66-869.16              & 0.115(69) & -0.052(103) &                             \\
  946.93$^m$-301.75          & 0.001(50) &  0.394(73)  & 2 - 2 - 0                   \\
                             &           &             &-10(-19,+4); 0.9             \\
 1106.20$^m$-450.60          & 0.01(8)   & -0.18(12)   & 3 - 4 - 2;                  \\
                             &           &             &+35(-27,+Inf; 0.2            \\
 1129.30$^m$-450.60          &-0.06(12)  & -0.08(17)   & 5 - 4 - 2;                  \\
                             &           &             &+10(-6,+Inf); 0.1            \\
                             &           &             &-0.02(18); 0.2               \\
 1209.70$^m$-301.75          &-0.091(27) & -0.077(42)  & 3 - 2 - 0                   \\
                             &           &             &+6(-1,+2); 0.1               \\
 1260.60-869.16              & 0.24(13)  & -0.19(20)   &                             \\
 1343.7$^m$-301.75           & 0.22(15)  &  0.14(24)   & 1 - 2 - 0                   \\
                             &           &             &-0.38(15); 1.3               \\
 1357.80$^m$-301.75          & 0.144(35) & -0.049(55)  & 2 - 2 - 0                   \\
                             &           &             &+0.14(5); 1.0                \\
 1381.62$^m$-301.75          &-0.154(43) &  0.071(61)  & 2 - 2 - 0                   \\
                             &           &             &+0.58(9); 0.1                \\
 1557.0$^m$-301.75           &-0.32(15)  & -0.11(23)   & 2 - 2 - 0                   \\
                             &           &             &+0.9(-3,+99); 1.4            \\
                             &           &             & 3 - 2 - 0                   \\
                             &           &             &-6(-99,+3); 0.1              \\
                             &           &             &-0.4(3); 0.2                 \\
 2130.0$^m$-301.75           & 0.193(76) &  0.100(108) & 2 - 2 - 0                   \\
                             &           &             &+0.08(10); 0.8               \\
 2629.0$^m$-301.75           &-0.11(15)  & -0.11(3)    & 1 - 2 - 0                   \\
                             &           &             &-0.13(13); 0.2               \\
                             &           &             & 2 - 2 - 0                   \\
                             &           &             &+0.45(30); 0.6               \\
\hline
\end{tabular}
\end{center}
\label{table_Nd148_ang}
\end{table}

\begin{figure*}
\centering
\scalebox{.84}{\includegraphics{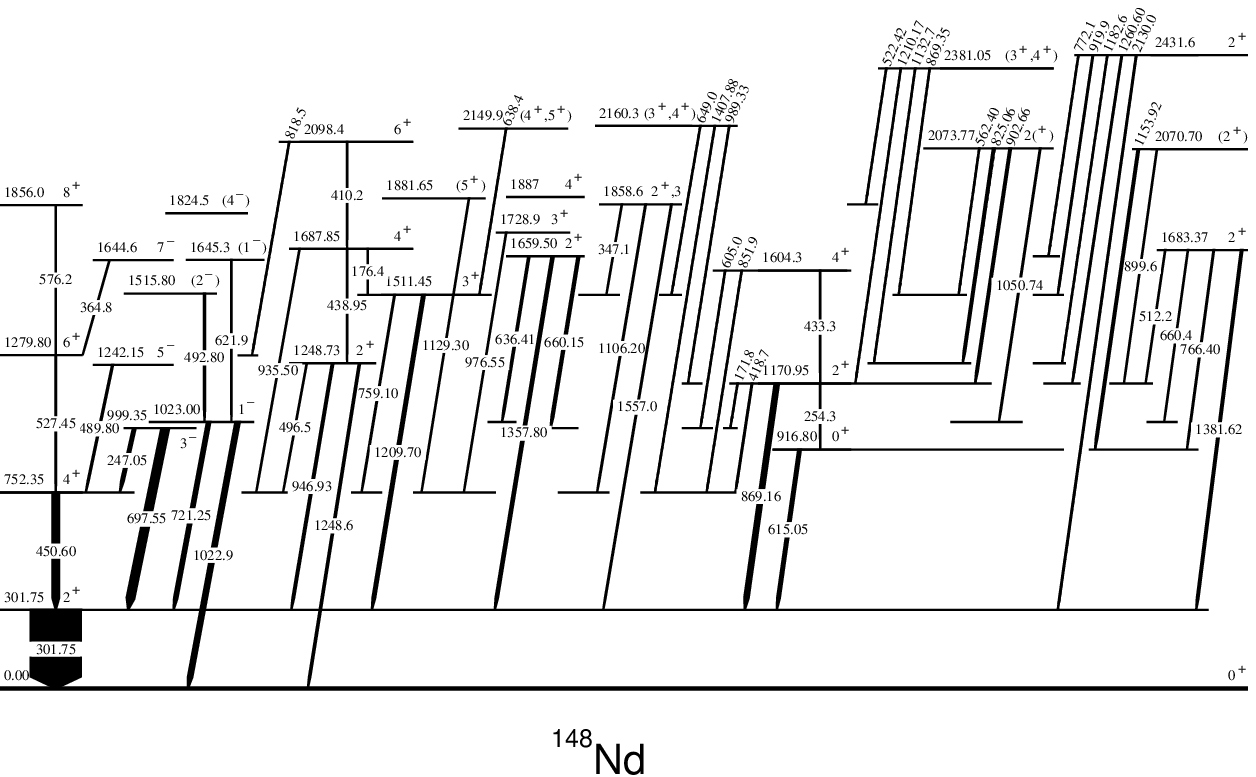}}
\caption{Partial level scheme of $^{148}$Nd populated in $\beta^-$ decay of $^{148}$Pr, as
         observed in the present work. Only most important levels and transitions are shown. See
         Tables \ref{table_Nd148_levels_I} - \ref{table_Nd148_levels_IV}  for all excited states
         and $\gamma$ decays in $^{148}$Nd, observed in the present work.}
\label{A150_Nd_148_scheme}
\end{figure*}

In our measurement $^{148}$Nd is populated predominantly in $\beta^-$ decay of $^{148}$Pr, as
seen in Fig. \ref{A150_Nd_population}. Figure \ref{A150_Nd_148_scheme} shows partial scheme
of $^{148}$Nd levels with the most important excitations, which are discussed in the text.

All levels and their $\gamma$ decays in $^{148}$Nd observed in the present work are listed in
Tables \ref{table_Nd148_levels_I} - \ref{table_Nd148_levels_IV}. They are determined using
double- and triple-$\gamma$ coincidences. Transitions reported in the compilation \cite{NDS148},
which are feeding the ground state but are not in cascade with other transitions, were not
analysed and are not shown in the tables. Previous $\beta^-$ decay results for $^{148}$Pr
\cite{Wal86,Kar88} are extended in the present work by 98 new excited levels and 175 new
$\gamma$ transitions. New data are marked by asterisks in Tables \ref{table_Nd148_levels_I} - \ref{table_Nd148_levels_III}. All levels and transitions in Table \ref{table_Nd148_levels_IV}
are newly observed.

The 2182.2-, 2545.0- an 3129.9-keV levels reported previously in $^{148}$Nd from $\beta^-$ decay
of $^{148}$Pr \cite{Wal86,Kar88} are not confirmed  and the 522.2-, 1156.6-,1521.8- and 2106.7-keV
transitions are placed differently in the level scheme.

The quasi-rotational ground-state band in $^{148}$Nd was observed up to medium spins in Refs.
\cite{Urb88,Ibo97}. In $\beta^-$ decay it is populated up to spin I=6 \cite{NDS148}. The
observation of the 8$^+$ level at 1856.0 keV in the present work indicates that in our data
$^{148}$Nd is also weakly populated in prompt-$\gamma$ fission. However, this population is on
a percent level, only, and the $\gamma$ intensities shown in Tables \ref{table_Nd148_levels_I} -
\ref{table_Nd148_levels_IV} reflect the population in $\beta^-$ decay, except for the 1279.80-,
1644.6-, 1856.0- and 2098.4-keV levels.

For the strongest $\gamma\gamma$ cascades we have determined angular correlations. Using the results
listed in Table \ref{table_Nd148_ang}, the observed decay branchings and the previous and new
$log ft$ estimates we confirm previously reported and propose several new spin-parity assignments to
levels in $^{148}$Nd, as shown in Tables. \ref{table_Nd148_levels_I} - \ref{table_Nd148_levels_IV}.

An important new results is the observation of the 254.3-keV decay branch from the 2$^+$ level at
1170.95 keV the to 0$^+$ level at 916.80 keV and the 433.3-keV decay branch from the 4$^+$
level at 1604.3 keV. The new in-band transitions firmly establish the rotational band on top of
the 0$^+_2$, 916.80-keV level. The newly obtained branching ratio, together with the known
T$_{1/2}$=1.4(1) ps of the 1170.95-keV level \cite{NDS148}, allowed to estimate the B(E2)=40(15)
W.u. rate for the 254.3-keV, in-band transition, indicating its collective character.

Other important observations concern the build-up of $\gamma$ collectivity in the Nd chain. The
spin of the 1687.85-keV level is not I=5 reported in \cite{NDS148} because of its newly observed,
438.95-keV decay to the 2$^+$ level at 1248.73 keV. For the 1687.85-keV level we propose
spin-parity I$^{\pi}$=4$^+$. This is further supported by its new, 176.4-keV decay to the 3$^+$
level at 1511.45 keV. We also found that the 1683.37-keV level, reported previously as the
I$^{\pi}$=(4$^+$) member of $\gamma$ band in $^{148}$Nd has spin-parity 2$^+$ because of its
766.40-keV decay to the 0$^+_2$ level at 916.80 keV.

The 2149.0(6)-keV level proposed to be the 6$^+$ member of the 0$^+_2$ band is not seen in our
work. Instead we observe the 2149.9(3)-keV level, for which tentative spin-parity I$^{\pi}$=
(4$^+$,5$^+$) is proposed becaues of its decay to the 3$^+$ level at 1511.45 keV.  The revised
$\gamma$ band in $^{148}$Nd proposed in the present work comprises the 1248.73-, 1511.45-,
1687.85-, 1881.65- and 2098.4-keV levels as the 2$^+$, 3$^+$, 4$^+$, (5$^+$) and 6$^+$ band
members, respectively.

The 1659.50-, 1728.9-, 1887- and 2149.9-keV levels are candidates for 2$^+$, 3$^+$, 4$^+$ and
5$^+$ members of the second $\gamma$ band in $^{148}$Nd, as discussed later in the
text (one notes significant difference between energy of the 1659.50(5)-keV level determined in
the present work and the 1659.92(5)-keV energy reported previously \cite{NDS148}). The 1887-keV
level is drawn in Fig. \ref{A150_Nd_148_scheme} after Ref. \cite{NDS148}. It is not observed in
our work.

The 2073.77- and 2381.05-keV levels are candidates for members of a $\gamma$ band build on top
of the 0$^+_2$ level at 916.80 keV, considering their decay branches.

Angular correlations reject spin I=4 for the 1683.37-keV level and indicate spin I=2. Positive
parity is proposed becaues of the 766.40-keV decay to the 0$^+$ level at 916.80 keV. Together with
the 1577(2)-keV, 2$^+$level, reported in Ref. \cite{NDS148} (not seen in the present work) they
are candidates for spherical, seniority-type excitations, as discussed in Sec. III.A.1.

The 1515.80-, 1645.3- and 1824.5-keV levels are probably due to octupole collectivity. Excitation
energies of the 1515.80- and 1824.5-keV levels fit that expected for the 2$^-$ and 4$^-$
excitations, respectively. The observed decay and feeding of the 1645.3-keV level suggests its
(1$^-$) spin parity.

\subsection{Results for $^{150}$Nd}

Low spin excitations in the $^{150}$Nd nucleus were previously measured in $\beta^-$ decay in Ref. \cite{Fog86,Kar88} and in numerous reactions listed in the compilation \cite{NDS150} whereas medium
spin states were observed in fission of $^{252}$Cf \cite{Zhu95} and Coulomb excitations
\cite{Kru02}.

\begin{figure*}
\centering
\scalebox{.87}{\includegraphics{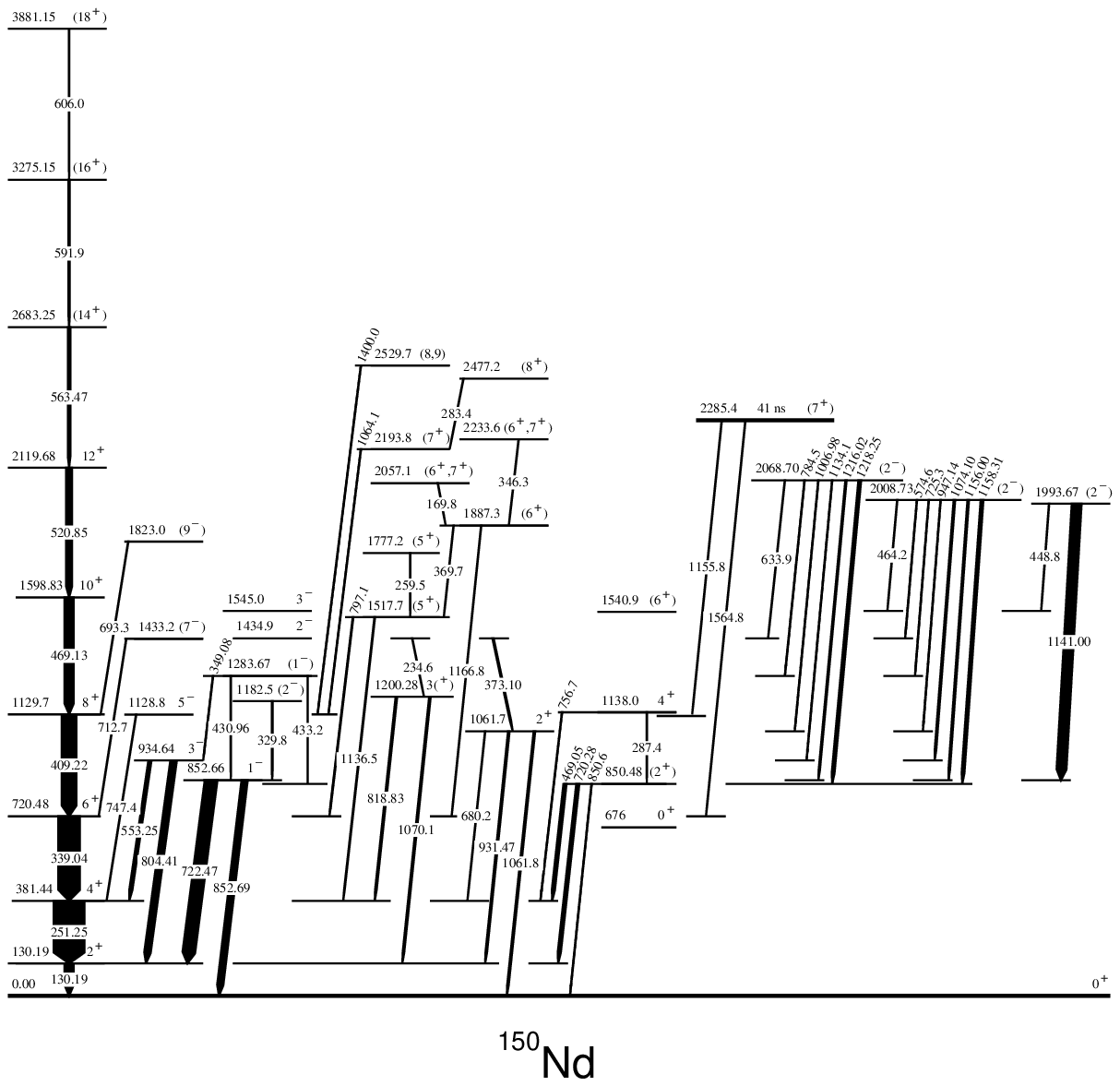}}
\caption{Partial excitation scheme of $^{150}$Nd, as observed in the present work.}
\label{A150_Nd_150_scheme}
\end{figure*}

In our data the $^{150}$Nd nucleus is strongly populated in fission of $^{252}$Cf, though we
could also see most of the levels reported in $\beta^-$ decay of $^{150}$Pr \cite{Fog86,Kar88}.
Figure \ref{A150_Nd_150_scheme} display partial excitation scheme of $^{150}$Nd observed in the
present work showing excitations which are discussed in the text. The 676- and 1540.9-keV levels
are shown in the scheme after Ref. \cite{NDS150} to assist the discussion.

All levels and their $\gamma$ decays in $^{150}$Nd observed in the present work are listed in
Tables \ref{table_Nd150_levels_I} and \ref{table_Nd150_levels_II}. They have been determined using
double- and triple-$\gamma$ coincidences. Transitions reported in the compilation \cite{NDS150},
which are feeding the ground state but are not in coincidence with other transitions, were not
analysed and are not shown in the tables. Previously reported results for $^{150}$Nd \cite{NDS150}
are extended in the present work by 17 new excited states and 25 new $\gamma$ transitions as marked
in Tables \ref{table_Nd150_levels_I} and \ref{table_Nd150_levels_II}.

For the strongest $\gamma\gamma$ cascades in $^{150}$Nd we have analysed angular correlations. Using
the results from Table \ref{table_Nd150_ang}, the observed decay branchings and the previous $log~ft$
estimates \cite{NDS150} we confirmed previous and proposed several new spin-parity assignments to
levels in $^{150}$Nd, as shown in Tables \ref{table_Nd150_levels_I} and \ref{table_Nd150_levels_II}.

\begin{table}[]
\caption{Excited levels and their $\gamma$ decays in $^{150}$Nd populated following fission of
         $^{252}$Cf, as observed in the present work. New data are indicated by the star symbol.}
\begin{center}
\begin{tabular}{l c r c r }
\hline
~~Initial      &level    &~~~${\gamma}$~~~~-&decay& Final level~~~                           \\
E$_{exc}$ (keV)&I$^{\pi}$&~~~E$_{\gamma}$(keV)& I$_{\gamma}$(rel)&E$_{exc}$ (keV), I$^{\pi}$ \\
\hline
   130.19(5)   &   2$^+$ &   130.19(5)        & $\geq$110 &     0.00,  0$^+$           \\
   381.44(6)   &   4$^+$ &   251.25(5)        & 100(5)    &   130.19,  2$^+$           \\
   720.48(7)   &   6$^+$ &   339.04(3)        &  70(5)    &   381.44,  4$^+$           \\
   850.48(6)   &   2$^+$ &   469.05(3)        &  15(3)    &   381.44,  4$^+$           \\
               &         &   720.28(6)        &  22(3)    &   130.19,  2$^+$           \\
               &         &   850.6(2)         &   3(1)    &     0.00,  0$^+$           \\
   852.66(7)   &   1$^-$ &   722.47(3)        &  41(4)    &   130.19,  2$^+$           \\
               &         &   852.69(5)        &  20(3)    &     0.00,  0$^+$           \\
   934.64(5)   &   3$^-$ &   553.25(3)        &  32(3)    &   381.44,  4$^+$           \\
               &         &   804.41(3)        &  21(3)    &   130.19,  2$^+$           \\
  1061.7(1)    &   2$^+$ &   680.2(1)         &   6(2)    &   381.44,  4$^+$           \\
               &         &   931.47(5)        &  19(3)    &   130.19,  2$^+$           \\
               &         &  1061.8(1)         &   5(2)    &     0.00,  0$^+$           \\
  1128.8(3)    &   5$^-$ &   747.4(2)         &   3(1)    &   381.44,  4$^+$           \\
  1129.7(1)    &   8$^+$ &   409.22(5)        &  48(3)    &   720.48,  6$^+$           \\
  1138.0(2)    &   4$^+$ &   287.4(2)         &   0.2(1)  &   850.48,  2$^+$           \\
               &         &   756.7(1)         &   4(1)    &   381.44,  4$^+$           \\
  1182.5(2)    &  (2$^-$)&   329.8(1)         &   5(1)    &   852.66,  1$^-$           \\
  1200.28(8)   &3$^{(+)}$&   818.83(5)        &   5(1)    &   381.44,  4$^+$           \\
               &         &  1070.1(2)         &  10(2)    &   130.19,  2$^+$           \\
  1283.67(6)   &  (1$^-$)&   349.08(5)        &   6(2)    &   934.64,  3$^-$           \\
               &         &   430.96(5)        &   7(1)    &   852.66,  1$^-$           \\
               &         &   433.2(1)         &   2.0(6)  &   850.48,  2$^+$           \\
  1433.2(3)    &  (7$^-$)&   712.7(2)         &   3(1)    &   720.48,  6$^+$           \\
  1434.9(1)    &   2$^-$ &   234.6(1)         &   0.6(2)  &  1200.28, 3$^{(+)}$        \\
               &         &   373.10(6)        &   2.2(7)  &  1061.7,   2$^+$           \\
  1483.5(2)    &   3$^-$ &   283.2(2)         &   0.5(2)  &  1200.28, 3$^{(+)}$        \\
               &         &  1102.0(2)         &   6(1)    &   381.44,  4$^+$           \\
  1517.7(2)    & (5$^+$)*&   797.1(2)         &   3(1)    &   720.48,  6$^+$           \\
               &         &  1136.5(2)         &   7(2)    &   381.44,  4$^+$           \\
  1545.0(2)    &   3$^-$ &  1414.8(1)         &   5(1)    &   130.19,  2$^+$           \\
  1565.5(3)    &   4$^-$ &   365.2(2)         &   0.7(2)  &  1200.28,  3$^{(+)}$       \\
  1578.8(2)*   &         &   726.0(1)*        &   1.5(4)  &   852.66,  1$^-$           \\
  1579.8(2)    &  3$^-$  &  1198.4(1)         &   6(1)    &   381.44,  4$^+$           \\
  1598.83(10)  &  10$^+$ &   469.13(5)        &  30(3)    &  1129.7,   8$^+$           \\
  1738.2(3)    &         &   803.6(2)*        &   3(1)    &   934.64,  3$^-$           \\
               &         & (1608)             &           &   130.19,  2$^+$           \\
  1777.2(3)    & (5$^+$)*&   259.5(2)*        &   2(1)    &  1517.7,  (5$^+$)          \\
  1823.0(3)*   & (9$^-$)*&   693.3(2)*        &   2(1)    &  1129.7,   8$^+$           \\
  1887.3(2)*   & (6$^+$) &   369.7(2)*        &   6(2)    &  1517.7,  (5$^+$)          \\
               &         &  1166.8(1)*        &   4(1)    &   720.48,  6$^+$           \\
  1993.67(7)   & (2$^-$)*&   448.8(1)*        &   6(2)    &  1545.0    3$^-$           \\
               &         &  1141.00(5)        &  32(4)    &   852.66,  1$^-$           \\
  2004.3(3)*   &         &   804.0(1)*        &   1.1(3)  &  1200.28,  3$^{(+)}$       \\
  2008.73(7)   & (2$^-$)*&   464.2(1)*        &   5(2)    &  1545.0,   3$^-$           \\
               &         &   574.2(3)*        &   1.2(3)  &  1434.9,   2$^-$          \\
               &         &   725.5(4)*        &   1.5(4)  &  1283.67, (1$^-$)          \\
               &         &   947.14(4)        &   3(1)    &  1061.7,   2$^+$           \\
               &         &  1074.10(3)        &  17(2)    &   934.64,  3$^-$           \\
               &         &  1156.00(9)        &   4(1)    &   852.66,  1$^-$           \\
               &         &  1158.31(5)        &  10(2)    &   850.48,  2$^+$           \\
  2057.1(3)*   &(6$^+$,7$^+$)*&169.8(2)*      &   2.4(6)  &  1887.3,  (6$^+$)          \\
\hline
\end{tabular}
\end{center}
\label{table_Nd150_levels_I}
\end{table}

\begin{table}[]
\caption{ continuation of Table \ref{table_Nd150_levels_I}.}
\begin{center}
\begin{tabular}{l c r c r }
\hline
~~Initial      &level    &~~~${\gamma}$~~~~-&decay& Final level~~~                           \\
E$_{exc}$ (keV)&I$^{\pi}$&~~~E$_{\gamma}$(keV)& I$_{\gamma}$(rel)&E$_{exc}$ (keV), I$^{\pi}$ \\
\hline
  2068.70(5)   & (2$^-$)*&   633.9(2)*        &   2.1(5)  &  1434.9,   2$^-$          \\
               &         &   784.5(2)*        &   1.0(3)  &  1283.67, (1$^-$)          \\
               &         &  1006.98(3)        &   4(1)    &  1061.7,   2$^+$           \\
               &         &  1134.1(1)*        &   2.4(5)  &   934.64,  3$^-$           \\
               &         &  1216.02(5)        &  12(2)    &   852.66,  1$^-$           \\
               &         &  1218.25(8)        &   8(2)    &   850.48,  2$^+$           \\
  2119.68(12)  &  12$^+$ &   520.85(5)        &  19(2)    &  1598.83, 10$^+$           \\
  2193.8(3)*   & (7$^+$)*&  1064.1(2)*        &   0.8(3)  &  1129.7,   8$^+$           \\
  2233.6(3)*   &(6$^+$,7$^+$)*&346.3(2)*      &   3(1)    &  1887.3,  (6$^+$)          \\
  2285.4(2)*   & (7$^+$)*&  1155.8(1)*        &   0.9(3)  &  1129.7,   8$^+$           \\
               &         &  1564.8(2)*        &   1.1(3)  &   720.48,  6$^+$           \\
  2340.9(3)*   &         &  1279.2(2)*        &   0.5(2)  &  1061.7,   2$^+$           \\
  2360.9(3)*   &         &  1508.2(2)*        &   0.9(3)  &   852.66,  1$^-$           \\
  2477.2(4)*   & (8$^+$)*&   283.4(2)*        &   0.5(2)  &  2193.8,  (7$^+$)          \\
  2503.5(3)*   &         &  1302.4(2)*        &   0.7(2)  &  1200.28, 3$^{(+)}$        \\
               &         &  1441.5(2)*        &   0.7(2)  &  1061.7 ,  2$^+$           \\
  2529.7(3)*   & (8,9)   &  1400.0(2)*        &   1.6(4)  &  1129.7,   8$^+$           \\
  2683.25(14)  &  14$^+$ &   563.47(7)        &   8(1)    &  2119.68, 12$^+$           \\
  2690.0(3)*   &         &  1837.3(2)         &   1.3(4)  &   852.66,  1$^-$           \\
  2716.4(2)*   &         &  1781.8(1)         &   1.8(4)  &   934.64,  3$^-$           \\
  2817.2(3)*   &         &  1964.5(2)*        &   0.8(3)  &   852.66,  1$^-$           \\
  3275.15(16)  & (16$^+$)&   591.9(1)         &   2.9(5)  &  2683.25, 14$^+$           \\
  3881.15(16)* &(18$^+$)*&   606.0(1)         &   1.5(5)  &  3275.15,(16$^+$)          \\

\hline
\end{tabular}
\end{center}
\label{table_Nd150_levels_II}
\end{table}

\begin{table}[h]
\caption{Angular correlation results for $\gamma \gamma$ cascades in $^{150}$Nd populated
in fission of $^{252}$Cf.}
\begin{center}
\begin{tabular}{ c c c c }
\hline
E${\gamma_1}$ - E${\gamma_2}$& $A_2/A_0$ & $A_4/A_0$  &     Spins in cascade         \\
      (keV)                  &   exp.    &    exp.    & and $\delta_{exp}$; $\chi^2$ \\
\hline
  339.04-251.25              & 0.104(14) & 0.027(22)   & 6-4-2                       \\
                             &           &             &   $\chi^2$=0.7              \\
  469.13-409.22              & 0.106(18) & 0.011(27)   & 10-8-6                      \\
                             &           &             &   $\chi^2$=0.6              \\
  520.85-469.13              & 0.087(32) & -0.051(56)  & 12-10-8                     \\
                             &           &             &   $\chi^2$=0.9              \\
\hline
\end{tabular}
\end{center}
\label{table_Nd150_ang}
\end{table}

The ground-state band is extended up to spin (18$^+$). The 381.44(6) keV energy of the 4$^+_1$ level
found in the present agrees with the 381.46(11)-keV value reported in Refs. \cite{Fog86,Kar88} but
is inconsistent with the Adopted ENSDF value of 381.10(8) \cite{NDS150} (the 251.25(5)-keV energy of
the 4$^+ \rightarrow 2^+$ transition was mistakenly reported as 250.25(5) keV in Table I of Ref.
\cite{Urb20}).

The new level at 1823.0 keV is a candidate for the 9$^-$ member of the octupole band in $^{150}$Nd.
The 1182.5-keV level, strongly populated in $\beta^-$ decay of the (1)$^-$ g.s. of
$^{150}$Pr, is similar to the (2$^-$), 1515.80-keV level in $^{148}$Nd.

The newly observed levels at 1517.7, 1777.2, 1887.3, 2057.1, 2193.8, 2233.6, 2477.2 and 2529.7 keV
are candidates for members of $\gamma$ bands. The 4$^+$, 1435.16-keV level reported in $\beta^-$
decay \cite{Fog86,Kar88} is shown as 2$^-$, 1435.03(9)-keV in the compilation \cite{NDS150}. Our
energy is 1434.9(1) keV and we propose tentative, (4$^+$) spin-parity to this level.

The (6$^+$), 1540.9-keV member of the 0$^+_2$ band could not be confirmed. Its 1159.8-keV
decay \cite{NDS150} is masked by the strong, 1158.31-keV decay of the 2008.73-keV level.

We do not confirm the 1911.5- and 1967.5-keV levels reported previously \cite{Fog86,Kar88}. The
1781.8- and 1837.3-keV transitions are moved to higher locations and define new levels at 2716.4
and 2690.0 keV, respectively.

The 2$^+$ spin-parity of the 2069.21-keV level has been proposed because of the 634.1-keV decay to
the 4$^+$ level at 1435.16 keV \cite{Fog86,Kar88}. This is inconsistent with the $log ft$=5.7
reported for this level \cite{Fog86,Kar88}. The compilation \cite{NDS150} reports
2$^-$ spin-parity for the 1435.03-keV level and still 2$^+$ spin-parity for the 2069.12-keV
level (2068.70-keV in the present work). Our data are consistent with spin-parity 2$^-$ for the
2068.70-keV level.

The new 803.6(2)-keV transition feeding the 3$^-$ level at 934.64 keV defines a level at 1738.2(3)
keV. If this is the same level as the 0$^+$, 1738.3(4)-keV one reported in the compilation
\cite{NDS150} then its 0$^+$ spin-parity can be questioned. In our data the 1608-keV decay of this
level is strongly contaminated.

\begin{figure}
\centering
\scalebox{.5}{\includegraphics{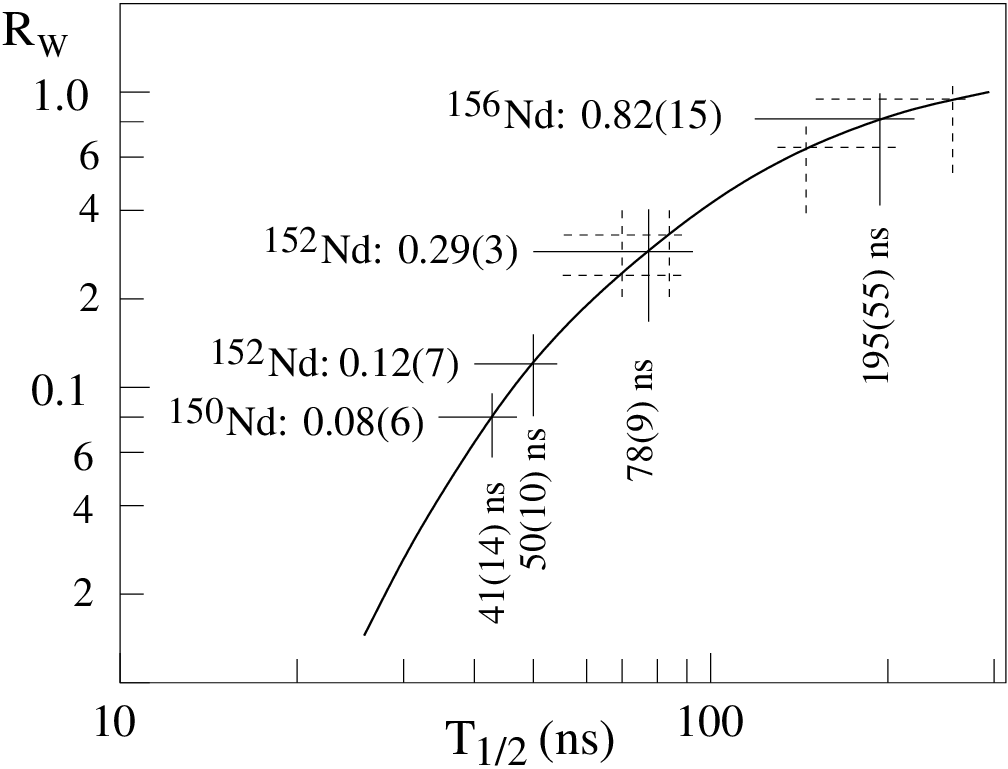}}
\caption{Half lives analysis of isomers in Nd isotopes using the R$_W$ techniques \cite{Urb09b}.
         See text for further explanations.}
\label{A150_Nd_150_Rw_time}
\end{figure}

Analysis of triple-$\gamma$ coincidences sorted with time-delayed windows revealed new isomeric
state at 2285.4 keV in $^{150}$Nd. Its weak population does not allow any sufficient time-delayed
spectrum to be constructed. To determine its half life we employed the technique described in Ref.
\cite{Urb09b}, analysing the intensity ratio, R$_W$, of triple-$\gamma$ cascades deexciting an
isomer, observed in two 3D $\gamma\gamma\gamma$ histograms sorted with different time-delayed
window, one sorted using a 170 ns wide window extending from 40 ns to 210 ns and the other sorted
using 340 ns wide window extending from 210 ns to 550 ns after time ``0'' \cite{Urb09b}. There is
a systematic error due to random coincidences, which are higher in the longer of the two time
windows. Therefore a correction factor of 0.9(1) was estimated by adjusting to the well known,
T$_{1/2}$=164.1(9) ns isomer at 1691.34 keV in $^{134}$Te \cite{NDS103}.

For the 2285.4-keV isomer in $^{150}$Nd  the analysis provided the ratio R$_W$=0.08(6) yielding
half life T$_{1/2}$=41(14) ns, as illustrated in Fig. \ref{A150_Nd_150_Rw_time}. Decays to the
6$^+$ and 8$^+$ levels suggests spin-parity (7$^+$) for the 2285.4-keV isomer.

We also reevaluated the rather imprecise, T$_{1/2}$=365(145) half life of the 1431.2-keV
isomer in $^{156}$Nd \cite{Sim09} applying the correction factor and the improved analysis proposed
in Ref. \cite{Urb09b}, which imposes an extra requirement that the delayed triple-$\gamma$ cascade
is observed within 30 ns. The improved half life of the 1431.2-keV isomer in $^{156}$Nd is
T$_{1/2}$=195(55) ns, as illustrated in Fig. \ref{A150_Nd_150_Rw_time}. The new value overlaps with
the 135(40) ns half life reported in Ref. \cite{Gau98} for this isomer.

\subsection{Results for $^{152}$Nd}

The $^{152}$Nd nucleus is observed in our data predominantly in prompt-$\gamma$ fission.
Previous prompt-$\gamma$ studies of $^{152}$Nd reported, among others, a 2-q.p. isomeric state
at 2241 keV \cite{Gau98} and new negative-parity levels \cite{Zha98}, which were studied later
in Ref. \cite{Yeo10}.

The $^{152}$Nd nucleus is also populated in $\beta^-$ decay of $^{152}$Pr produced in fission
of $^{252}$Cf (about one third of the intensity of the 164.15-keV transition in $^{152}$Nd is
due to $\beta^-$ decay). Previous $\beta^-$ decay studies of $^{152}$Nd agree on excitation
energies but report different spin-parity assignments, resulting from different spin-parity of
the ground state of $^{152}$Pr, reported as 4$^{\pm}$ in Ref. \cite{Kar88}, 4$^-$ in Refs.
\cite{Hel92,Toh98}, 4$^+$ in Ref. \cite{Tag90,Yam95,Toh99} and (3$^+$) in Ref. \cite{Ale18}.
Consequently, the 1826.8-keV level in $^{152}$Nd, populated by Gamow-Teller transition, was
reported with spin-parity 3$^-$ in Ref. \cite{Hel92} and 3$^+$ in Ref. \cite{Toh99}. This level
is expected to have a dominating 2-q.p. contribution related to that of the ground state of
$^{152}$Pr and may provide useful information on neutron and proton orbitals near the Fermi
surface once its spin-parity is firmly determined. We note that the advanced $\beta^-$ decay
measurement \cite{Toh99} reported 0.5$\%$ and 15.5$\%$ decay branches to the 2$^+_1$ and 4$^+_1$,
levels in $^{152}$Nd, respectively, which strongly supports spin I=4 for the ground state of
$^{152}$Pr.

\begin{figure*}
\centering
\scalebox{.84}{\includegraphics{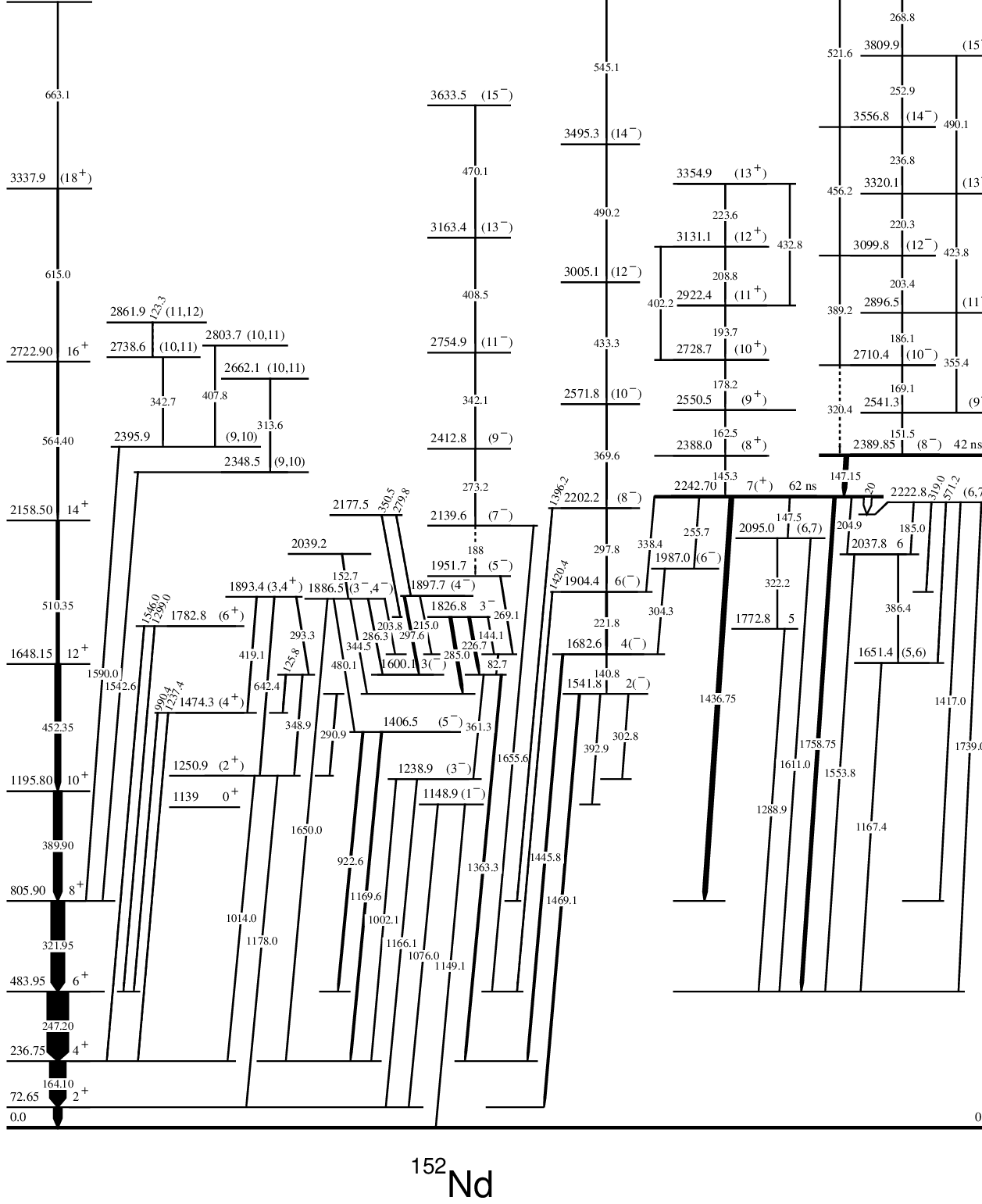}}
\caption{The excitation scheme of $^{152}$Nd, as observed in the present work. Newly observed
levels and transitions are marked by asterisk in Tables \ref{table_Nd152_levels_I} and
\ref{table_Nd152_levels_II}. Some transitions listed in the Tables are not drawn in the scheme
to keep it more readable.}
\label{A150_Nd_152_scheme}
\end{figure*}

Excited levels in $^{152}$Nd observed in the present work are drawn in Fig. \ref{A150_Nd_152_scheme}
and are listed in Tables \ref{table_Nd152_levels_I} and \ref{table_Nd152_levels_II}. We have also
analysed angular correlations for $\gamma\gamma$ cascades in $^{152}$Nd, using the technique
described in Ref. \cite{Nai17}. The results are listed in Table \ref{table_Nd52_ang}.

The results of $\beta$ decay work \cite{Toh99} are confirmed except the 1672-, 1990-, 2419 and
2702-keV levels and several weak decay, which are not seen in our data. To the results reported
in fission works \cite{Zha98,Yeo10} we add 29 levels and 38 transitions and make 35 spin-parity
assignments, marked as new in Tables \ref{table_Nd152_levels_I} and \ref{table_Nd152_levels_II}.

In the ground-state band we have determined precise transition energies confirming those reported
in Ref. \cite{Yeo10} rather than in \cite{Zha98}.

The new 1782.8-keV level with tentative spin-parity (6$^+$) suggested by its decay branching is
a likely  member of the band build on top of the 0$^+_2$ level at 1139 keV.

The 0$^+$, 868-keV level reported in the compilation \cite{NDS152} could not be confirmed in the
present work.

The band head of the new cascade above the 2139.6-keV level is not known. There is a tentative,
188-keV transition to the 1951.7-keV level but further decay to the 1826.8-keV level is not seen.
The 1826.8-keV level is about 20 keV below the expected position in this, otherwise, very
regular sequence. Angular correlations indicate spin I=7 spin for the 2139.6-keV level,
excluding spins 6 and 8, but negative parity of this level remains tentative.

Angular correlations indicate spin I=6 for the 1904.4-keV level and spin I=8 for the 2202.2-keV
level in the cascade on top of the 1541.8-keV level supporting spin I=2 for the 1541.8-keV level.

For the 1600.1-, 1826.8- and 1897.7-keV levels, strongly populated in $\beta^-$ decay, precise
angular correlations have been determined using clean spectra collected off the prompt-$\gamma$
radiation from fission:

- angular correlations in the 1363.3-164.10-keV cascade indicate spin I=3 for the 1600.1-keV
level, excluding spins 2 and 4. Small mixing ratio, $\delta$=0.10, of the 1363.3-keV transition
is consistent with negative parity for this level.

- angular correlations in the 226.7-1363.3-keV cascade indicate spin I=3 for the 1826.8-keV level
with spins 2 and 4 excluded. Large $\delta$ of the 226.7-keV transition, suggests negative parity
for this level, contrary to Refs. \cite{Toh99,Ale18} reporting its positive parity.

- angular correlation in the 297.6-1363.3-keV cascade is consistent with the I$^{\pi}$=4$^-$
spin parity of the 1897.7-keV level adopted in the compilation \cite{NDS152}.

\begin{table}[]
\caption{Excited levels and their $\gamma$ decays in $^{152}$Nd as observed in the present
work following fission of $^{252}$Cf. New data are marked with ``*'' symbol.}
\begin{center}
\begin{tabular}{l c r c c c }
\hline
~~~~Initial    &level     &${\gamma}$~~~-~~&decay& Final&level                                \\
E$_{exc}$ (keV)&I$^{\pi}$ &~~~E$_{\gamma}$(keV)&~~I$_{\gamma}$(rel)&E$_{exc}$(keV)&I$^{\pi}$  \\
\hline
   72.65(7)    &  2$^+$   &              72.65(5)       &  150(30) &    0.0      &  0$^+$     \\
  236.75(8)    &  4$^+$   &             164.10(5)       &  790(40) &   72.65     &  2$^+$     \\
  483.95(9)    &  6$^+$   &             247.20(5)       & 1000(30) &  236.75     &  4$^+$     \\
  805.90(10)   &  8$^+$   &             321.95(5)       &  510(30) &  483.95     &  6$^+$     \\
 1148.9(2)     & (1$^-$)  &            1076.0(2)        &   10(5)  &   72.65     &  2$^+$     \\
               &          &            1149.1(2)        &    8(4)  &    0.0      &  0$^+$     \\
 1195.80(11)   & 10$^+$   &             389.90(5)       &  340(20) &  805.90     &  8$^+$     \\
 1238.9(2)     & (3$^-$)  &            1002.1(1)        &   19(11) &  236.75     &  4$^+$     \\
               &          &            1166.1(2)        &   10(5)  &   72.65     &  2$^+$     \\
 1250.9(2)     & (2$^+$)  &            1014.0(1)        &  115(20) &  236.75     &  4$^+$     \\
               &          &            1178.0(2)        &   20(10) &   72.65     &  2$^+$     \\
 1406.5(2)     & (5$^-$)  &             922.6(1)        &   16(2)  &  483.95     &  6$^+$     \\
               &          &            1169.6(1)        &   21(9)  &  236.75     &  4$^+$     \\
 1474.3(2)     & (4$^+$)  &             990.4(1)        &   20(2)  &  483.95     &  6$^+$     \\
               &          &            1237.4(1)        &   21(7)  &  236.75     &  4$^+$     \\
 1541.8(1)     &2$^{(-)}$ &             290.9(2)        &   10(5)  & 1250.9      & (2$^+$)    \\
               &          &             302.8(1)        &   10(5)  & 1238.9      & (3$^-$)    \\
               &          &             392.9(1)        &   15(7)  & 1148.9      & (1$^-$)    \\
               &          &            1469.1(1)        &  220(40) &   72.65     &  2$^+$     \\
 1600.1(1)     &3$^{(-)}$ &             125.8(2)        &    2(1)  & 1474.3      & (4$^+$)    \\
               &          &             348.9(1)        &    4(2)  & 1250.9      & (2$^+$)    \\
               &          &             361.3(1)        &    5(2)  & 1238.9      & (3$^-$)    \\               &          &            1363.3(1)        &  150(20) &  236.75     &  4$^+$     \\
 1648.15(12)   & 12$^+$   &             452.35(5)       &  180(15) & 1195.80     & 10$^+$     \\
 1651.4(2)     & (5,6)    &            1167.4(1)        &   10(2)  &  483.95     &  6$^+$     \\
 1682.6(2)     &4$^{(-)}$ &              82.7(3)        &   10(5)  & 1600.1      &  3$^{(-)}$ \\
               &          &             140.8(2)        &   30(10) & 1541.8      &  2$^{(-)}$ \\
               &          &            1445.8(2)        &   35(5)  &  236.75     &  4$^+$     \\
 1772.8(2)     & 5 *      &            1288.9(1)        &   22(3)  &  483.95     &  6$^+$     \\
 1782.8(2)     & (6$^+$)* &            1299.0(2)        &    7(2)  &  483.95     &  6$^+$     \\
               &          &            1546.0(2)        &   50(15) &  236.75     &  4$^+$     \\
 1826.8(2)     &  3$^-$*  &             144.1(2)        &   20(10) & 1682.6      &  4$^{(-)}$ \\
               &          &             226.7(1)        &  130(20) & 1600.1      &  3$^{(-)}$ \\
               &          &             285.0(1)        &  260(30) & 1541.8      &  2$^{(-)}$ \\
 1886.5(2)     &(3$^-$,4$^-$)&          203.8(2)        &   15(7)  & 1682.6      &  4$^{(-)}$ \\
               &          &             286.3(1)        &   10(5)  & 1600.1      &  3$^{(-)}$ \\
               &          &             344.5(3)        &   30(10) & 1541.8      &  2$^{(-)}$ \\
               &          &             480.1(2)        &   20(10) & 1406.5      & (5$^-$)    \\
               &          &            1650.0(2)        &    5(2)  &  236.75     &  4$^+$     \\
 1893.4(2)     &(3,4$^+$) &             293.3(1)        &    6(2)  & 1600.1      &  3$^{(-)}$ \\
               &          &             419.1(1)        &   20(10) & 1474.3      &  (4$^+$)   \\
               &          &             642.4(2)        &   30(10) & 1250.9      &  (2$^+$)   \\
 1897.7(2)     & (4$^-$)  &             215.0(2)        &   50(10) & 1682.6      &  4$^{(-)}$ \\
               &          &             297.6(1)        &   90(20) & 1600.1      &  3$^{(-)}$ \\
 1904.4(2)     &6$^{(-)}$ &             221.8(1)        &  165(15) & 1682.6      &  4$^{(-)}$ \\
               &          &            1420.4(1)        &   25(3)  &  483.95     &  6$^+$     \\
 1951.7(3)     & (5$^-$)  &             269.1(1)        &   30(10) & 1682.6      &  4$^{(-)}$ \\
 1987.0(2)     & (6-)     &             304.3(2)        &   30(10) & 1682.6      &  4$^{(-)}$ \\
 2037.8(1)     &  6 *     &             386.4(2)        &    7(2)  & 1651.4      &  (5,6)     \\
               &          &            1553.8(1)        &    9(2)  &  483.95     &  6$^+$     \\
 2039.2(3)     &          &             152.7(2)        &   25(10) & 1886.5     &(3$^-$,4$^-$)\\
 2095.0(3)*    &  (6,7)*  &             322.2(3)*       &    8(4)  & 1772.8      &  5         \\
               &          &            1611.0(3)        &    6(2)  &  483.95     &  6$^+$     \\
 2139.6(2)*    & (7$^-$)* &            1655.6(1)*       &   11(2)  &  483.95     &  6$^+$     \\
 2158.50(13)   & 14$^+$   &             510.35(5)       &  125(15) & 1648.15     & 12$^+$     \\
\hline
\end{tabular}
\end{center}
\label{table_Nd152_levels_I}
\end{table}

\begin{table}
\caption{Continuation of Table \ref{table_Nd152_levels_I}.}
\begin{center}
\begin{tabular}{l c r c c c }
\hline
~~~~Initial    &level     &${\gamma}$~~~-~~&decay& Final&level                                \\
E$_{exc}$ (keV)&I$^{\pi}$ &~~~E$_{\gamma}$(keV)&~~I$_{\gamma}$(rel)&E$_{exc}$(keV)&I$^{\pi}$  \\
\hline
 2177.5(3)     &          &             279.8(2)        &   20(7)  & 1897.7      &  4$^-$     \\
               &          &             350.5(2)        &   25(10) & 1826.8      &  3$^-$     \\
 2202.2(2)     & (8$^-$)  &             297.8(1)        &   98(15) & 1904.4      &  6$^{(-)}$ \\
               &          &            1396.2(1)        &    8(2)  &  805.90     &  8$^+$     \\
 2222.8(2)     & (6,7)    &             185.0(2)        &    12(3) & 2037.8      &  6         \\
               &          &             319.0(2)        &    5(2)  & 1904.4      &  6$^{(-)}$ \\
               &          &             571.2(2)        &    4(2)  & 1651.4      & (5,6)      \\
               &          &            1417.0(2)        &    6(2)  &  805.90     &  8$^+$     \\
               &          &            1739.0(3)        &    9(2)  &  483.95     &  6$^+$     \\
 2242.70(9)    &7$^{(+)}$*&             147.5(2)*       &    4(2)  & 2095.0      & (6,7)      \\
               &          &             204.9(2)        &   10(2)  & 2037.8      &  6         \\
               &          &             255.7(2)        &    3(1)  & 1987.0      &  (6$^-$)   \\
               &          &             338.4(1)        &   12(2)  & 1904.4      &  6$^{(-)}$ \\
               &          &            1436.75(5)       &   16(3)  &  805.90     &  8$^+$     \\
               &          &            1758.75(15)      &   14(3)  &  483.95     &  6$^+$     \\
 2348.5(3)*    & (9,10)*  &            1542.6(2)*       &    5(1)  &  805.90     &  8$^+$     \\
 2388.0(2)*    & (8$^+$)* &             145.3(15)*      &   11(3)  & 2242.70     &  7$^{(+)}$ \\
 2389.85(15)   & (8$^-$)* &             147.15(9)       &   18(3)  & 2242.70     &  7$^{(+)}$ \\
 2395.9(2)*    & (9,10)*  &            1590.0(1)*       &    9(1)  &  805.90     &  8$^+$     \\
 2412.8(3)*    & (9$^-$)* &             273.2(2)*       &    9(2)  & 2139.6      &  7$^{(-)}$ \\
 2541.3(2)*    & (9$^-$)* &             151.5(1)*       &   14(2)  & 2389.85     & (8$^-$)    \\
 2550.5(3)*    & (9$^+$)* &             162.5(2)*       &    9(3)  & 2388.0      & (8$^+$)    \\
 2571.8(3)*    & (10$^-$)*&             369.6(1)*       &   80(15) & 2202.2      & (8$^-$)    \\
 2662.1(4)*    & (10,11)* &             313.6(2)*       &    7(2)  & 2348.5      &  (9,10)    \\
 2710.4(3)     & (10$^-$)*&             169.1(2)        &   12(2)  & 2541.3      & (9$^-$)    \\
               &          &             320.4(3)*       &   $<$ 1  & 2389.85     & (8$^-$)    \\
 2722.90(14)   & 16$^+$   &             564.40(5)       &   65(10) & 2158.50     & 14$^+$     \\
 2728.7(4)*    & (10$^+$)*&             178.2(2)*       &    6(2)  & 2550.5      & (9$^+$)    \\
 2738.6(3)*    & (10,11)* &             342.7(1)*       &    6(2)  & 2395.9      & (9,10)     \\
 2754.9(3)*    & (11$^-$)*&             342.1(2)*       &    6(2)  & 2412.8      & (9$^-$)    \\
 2803.7(4)*    & (10,11)* &             407.8(3)*       &    4(1)  & 2395.9      & (9,10)     \\
 2861.9(4)*    & (11,12)* &             123.3(1)*       &    4(1)  & 2738.6      & (10,11)    \\
 2896.5(3)*    & (11$^-$)*&             186.1(1)        &    8(2)  & 2710.4      & (10$^-$)   \\
               &          &             355.4(2)*       &    1(1)  & 2541.3      & (9$^-$)    \\
 2922.4(4)*    & (11$^+$)*&             193.7(2)*       &    5(2)  & 2728.7      & (10$^+$)   \\
 3005.1(3)*    & (12$^-$)*&             433.3(2)*       &   55(10) & 2571.8      & (10$^-$)   \\
 3099.8(3)*    & (12$^-$)*&             203.4(2)*       &   11(3)  & 2896.5      & (11$^-$)   \\
               &          &             389.2(3)*       &    2(1)  & 2710.4      & (10$^-$)   \\
 3131.1(4)*    & (12$^+$)*&             208.8(2)*       &    4(1)  & 2922.4      & (11$^+$)   \\
               &          &             402.2(2)*       &    1(1)  & 2728.7      & (10$^+$)   \\
 3163.4(4)*    & (13$^-$)*&             408.5(2)*       &    4(2)  & 2754.9      & (11$^-$)   \\
 3320.1(3)*    & (13$^-$)*&             220.3(1)*       &    6(2)  & 3099.8      & (12$^-$)   \\
               &          &             423.8(2)*       &    2(1)  & 2896.5      & (11$^-$)   \\
 3337.9(2)     & (18$^+$) &             615.0(1)        &   16(3)  & 2722.90     &  16$^+$    \\
 3354.9(4)*    & (13$^+$)*&             223.6(2)*       &    2(1)  & 3131.1      & (12$^+$)   \\
               &          &             432.8(2)*       &    1(1)  & 2922.4      & (11$^+$)   \\
 3495.3(4)*    & (14$^-$)*&             490.2(2)*       &   25(10) & 3005.1      & (12$^-$)   \\
 3556.8(4)*    & (14$^-$)*&             236.8(1)*       &    6(2)  & 3320.1      & (13$^-$)   \\
               &          &             456.2(3)*       &    2(1)  & 3099.8      & (12$^-$)   \\
 3633.5(5)*    & (15$^-$)*&             470.1(2)*       &    2(1)  & 3163.4      & (13$^-$)   \\
 3809.9(4)*    & (15$^-$)*&             252.9(1)*       &    4(2)  & 3556.8      & (14$^-$)   \\
               &          &             490.1(2)*       &    4(2)  & 3320.1      & (13$^-$)   \\
 4001.0(3)     & (20$^+$) &             663.1(2)        &    7(2)  & 3337.9      & (18$^+$)   \\
 4040.4(5)*    & (16$^-$)*&             545.1(2)*       &   10(5)  & 3495.3      & (14$^-$)   \\
 4078.6(5)*    & (16$^-$)*&             268.8(2)*       &    1(1)  & 3809.9      & (15$^-$)   \\
               &          &             521.6(2)*       &    2(1)  & 3556.8      & (14$^-$)   \\
\hline
\end{tabular}
\end{center}
\label{table_Nd152_levels_II}
\end{table}

- angular correlation for the 285.0-1469.1-keV cascade is consistent with spin I=3 of the
1826.8-keV level taking spin-parity 2$^-$ for the 1541.8-keV level (shown above) and small
$\delta$ ratio for the 1469.1-keV, E1+M2 transition.

\begin{table}[]
\caption{Angular correlation results for $\gamma \gamma$ cascades in $^{152}$Nd. Label ``sum''
marks summed correlations with quadrupole transitions below the E$_{\gamma 1}$. Label ``u''
denotes unobserved, stretched quadrupole transition in the cascade. Superscript ``m'' indicates
mixed transition for which $\delta_{exp}$ value, shown in the last column with the corresponding
$\chi^2$ of the fit, is determined.}
\begin{center}
\begin{tabular}{ c c c c }
\hline
E${\gamma_1}$ - E${\gamma_2}$& $A_2/A_0$  & $A_4/A_0$   &     Spins in cascade         \\
      (keV)                  &   exp.     &    exp.     & and $\delta_{exp}$; $\chi^2$ \\
\hline
  226.7$^m$-1363.3           &  0.024(33) & -0.084(50)  & 3-3-4                        \\
                             &            &             & -5.9(-66,+20); 0.1           \\
  247.20-164.10              &  0.80(15)  & -0.030(21)  & 6-4-2                        \\
                             &            &             & $\chi^2$=2.5                 \\
  285.0$^m$-1469.1           &  0.015(21) & -0.034(33)  & 3-2-2                        \\
                             &            &             & 5.1(-18,+49); 0.1            \\
  297.6$^m$-1363.3           & -0.102(37) & -0.045(59)  & 4-3-4                        \\
                             &            &             & 0.34(11); 0.7                \\
                             &            &             & 2.1(5); 1.3                  \\
  321.95-247.20              & 0.101(11)  &  0.019(15)  & 8-6-4                        \\
                             &            &             & $\chi^2$=0.6                 \\
  389.90-321.95              & 0.102(11)  &  0.007(16)  & 10-8-6                       \\
                             &            &             & $\chi^2$=0.4                 \\
  452.35-389.90              & 0.120(15)  &  0.027(22)  & 12-10-8                      \\
                             &            &             & $\chi^2$=2.2                 \\
  564.40-452.35              & 0.121(36)  &  0.076(56)  & 16-14-u-12-10                \\
                             &            &             & $\chi^2$=1.7                 \\
  1288.9$^m$-164.10          & -0.260(32) &  0.000(46)  & 5-6-u-4-2                    \\
                             &            &             & 0.21(6); 0.2                 \\
  1363.3$^m$-164.10          & -0.049(15) & -0.013(22)  & 3-4-2                        \\
                             &            &             & -0.10(2); 0.3                \\
  1396.2$^m$-sum             &  0.192(98) & -0.042(145) & 8-8-6                        \\
                             &            &             & -0.13(70); 0.2               \\
                             &            &             & 7-6-4                        \\
                             &            &             & 0.26(12); 0.1                \\
  1420.4$^m$-sum             &  0.081(58) & -0.032(85)  & 6-6-4                        \\
                             &            &             & 0.36(20); 0.3                \\
                             &            &             & 7-6-4                        \\
                             &            &             & 0.26(12); 0.1                \\
  1436.75$^m$-321.95         &  0.195(99) & -0.039(138) & 7-8-6                        \\
                             &            &             & -0.51(-48,+21); 0.1          \\
                             &            &             & -2.6(-28,+15); 0.2           \\
                             &            &             & 8-8-6                        \\
                             &            &             & 0.15(65); 0.1                \\
  1553.8$^m$-sum             & -0.01(8)   &  0.11(11)   & 6-6-4                        \\
                             &            &             & -2.4(-49,+10); 0.1           \\
                             &            &             & 0.75(-30,+50); 0.4           \\
  1655.6$^m$-sum             & -0.05(10)  & -0.11(15)   & 7-6-4                        \\
                             &            &             & 5.1(-23,+110); 0.9           \\
                             &            &             & 0.03(14); 2.1                \\
  1758.75$^m$-247.20         & -0.304(84) & -0.033(127) & 7-6-4                        \\
                             &            &             & -2.7(-25,+13); 0.1           \\
                             &            &             & -0.42(-33,+18); 0.2          \\
\hline
\end{tabular}
\end{center}
\label{table_Nd52_ang}
\end{table}

Angular correlations indicate spin I=7 for the 2242.70-keV level (angular correlations for the 1758.75-247.20-keV cascade exclude spins I=6 and I=8). Large $\delta$ ratios for 1436.8- and
1758.8-keV transitions suggest positive parity for the 2242.70-keV level.

\begin{figure}
\centering
\scalebox{.65}{\includegraphics{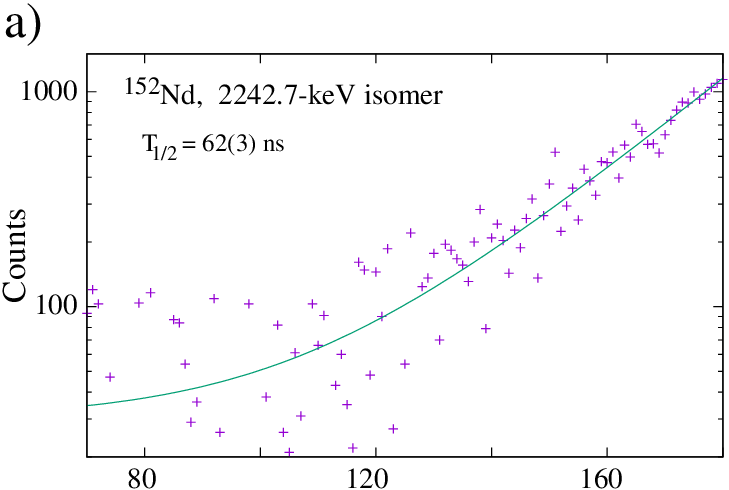}}
\scalebox{.65}{\includegraphics{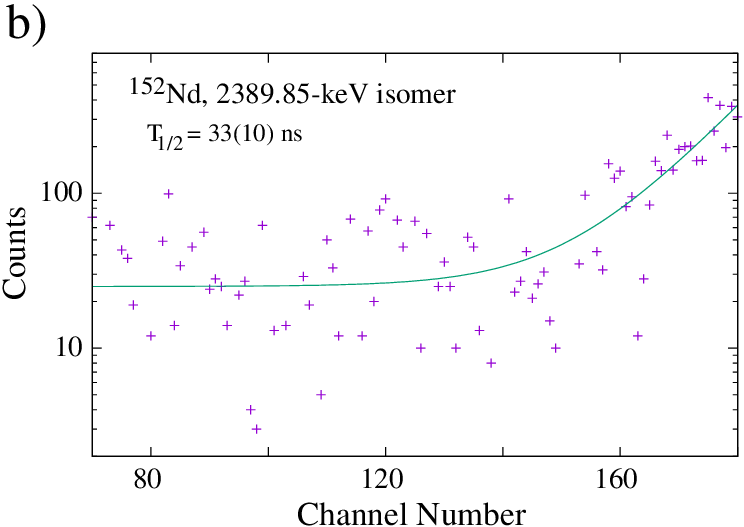}}
\caption{Off-prompt fragments of time-decay spectra for (a) the 2242.70-keV isomer and (b) the
         2389.85-keV isomer in $^{152}$Nd. Time calibration is 4.4 ns per channel. Green lines
         represent exponent-plus-constant-background fits.}
\label{A150_Nd152_time}
\end{figure}

The 2242.70-keV isomer in $^{152}$Nd was reported previously with an average half life of 73(7)
ns \cite{Gau98}. In Ref.\cite{Yeo10} its half life of 63(7) ns was determined by observing
intensity of $\gamma$ decay of the isomer in various time windows. An analogous method applied
in the present work provided half life of 78(9) ns, as shown in Fig. \ref{A150_Nd_150_Rw_time}.
Thanks to high population of this isomer it was also possible to construct time-decay spectrum
gated on the 147.15-keV $\gamma$ line above the isomer ($time~start$) and 1436.75-, 321.95- and
247.20-keV $\gamma$ lines below the isomer ($time~stop$), as shown in Fig. \ref{A150_Nd152_time}
(a), which provided more accurate half life of T$_{1/2}$=62(3) ns.

The 2242.70-keV isomer has a 20-keV decay branch to the 2222.8-keV level, as indicated by
coindidence data. The corresponding $\gamma$ line is not seen. Total intensity of this decay
branch amounts to 15(5) relative units of Table \ref{table_Nd152_levels_I}, estimated from spectra
of delayed $\gamma$ decays of the isomer. Further analysis has shown that about 60$\%$ of its
population the 2222.8-keV level receives in direct, prompt side feeding from fission.

The band structure above the 2242.70-keV isomer, comprising two bands on top of the 2242.70- and
2389.85-keV levels, is significantly changed and extended compared to the single band above the
isomer reported previously \cite{Yeo10}. We also found that the new  2389.85-keV band head is
isomeric with a half life T$_{1/2}$=42(8) ns. This is an average of the 34(10) ns value obtained
from the time-decay spectrum gated on the 151.5-keV $\gamma$ line above ($time~start$) and the
147.15-keV 1 line below the 2389.85-keV isomer isomer ($time~stop$), as shown in Fig.
\ref{A150_Nd152_time} (b) and the 50(10) ns value determined in Fig. \ref{A150_Nd_150_Rw_time}.
The isomeric nature of the 147.15-keV transition suggests its E1 multipolarity supporting opposite
parities of 2242.70-keV and 2389.85-keV levels.

The observed properties of the two bands above the 2242.70- and 2389.85-keV isomers suggest
that they are high-K, $\Delta$I=1 M1+E2, prompt-$\gamma$ cascades.

\newpage

\section{Discussion}

The energy of the 2$^+_1$ level, the lowest excitation in most of spherical, even-even nuclei,
traditionally associated with a ``surface vibration'', is used as a measure of the quadrupole
collectivity in a nucleus.

\begin{figure}[]
\centering
\scalebox{.38}{\includegraphics{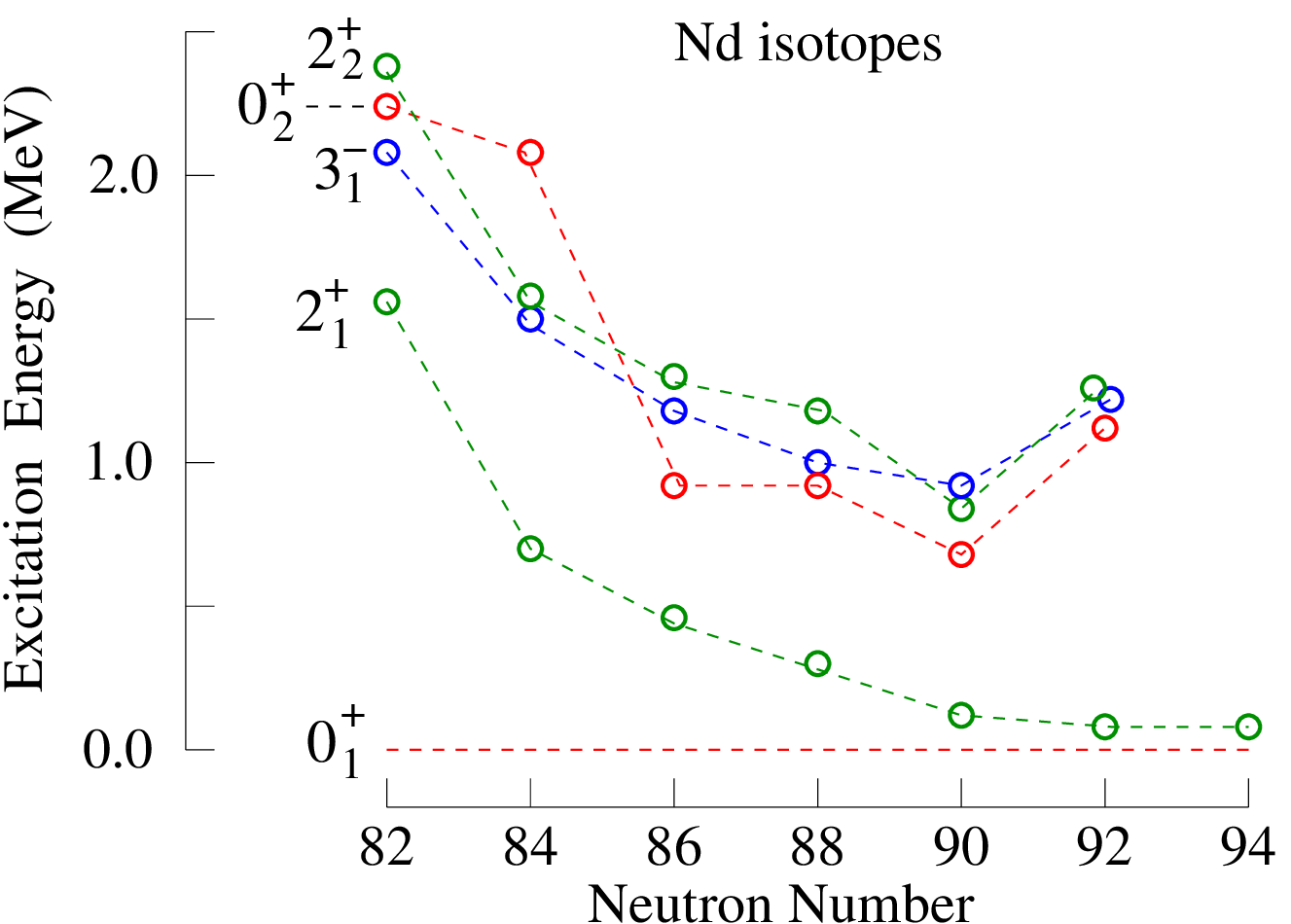}}
\caption{Energies of low-spin levels in Nd isotopes. The data are taken from
         the data base \cite{ENSDF}. Dashed lines are drawn to guide the eye.}
\label{A150_Nd_low_exc}
\end{figure}

Around twice the energy of the 2$^+_1$ level one encounters a 0$^+_2$ level, often interpreted
as coupling of two quadrupole excitations. It may evolve into, so called, $\beta$ vibrations when
the number of valence particle increases (see Fig. 4 in Ref. \cite{Urb13}).

Another, ``early-bird'' collectivity above a closed shell is due to non-axial vibrations, as
discussed in our works on N=86 isotones \cite{Nai17,Urb16}. Such excitations, usually associated
with the 2$^+_2$ level in even-even nuclei, are due to the proximity of low-$\Omega$, prolate
and high-$\Omega$, oblate orbitals, mixing of which induces non-axial distortion of the nuclear
potential.

Finally just above closed shells one observes in even-even nuclei characteristic 3$^-$,
excitations. The adjacent major shells have opposite parities enabling low-energy,
octupole collectivity in spherical and transitional nuclei.

Figure \ref{A150_Nd_low_exc} shows known, low-spin 0$^+_2$, 2$^+_1$, 2$^+_2$ and 3$^-_1$
excitations in even-even Nd isotopes above the N=82 closed shell. From this ``typical-style''
systematics one can draw a general conclusion that the energy of the excitations drops with
the increasing number of valence neutrons, due to increasing contribution of emerging
collective effects. Above N=88 the 2$^+_1$ level evolves smoothly into very collective,
rotational excitation. The ``collective'' 0$^+_2$, 2$^+_2$ and 3$^-_1$ levels stick together,
reaching minimum of excitation energy at N=90. The 0$^+_2$ and 2$^+_2$ levels are not at twice
the excitation energy of the 2$^+_1$ level, which questions their two-phonon nature.

In the following low-energy excitations in even-even Nd isotopes will be discussed in more
detail, using another-style systematics, applied successfully in the A$\approx$100 region
\cite{Urb19,Urb21,Wis23}.

\subsection{2$^+$ levels, $\gamma$ excitations}

\subsubsection{K=0, 2$^+$ levels}

\begin{figure}
\centering
\scalebox{.45}{\includegraphics{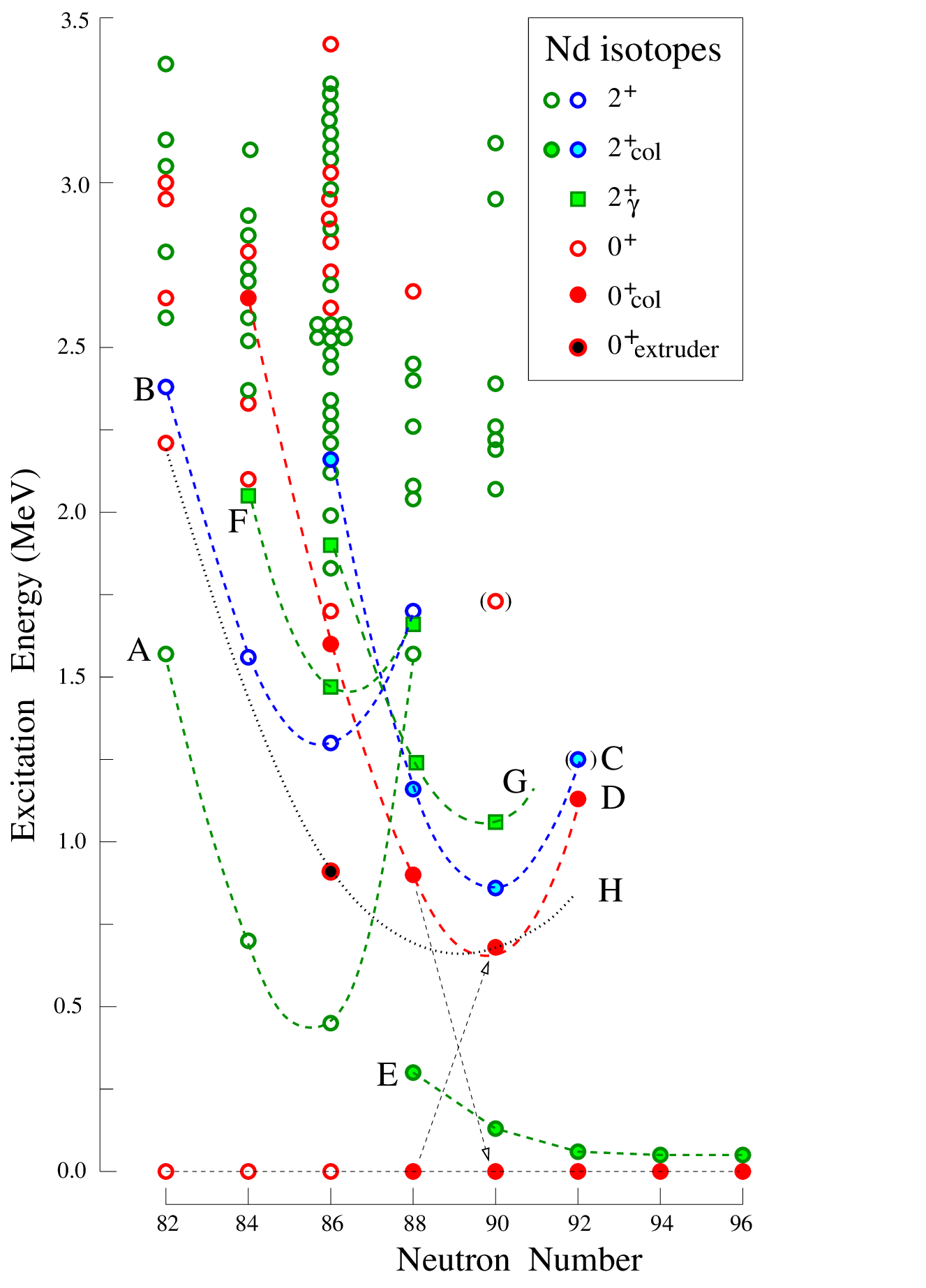}}
\caption{Energies of 0$^+$ and 2$^+$ levels in Nd isotopes. The data are taken from
         Ref. \cite{ENSDF} and the present work. Points in parenthesis have tentative
         spin-parity assignment. Dashed lines are drawn to guide the eye.
         See text for the explanation of labels A - H.}
\label{A150_Nd_0plus_2plus}
\end{figure}

Figure \ref{A150_Nd_0plus_2plus} shows known 0$^+$ and 2$^+$ levels up to 3.5 MeV in even-even
Nd isotopes with 82$\leq$N$\leq$96. The ``U''shaped curve marked A links 2$^+$ levels expected
to be due to excitations within the $\nu f_{7/2}$ shell, populated just above the N=82 closure.
Analogous 2$^+_1$ levels in even-even Sr isotopes with  50$\leq$N$\leq$56 were shown by detailed
large-scale, shell-model (LSSM) calculations to be due to population of the $d_{5/2}$ shell just
above N=50 closure (see Figs. 13 and 14 in Ref. \cite{Urb21}).

Curve A links 2$^+_1$ levels at N$<$88 and the 2$^+_4$ level at N=88. In these excitations one
expects a lowest-energy proton configuration. Curve B connects 2$^+_2$ levels at N$<$88 and
the 2$^+_6$ level at N=88, which most likely have the same neutron configuration but are
coupled to higher-energy proton configuration, analogously to what is observed in Sr isotopes
\cite{Urb21}.

With eight valence neutrons the $^{150}$Nd, having deformed 2$^+_1$ and 2$^+_2$ levels, is more
collective than its analog, $^{96}$Sr which at N=58 is still spherical due to the presence of
the spherical $\nu s_{1/2}$ shell. In the Nd isotopes the analogous spherical $\nu p_{3/2}$
shell is located above the deformation-driving $\nu i_{13/2}$ shell and the deformation sets
already at N=88. This is further helped by higher spin of the $\nu f_{7/2}$ and $\nu h_{9/2}$
shells in the A$\approx$150 region, compared to analogous $\nu d_{5/2}$ and $\nu g_{7/2}$
shells in the A$\approx$100 region. Low-$\Omega$ orbitals of the $\nu f_{7/2}$ and $\nu h_{9/2}$
shells drive Nd nuclei towards deformation faster than low-$\Omega$ orbitals of the $\nu d_{5/2}$
and $\nu g_{7/2}$ shells do in the A$\approx$100 region.

\begin{figure}[]
\centering
\scalebox{.36}{\includegraphics{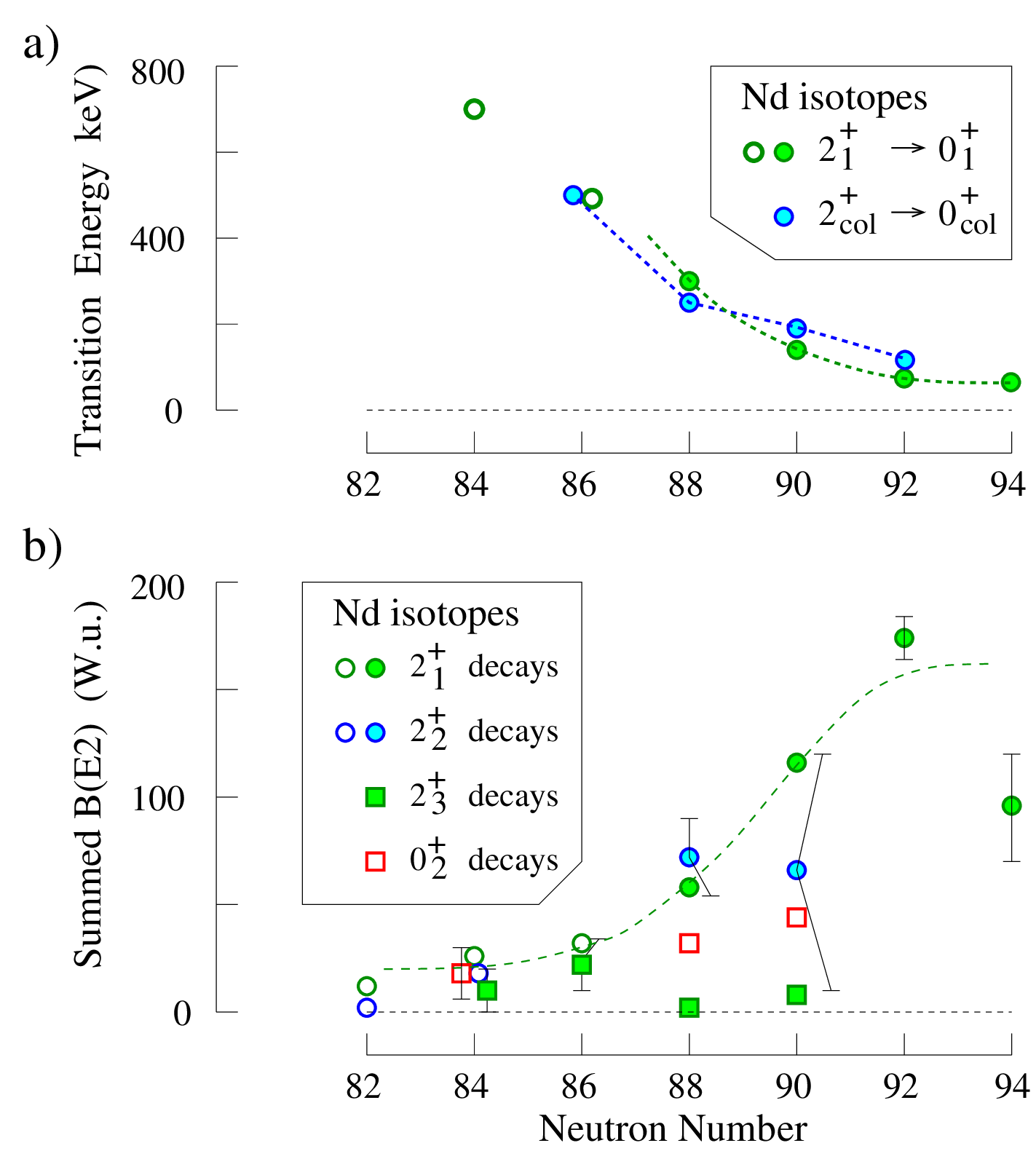}}
\caption{ a) Energies of 2$^+_1 - 0^+_1$ and 2$^+_{col} - 0^+_{col}$ transitions in Nd isotopes.
             Dashed lines are drawn to guide the eye.
          b) Sum of B(E2) rates of transitions deexciting the 2$^+_1$, 2$^+_2$, 2$^+_3$ and
             0$^+_2$, states in Nd isotopes. For data points without error bars the uncertainties
             are smaller than the symbol size. Green dashed line approximates the B(E2) for
             2$^+_1 \rightarrow 0^+_1$ transitions. The data are taken from the present work and
             from compilations \cite{NDS142,NDS144,NDS146,NDS148,NDS150,NDS152,NDS154}. See text
             for further explanations.}
\label{A150_Nd_BE2}
\end{figure}

Figure \ref{A150_Nd_0plus_2plus} reveals a significant increase in the number of 2$^+$ levels
above 2 MeV at N=86 compared to lower N, which is related to the population of both, $f_{7/2}$
and $h_{9/2}$ shells. Above N=88 this collectivity likely mixes into low-lying, 2$^+$ collective
states (see Fig. 1.10 in the textbook \cite{Cas00}) .

The 2$^+_{col}$ levels linked by curve C are members of bands on top of collective 0$^+_{col}$,
excitations on curve D (0$^+_2$ at N$>$86, 0$^+_3$ in $^{146}$Nd and 0$^+_4$ in $^{144}$Nd).
Figure \ref{A150_Nd_BE2} (a) compares energies of 2$^+_1$ levels to energies of 2$^+_{col}$
excitations on top of collective 0$^+_{col}$ levels. One notes similarity of both transition
energies. The collective character of 2$^+_{col}$ levels is confirmed by their enhanced B(E2)
decays shown in Fig. \ref{A150_Nd_BE2} (b) by blue circles.

Curve E connects 2$^+_1$ levels at N$>$86, which are members of deformed-ground-state rotational
bands. In $^{148}$Nd the ground-state band is already rather regular with the deformation
parameter $\beta_2$=0.20 in Ref. \cite{ENSDF}. The 0$^+_1$, g.s. level at N=86 may be weakly
deformed as suggested by the B(E2; 2$^+_1 \rightarrow 0^+_1$)=31.9(4) W.u. \cite{NDS146} and
the low energy of the 2$^+_1$ level there.

No 2$^+_2$ levels are known above N=92 where strong quadrupole collectivity of the ground state
configuration dominates the nuclear structure, as shown by B(E2) decay rates of 2$^+_1$ levels
in  Fig.  \ref{A150_Nd_BE2} (b) represented by filled green circles. One should remeasure the
B(E2) decay rate of the 2$^+_1$ in $^{154}$Nd, as already hinted in Refs. \cite{Hel91,Hel93}.

\subsubsection{K=2, 2$^+$ levels}

The 2$^+$ levels marked by squares in Fig. \ref{A150_Nd_0plus_2plus} and linked by curves F and
G belong to yet another structure. We propose that these levels represent $\gamma$ excitations,
which are due to coupling of high-$\Omega$ orbitals from $\nu f_{7/2}$ and $\nu h_{9/2}$ shells
with low-$\Omega$ orbitals of the $\nu p_{3/2}$ shell. At N=86 and N=88 one may expect two
$\gamma$ bands. The pronounced ``U'' shape of curves F and G suggests that these levels are
dominated by s.p. excitations within $\nu f_{7/2}$ and $\nu h_{9/2}$ shells and are weakly
admixed by $\gamma$ collectivity.

\begin{figure}
\centering
\scalebox{.40}{\includegraphics{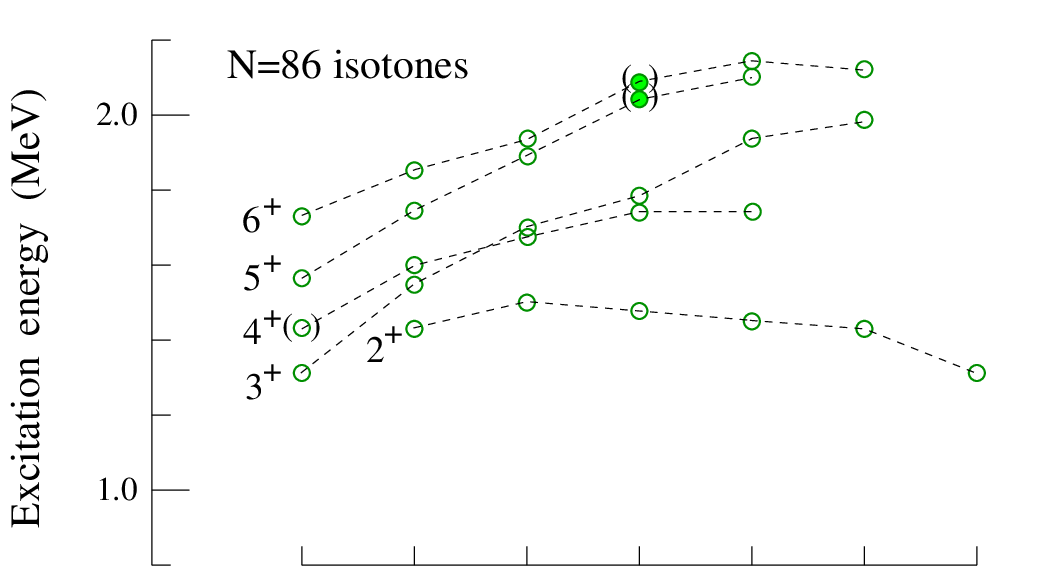}}
\scalebox{.40}{\includegraphics{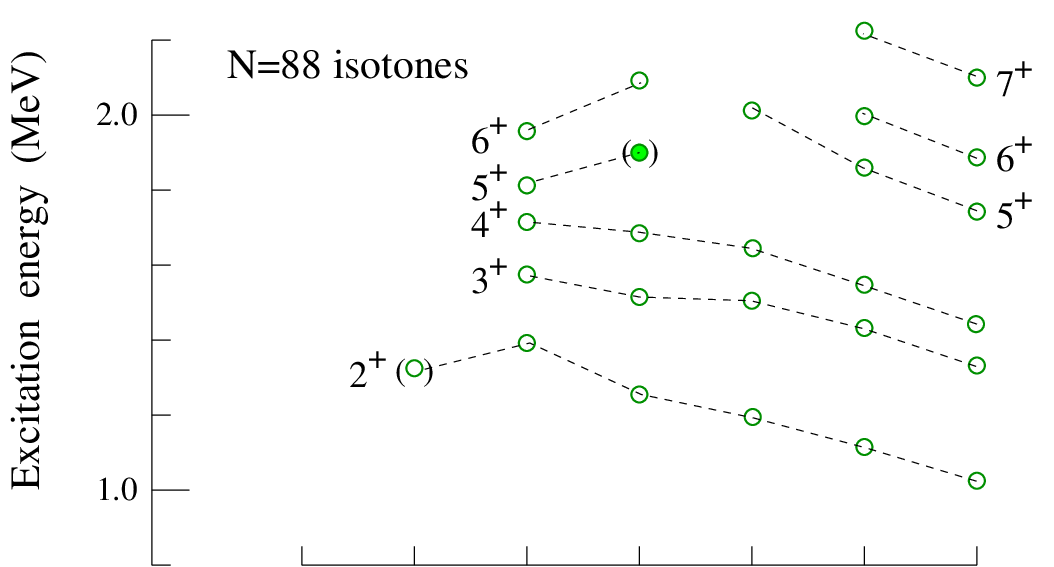}}
\scalebox{.40}{\includegraphics{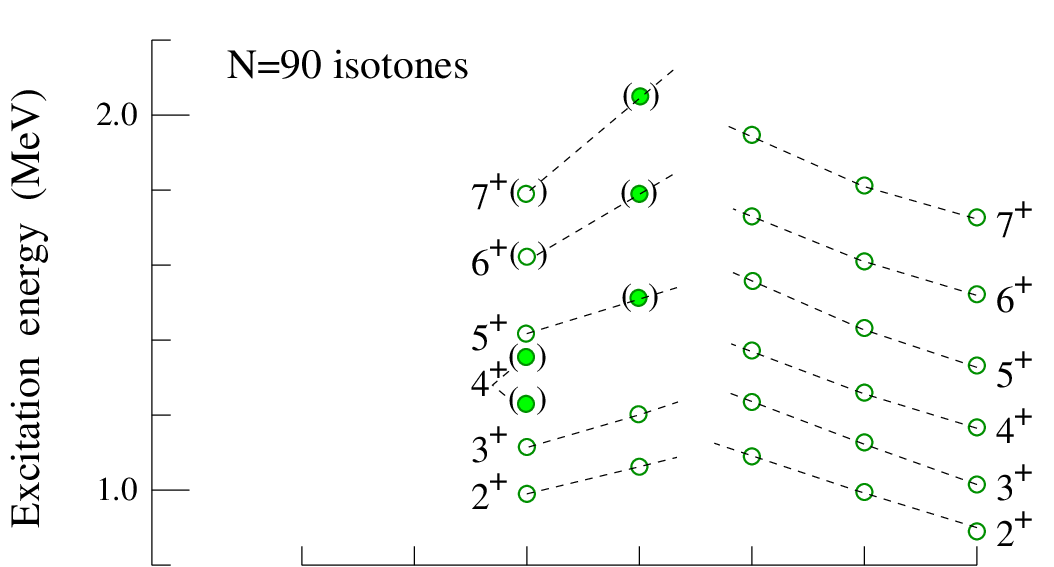}}
\scalebox{.40}{\includegraphics{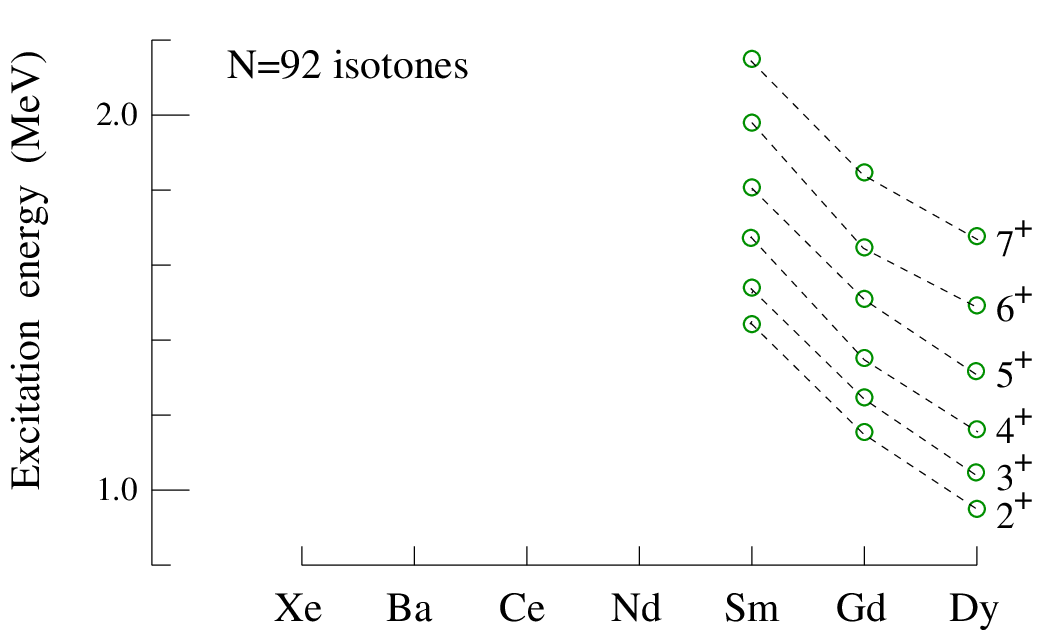}}
\caption{Excited levels in A$\approx$150 region interpreted as members of $\gamma$ bands. The
         data are taken from the compilation \cite{ENSDF} (open circles) and the present work
         (filled circles). Data points in parenthesis are tentative. Dashed lines are drawn to
         guide the eye.}
\label{A150_N92_gamma}
\end{figure}

In the N=86 isotones, $^{138}$Te, $^{140}$Xe, $^{142}$Ba and $^{144}$Ce this effect is clearly
present being most pronounced in $^{142}$Ba with a well developed $\gamma$ band  on top of
the 2$^+_2$ level at 1424.0 keV. Such structures are recognized by the characteristic, 3$^+$ and
5$^+$ band members. The 2$^+_3$, 1470.58-keV and 3$^+_1$, 1777.5-keV levels in $^{146}$Nd,
linked by the new, 307.0-keV transition, are members of a $\gamma$ band. We propose
that the 2045.70- and 2083.51-keV levels in $^{146}$Nd, reported previously \cite{NDS146}, are
the 5$^+$ and 6$^+$ members of this band, as shown in Fig. \ref{A150_N92_gamma}.

Figure \ref{A150_Nd_BE2} (b) shows that $\gamma$ collectivity in Nd isotopes is rather weak,
with the corresponding B(E2) rates of the order of 10 W.u., whereas in nuclei with strong $\gamma$
collectivity, like $^{98,100}$Mo, the 2$^+_2 \rightarrow$2$^+_1$ decay rates are of the order of
50 W.u. \cite{Che20,Sin21}. A candidate for the 2$^+_{\gamma}$ level at 1672.2 keV in $^{152}$Nd
proposed in \cite{NDS152} is not seen in the present work.

Figure \ref{A150_N92_gamma} suggests that there are two kinds of $\gamma$ bands, one in isotopes
below Z=62 and the other in isotopes with Z from 62 up, which probably differ in proton structure.
The former dominates at N=86 and is not seen at N=92 where the latter is well developed. The
staggering in $\gamma$ bands \cite{McC07} of Nd isotopes below N=90, with its minima at even
spins, is characteristic of a $\gamma$-soft structure. It changes to a triaxial structure at N=90.
The information available at N=92 for $^{154}$Sm \cite{NDS154} suggests weakly-deformed triaxial
band there.

We note the missing 4$^+$ member of $\gamma$ band in $^{150}$Nd. In the compilation \cite{NDS150}
a 3$^-$ level is reported at 1483.58 keV, which decay exclusively to (six) positive-parity states.
It was not reported in $\beta^-$ decay works \cite{Fog86,Kar88} but is clearly seen in our work.
Further work is needed to verify its spin and parity. It is also of high interest to search for
$\gamma$ excitations in $^{152}$Nd, where such levels have not been found to date.

Recent IBM calculations \cite{Kos22} suggest a decrease of axial asymmetry with increasing proton
number Z to zero value at $^{152}$Sm. The beyond mean field analysis \cite{Rod08} shows an
increasing $\gamma$ softness of nuclear potential in Nd isotopes in function of an increasing neutron
number (see Fig. 3 of Ref. \cite{Rod08}). The latter suggests a vibrational character of $\gamma$
bands in Z$>$60, N$>$90 isotopes, which does not agree with Fig. \ref{A150_N92_gamma}.

\subsubsection{Possible mixed-symmetry states}

\begin{figure}
\centering
\scalebox{.52}{\includegraphics{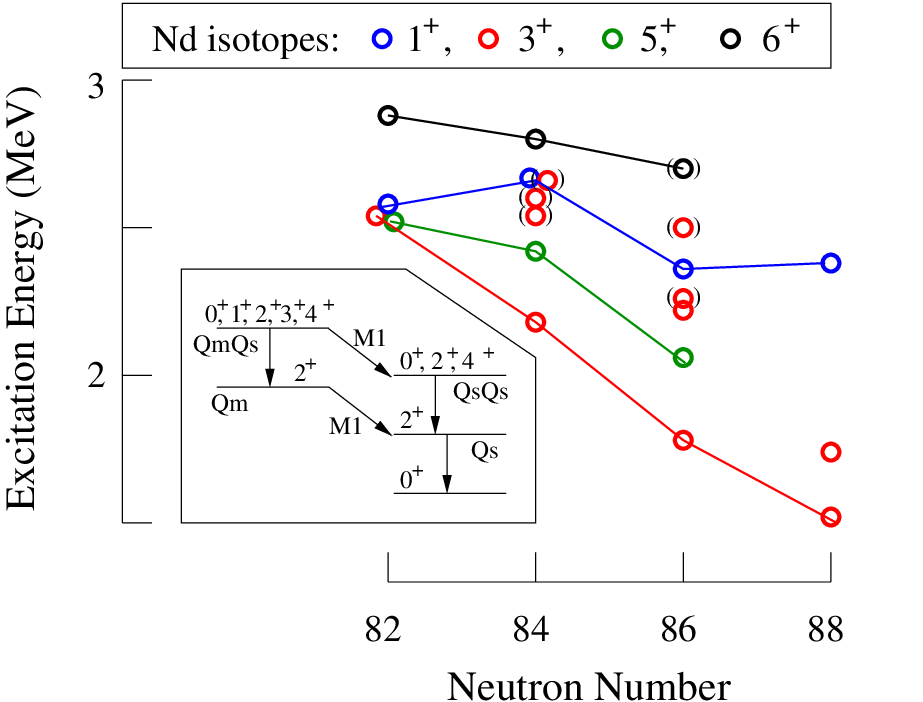}}
\caption{Low-energy, positive-parity excited states in A$\approx$150 region. The
         data are taken from the compilation \cite{ENSDF} and the present work. Data points
         in parenthesis are tentative. Lines are drawn to guide the eye.}
\label{A150_ms_states}
\end{figure}

There are more 2$^+$ and 3$^+$ level above 2 MeV and the 1$^+$ level at 2356.2 keV in $^{146}$Nd.
Low-energy 1$^+$ excitation is a characteristic member of the two-phonon multiplet, Q$_m$Q$_s$,
a coupling of symmetric, Q$_s$, and mixed-symmetry, Q$_m$, 2$^+$  vibrations \cite{Yat05}. Figure
\ref{A150_ms_states} shows 1$^+_1$ states in Nd isotopes and the figure inset illustrates the
Q$_m$Q$_s$ multiplet with its major, M1 decays. The approximately constant excitation energy of
1$^+$ states may signal a collective character of these excitations. The enhanced M1 decay of the
2601.7-keV, 4$^+$ level in $^{144}$Nd \cite{NDS144}, a possible member of the Q$_m$Q$_s$ multiplet,
supports its mixed-symmetry interpretation. However, the enhanced M1 decay of the 1$^+$, 2655.5-keV
level to the ground state in $^{144}$Nd \cite{NDS144} suggests that its structure is inconsistent
with that of the Q$_m$Q$_s$ multiplet member.

We note that there is a 1$^+$ level at 2585.6 keV in $^{142}$Nd \cite{NDS142}, a nucleus with no
valence neutrons expected to contribute to mixed-symmetry, Q$_m$ excitations. Such an effect was
also observed at N=50 in semi magic, $^{86}$Kr \cite{Urb16b} and $^{88}$Sr nuclei \cite{Urb21}. It
was suggested there that the levels, mimicking the Q$_m$Q$_s$ multiplet at N=50, are due to proton
$(f_{5/2}^{-1}p_{3/2})_{1^+,2^+,3^+,4^+}$ configuration, which may become a ``bone structure'' of
the Q$_m$Q$_s$ multiplet when enriched by the mixed-symmetry collectivity at N=52 and N=54
\cite{Urb16b,Urb21}.

By analogy, in the A$\approx$150 region one may expect the
$(g_{5/2}^{-1}d_{5/2})_{1^+,2^+,3^+,4^+,5^+,6^+}$ proton multiplet at the N=82 shell closure in
$^{142}$Nd. Figure \ref{A150_ms_states} shows 5$^+$ and 6$^+$ possible members above of this
multiplet. The multiplet seen at N=82 may become the ``bone structure'' of the Q$_m$Q$_s$
mixed-symmetry multiplet but also of the proto-$\gamma$ structure, the latter suggested by
quickly decreasing energies of 3$^+_1$ and 5$^+_1$ levels. A dedicated study is required to
verify these suggestions.

\subsection{0$^+$ excitations}

\subsubsection{General classification}

As seen in Fig. \ref{A150_Nd_0plus_2plus} there are four excited 0$^+$ levels located between
2 and 3 MeV in $^{142}$Nd. The same is observed in $^{144}$Nd but in $^{146}$Nd their energies
decrease rapidly, with the lowest three excited 0$^+$ levels dropping down by about 1 MeV. The
0$^+_2$ level in $^{146}$Nd has unusually low energy. The 0$^+_3$ and 0$^+_4$ levels are close
to each other indicating that their composition differs. This signals formation of new structures
at N=86, which at higher N develop into collective configurations.

The 0$^+_2$ levels at N$>$86, linked by the ``U''-shaped curve D in Fig. \ref{A150_Nd_0plus_2plus},
correspond to collective configurations, with 2$^+$ excitations on top of them as seen in Fig.
\ref{A150_Nd_BE2}. Similarly as energies of 2$^+_1$ levels in ground-state bands, the energies
of 2$^+_2\rightarrow 0^+_2$, in-band transitions decrease with the increasing neutron number.
The extension of curve D to N=86 includes the 0$^+_3$ level in $^{146}$Nd. The 2143.5-keV level
is probably the 2$^+$ excitation top of the 0$^+_3$level. It fits the energy expected from Fig.
\ref{A150_Nd_BE2} (a) and its 541.4-keV decay to the 0$^+_3$ level, found in the present work,
supports this proposition. We extend curve D to the 0$^+_4$ level N=84, as suggested by transfer
cross sections (see Section III.B.2).

Figure \ref{A150_Nd_0plus_2plus} suggests an avoided crossing at N=89 between the 0$^+_1$
and the 0$^+_2$ configurations. This is marked in the figure by dashed arrows indicating the
configuration exchange between 0$^+_1$ and 0$^+_2$ levels.

\begin{figure}
\centering
\scalebox{.35}{\includegraphics{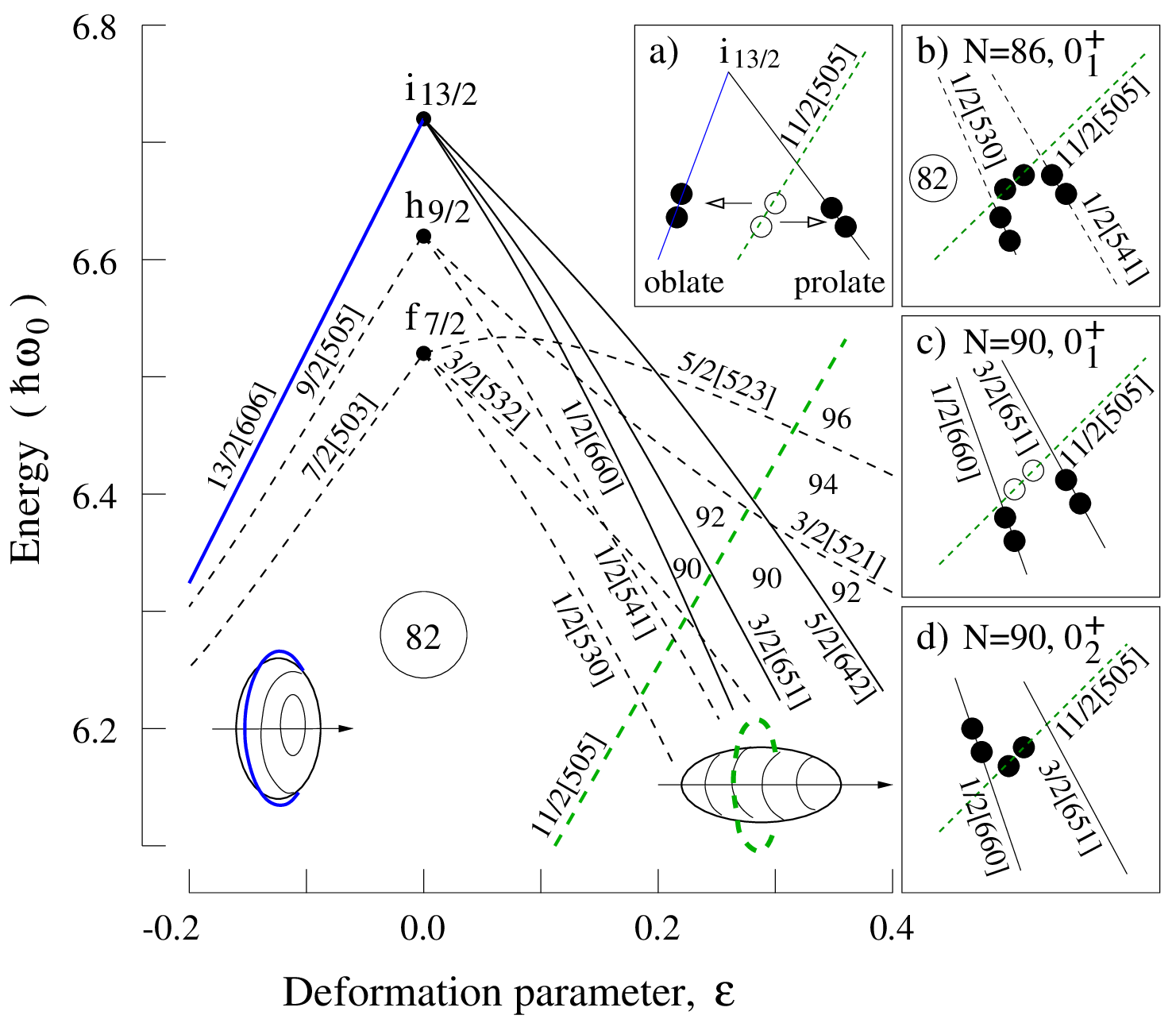}}
\caption{Energies of selected Nilsson levels for neutrons. The levels are drawn after Ref.
\cite{And76}.}
\label{A150_Nd_Nilsson_neutrons}
\end{figure}

The fact that the 0$^+_2$ levels are on a ``parabola'' similar to curves A and B linking 2$^+$
levels suggests that they are dominated by s.p. excitations. We propose that a neutron pair from
the $\nu 11/2^-$[505] extruder is involved, analogously to the 9/2$^+$[404] neutron extruder action
in Sr isotopes \cite{Urb21}. As illustrated in the inset (a) of Fig. \ref{A150_Nd_Nilsson_neutrons},
the $\nu 11/2^-$[505] extruder can pass its pair of neutrons to the low-$\Omega$, prolate or to
the high-$\Omega$, oblate down-slopping orbitals, both originating from the $i_{13/2}$ shell.
Such a transferred pair will polarize the potential producing prolate or oblate structures, as
sketched in the main panel of Fig.
\ref{A150_Nd_Nilsson_neutrons}.

\begin{figure}
\centering
\scalebox{.28}{\includegraphics{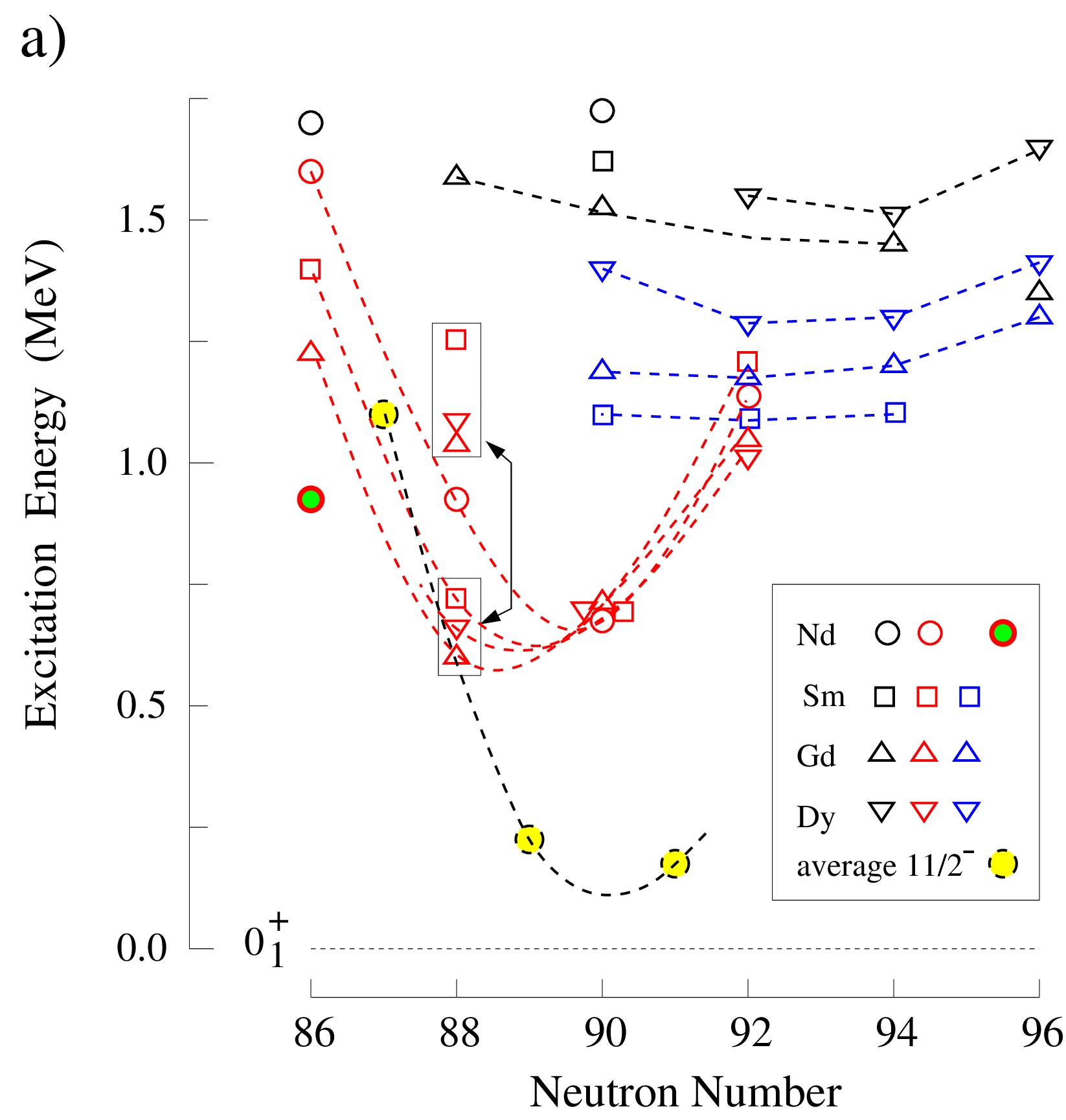}}
\scalebox{.355}{\includegraphics{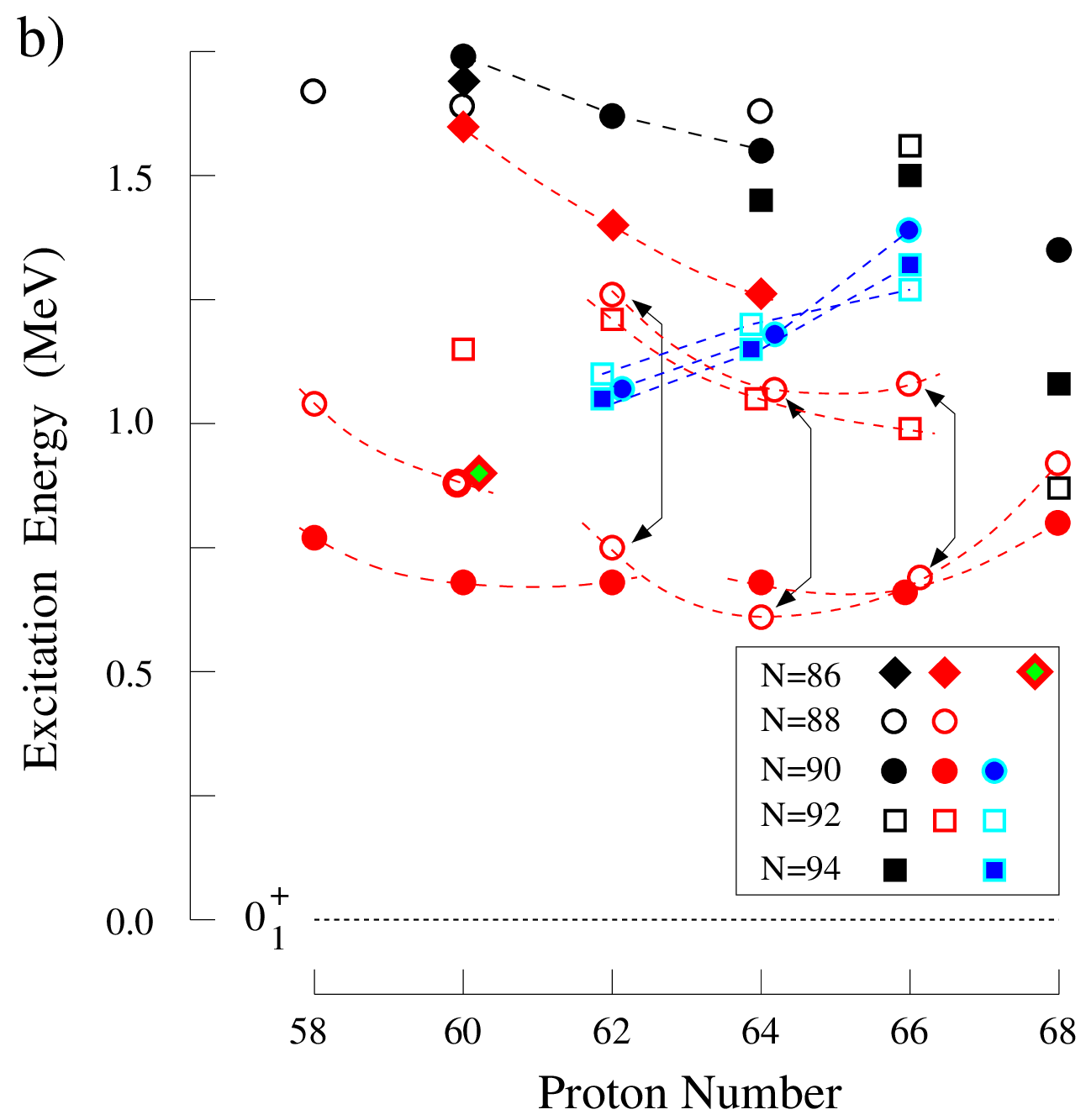}}
\caption{Excitation energies of $0^+$ levels up to 1.8 MeV in mass A$\approx$150 region shown
         a) for isotopes and b) for isotones. The data points are taken from the \cite{NNDC18}
         data base. See text for more comments.}
\label{A150_Nd_0plus_Z}
\end{figure}

Figure \ref{A150_Nd_0plus_Z} shows low-energy, 0$^+$ excitations in nuclei of the A$\approx$150
region along neutron number (part (a)) and proton number (part (b)). Below 1.7 MeV one expects
0$^+$ levels of a collective nature, albeit with leading two-particle components.

Excited 0$^+$ levels shown by red symbols are due to low-$\Omega$ orbitals of the $\nu i_{13/2}$
shell. At N=86 the 0$^+_1$ ground state is weakly deformed with no neutrons in the
deformation-driving orbitals of the $\nu i_{13/2}$ shell, as shown in the inset (b) of Fig.
\ref{A150_Nd_Nilsson_neutrons}.

When a pair of neutrons from the 11/2$^-$[505] extruder is passed to the low-$\Omega$,
1/2$^+$[660] orbital, a $prolate$ structure is formed. The 0$^+_3$ level in $^{146}$Nd
corresponds to such structure, which is more collective that the ground state in Fig.
\ref{A150_Nd_0plus_2plus}.

In Fig. \ref{A150_Nd_0plus_Z} (a) we show an average energy of 11/2$^-$ excitations from
Fig. \ref{A150_Nd_0plus_N} (c). Its trend is very similar to that of deformed (red), excited
0$^+$ levels, confirming the proposed involvement of the 11/2$^-$[505] neutron extruder.

A pair of neutrons from the 11/2$^-$[505] extruder can also be passed to the 13/2$^+$[606],
high-$\Omega$ orbital of the $\nu i_{13/2}$ shell forming an $oblate$ structure. We propose that
the 0$^+_2$ level in $^{146}$Nd shown in Figs. \ref{A150_Nd_0plus_Z} (a) and (b) by red-green
symbols corresponds to such configuration.

In the ENSDF compilation \cite{ENSDF} many of the 0$^+$ levels shown in red in Fig.
\ref{A150_Nd_0plus_Z}(a) are called $\beta$ vibrations, which, most likely they are not because
of significant variations of their energies. However, there are excited 0$^+$ levels around
1.3 MeV, shown by blue symbols, which are candidates for $\beta$ vibrational excitations because
of their weak dependence on the neutron number, expected for a collective, phonon-like
excitation.

The 0$^+$ levels shown in Fig. \ref{A150_Nd_0plus_Z} (a) are drawn in Fig. \ref{A150_Nd_0plus_Z}
(b) in function of proton number, Z, with the same the colours, i.e. red points from Fig.
\ref{A150_Nd_0plus_Z} (a) are shown in red in Fig. \ref{A150_Nd_0plus_Z} (b), etc. As in Fig.
\ref{A150_Nd_0plus_Z} (a) the corresponding 0$^+_2$ and 0$^+_3$ levels at N=88 in Sm, Gd and Dy
are linked by black arrows. Again, the red points show trends, which are different from trends
for the proposed $\beta$ vibrations (blue points). The 0$^+_2$ level at 1139
keV in $^{152}$Nd (red empty square) may alternatively be classified as $\beta$ vibration. In
this context an experimental search for 0$^+$ levels at similar energies in
$^{150}$Nd and $^{154}$Nd is of interest (see also Fig. \ref{A150_Nd_0plus_Z} (a)).

Figure \ref{A150_Nd_0plus_Z} (a) shows that the minimum of red ``parabolas'' for Sm, Gd and Dy
isotopes is shifted to lower N compared to Nd points. This may be due to extra lowering energy
of 0$^+_2$ levels in Sm, Gd and Dy at N=88 caused by repulsion with 0$^+_3$ levels. The
corresponding pairs of levels are within black rectangles in Fig. \ref{A150_Nd_0plus_Z} (a).
They are linked by black arrows in Figs. \ref{A150_Nd_0plus_Z} (a) and \ref{A150_Nd_0plus_Z} (b).

The 0$^+$ levels shown by black symbols in Fig. \ref{A150_Nd_0plus_Z} are not yet interpreted.
Their weak dependence on neutron number suggests their collective nature but their excitation
energies slowly decreasing with proton number distinguishes them from $\beta$ vibrations, which
increase their energies with increasing Z.

The proposed classifications of 0$^+$ levels allow to say that the ``wide parabolas'' of Fig.
\ref{A150_Nd_0plus_N} (a) appear to be artefacts. In Fig. \ref{A150_Nd_0plus_2plus} the lower
parabola from Fig. \ref{A150_Nd_0plus_N} (a), drawn as black dotted line, connects 0$^+$
levels which are of different origin, namely spherical 2-q.p. 0$^+_2$ levels at N=82 and N=84,
an oblate 0$^+_2$ level at N=86 and two deformed 0$^+_2$ at N=90 and N=92, which also appear
to have different structures, as discussed above.

\subsubsection{S$_{2n}$ separation energies and 2n-transfer cross sections.}

Two-neutron separation energies, S$_{2n}$, and two-neutron-transfer cross sections are
sensitive probes of the population of valence orbitals by pairs of neutrons. They can be used
to test scenarios of population of 0$^+$ levels in even-even nuclei of the A$\approx$150
region proposed above.

The S$_{2n}$ values, constantly being improved \cite{Jar24}, exhibit local variationsas seen
in the systematics available at NuDat data base \cite{NUDAT}. These variations were analysed
in the past in function of N or Z using the dS$_{2n}$ ``derivative'' of S$_{2n}$(Z,N)
\cite{Ang09,Spa25}). Such analysis reveals sudden changes of S$_{2n}$ at the magic neutron
numbers and at some other places (see Figs. 4 and 6 of Ref. \cite{Ang09}). To observe
variations over a wider nucleon range, such as span of the $\nu i_{13/2}$ shell, we propose
another approach.

As seen in the NuDat systematics \cite{NUDAT} (see also Fig. 7a of Ref. \cite{Ang09}), except
of local deviations, the S$_{2n}$ energy changes nearly linearly, increasing with growing Z and
decreasing with growing N (see also formula (6) in Ref. \cite{Gar05}). This allows to define
{\it local} reference plane, S$_{2n}^{ref}$(N,Z) = Z$\times$A$_n$ - N$\times$B$_n$ + C$_n$,
for the studied region, and use it to show the $\Delta$S$_{2n}$(N,Z) = S$_{2n}$(Z,N) -
S$_{2n}^{ref}$(N,Z) deviations from this reference over wider nucleon range in this region.

\begin{figure}
\centering
\scalebox{.36}{\includegraphics{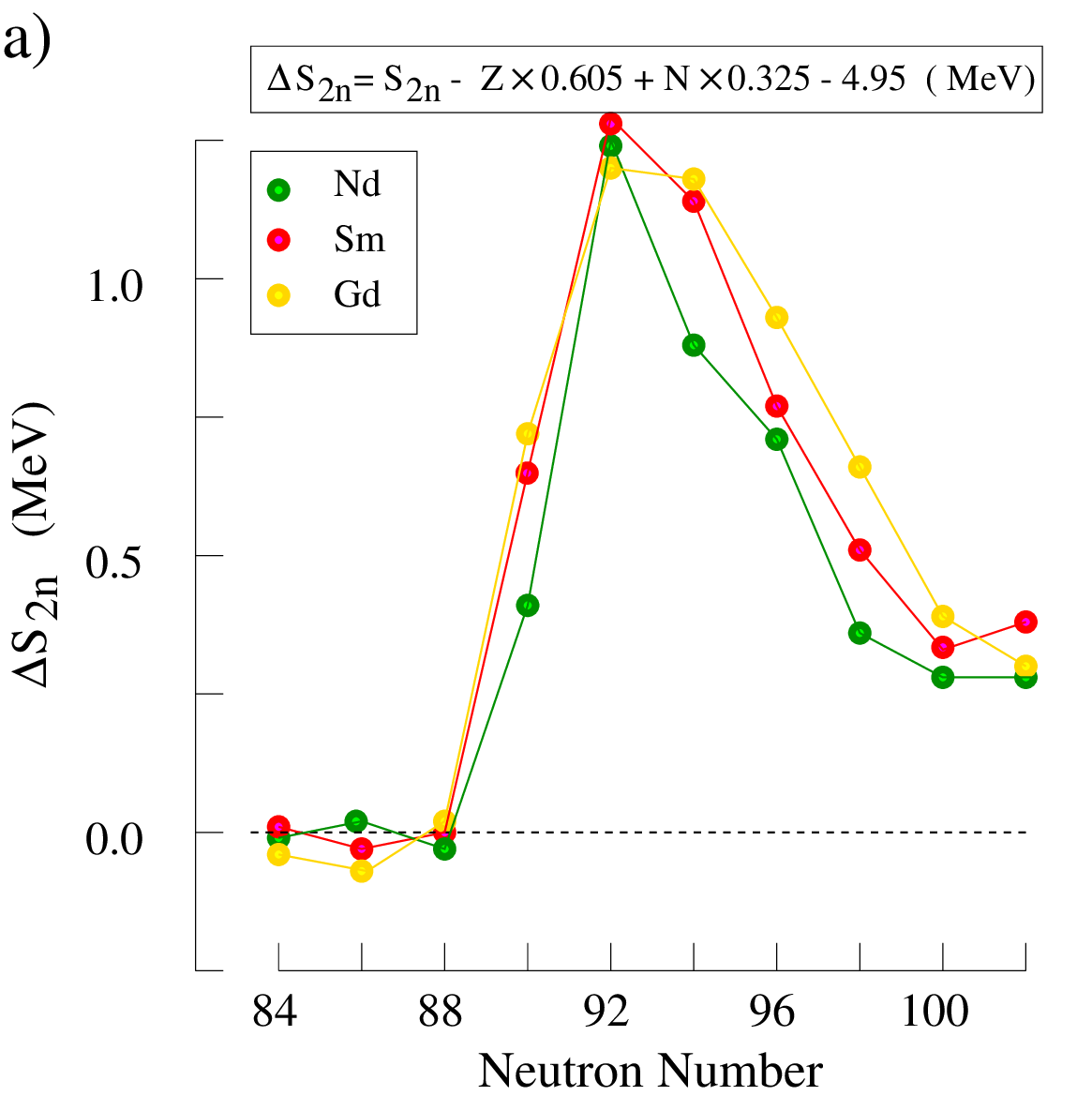}}
\scalebox{.36}{\includegraphics{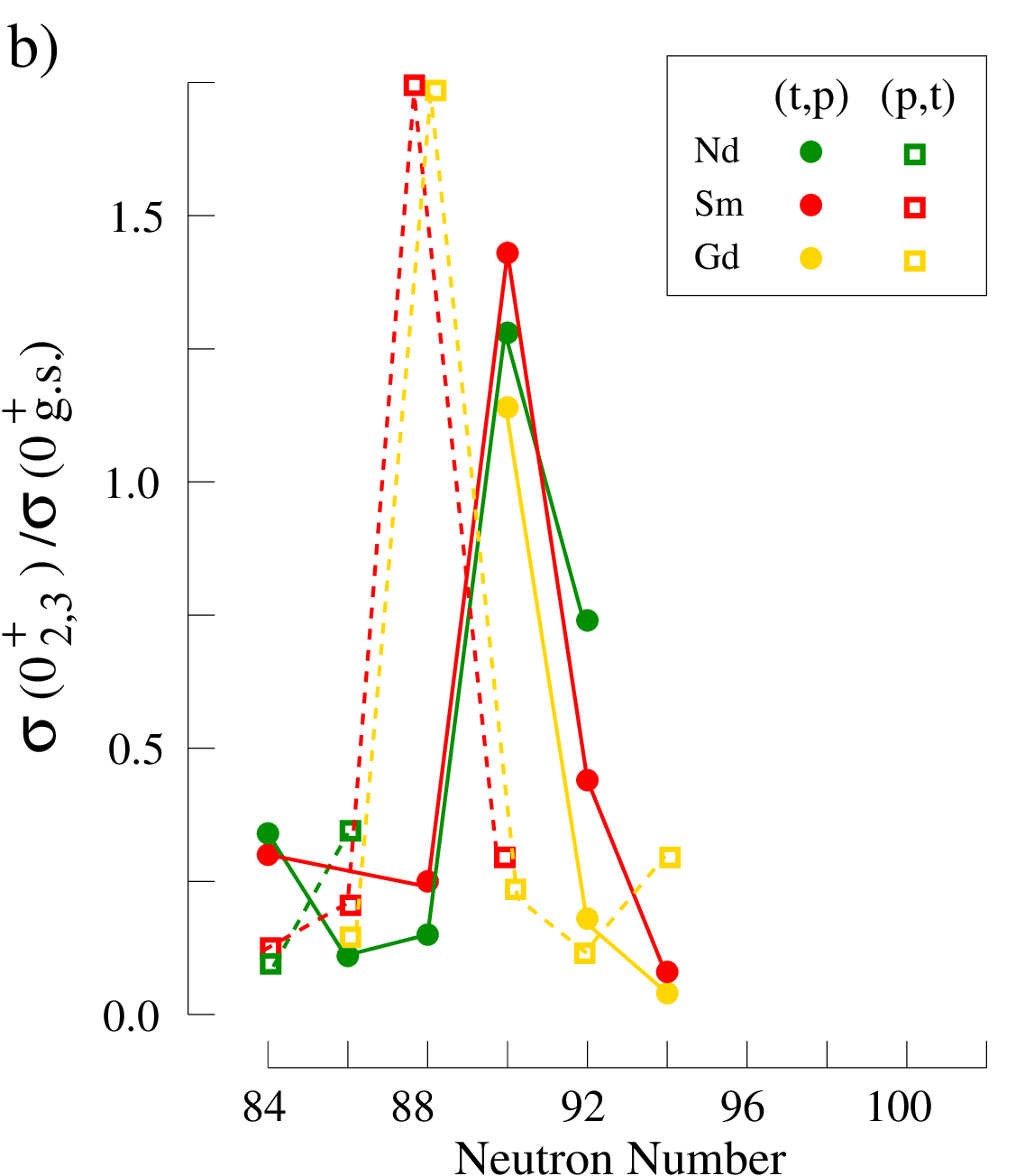}}
\caption{a) Local excess, $\Delta$S$_{2n}$(N,Z), of S$_{2n}$ binding energy in the A$\approx$150
         region. See text for the definition of $\Delta$S$_{2n}$(N,Z). The data are taken
         from the NuDat data base \cite{NUDAT}.\\
         b) Ratio of cross sections $\sigma$(0$^+_2$)/$\sigma$(0$^+_1$) for (t,p) and (p,t)
         transfer reactions. The data are taken from Refs. \cite{Bje66,McL69,Cha72,Apr25}.
}
\label{A150_Nd_S2n_transfer_excess}
\end{figure}

Figure \ref{A150_Nd_S2n_transfer_excess} (a) shows $\Delta$S$_{2n}$(N,Z) deviations in the
A$\approx$150 region. The coefficient A$_n$=0.605 MeV/Z has been obtained from fitting the
increase rate of the S$_{2n}$ separation energy along the N=82, 84, 86 and 88 isotonic lines in
the 58$\leq$Z$\leq$66 proton range. Nuclei in these proton and neutron ranges are expected to
be spherical and the A$_n$ values along the 84, 86 and 88 isotonic lines are consistent with
the A$_n$ coefficient determined along the N=82 line of semi magic, spherical isotones. The
coefficient B$_n$=0.325 MeV/N has been obtained from fitting the decrease rate of the S$_{2n}$
separation energy along the line of spherical, semi magic Pb isotopes in the neutron range
100$\leq$N$\leq$104 and along lines of Nd and Sm isotopes in the 84$\leq$N$\leq$88 neutron
range where nuclei are expected to have spherical shapes. The coefficient B$_n$ along the Pb
line is consistent with those along the Nd and Sm isotopic lines. The $local$ normalization
coefficient C$_n$=4.95 MeV has been adjusted to reproduce the average S$_{2n}$ value for
$^{144,146,148}$Nd and $^{146,148,150}$Sm nuclei.

Figure \ref{A150_Nd_S2n_transfer_excess} (a) reveals a sudden increase of the $\Delta$S$_{2n}$(N,Z)
deviation above N=88, where low-$\Omega$ neutron orbitals of the $\nu i_{13/2}$ shell start to be
populated (see Fig. \ref{A150_Nd_Nilsson_neutrons}). It drops later to about 0.4 MeV after the
low-$\Omega$ neutron orbitals of the $\nu i_{13/2}$ shell are filled. One notes the high regularity
and similarity of the trends in the three isotopic chains shown. The $\Delta$S$_{2n}$(N,Z) analysis
hown in Fig. \ref{A150_Nd_S2n_transfer_excess} (a) provides extended information compared to the
dS$_{2n}$ ``derivative'' analysis.

The S$_{2n}$ deviation around N=90 has been noted before in Ref. \cite{Bar64} though their
conclusion of the effect being much smaller for neodymium is not correct, as shown by our
$\Delta$S$_{2n}$(N,Z) analysis. In Ref. \cite{Mac70} a concept of nuclear deformation energy was
introduced, as the difference between M(Z,N) mass of a deformed nucleus A(Z,N) and M$_s$(Z,N) mass
of this nucleus if it was spherical. Using their language one may say that the our
S$_{2n}^{ref}$(N,Z) reference plain represents two-neutron separation energy in nuclei if they
were spherical whereas the $\Delta$S$_{2n}$(N,Z) quantity relates to deformation energy although
the analogy with equation (1) of Ref. \cite{Mac70} is not exact.

\begin{figure*}
\centering
\scalebox{.35}{\includegraphics{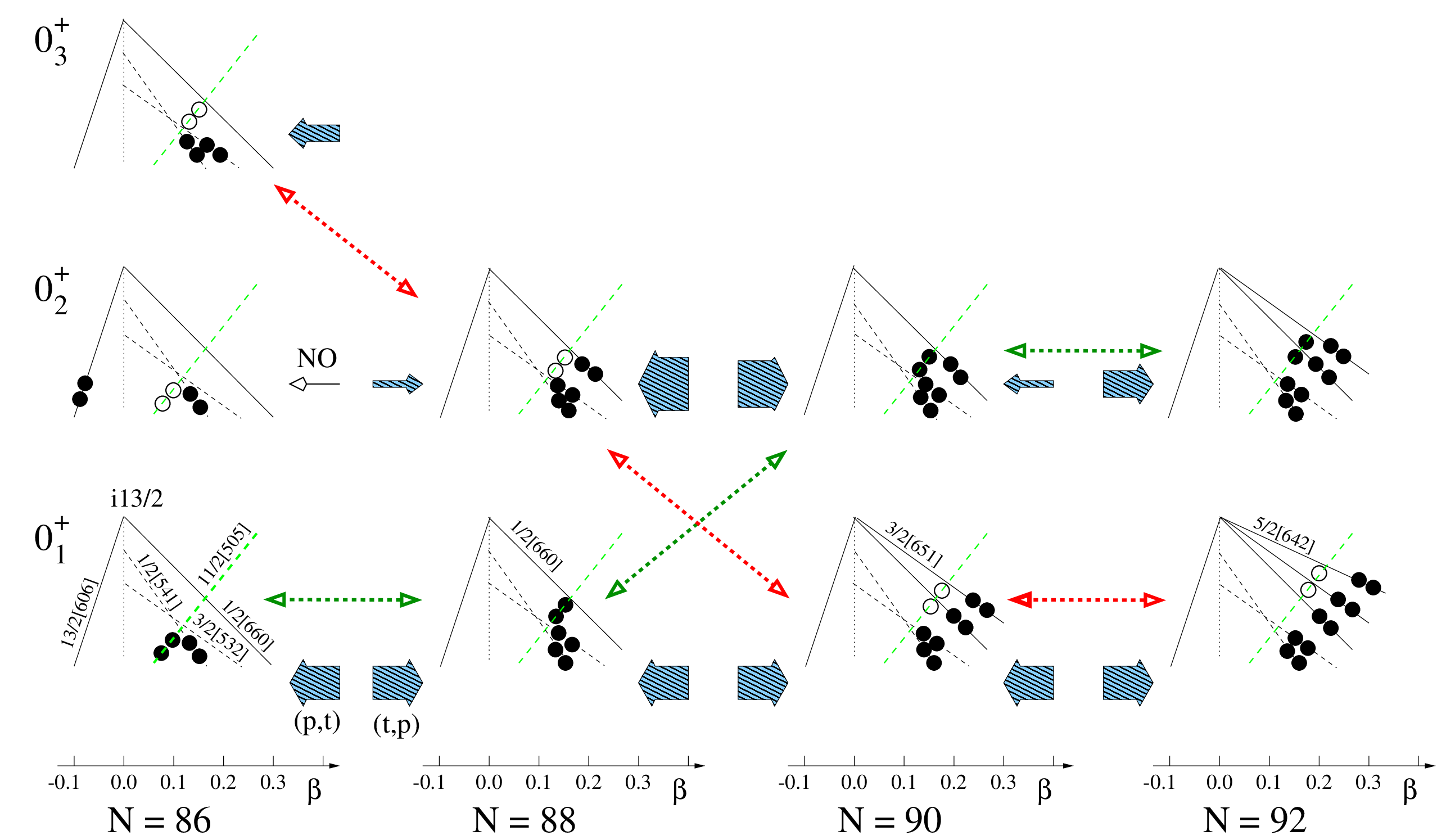}}
\caption{Schematic representation of dominating configurations in $0^+_1$, $0^+_2$ and $0^+_3$
         levels in 86$\leq$N$\leq$92 isotones and the strength of two-neutron transfer reactions.
         The transfer-strength data are taken from Refs. \cite{,Bje66,McL69,Cha72,Apr25}.
         See text for further comments.}
\label{A150_Nd_transfer}
\end{figure*}

Figure \ref{A150_Nd_S2n_transfer_excess} (b), similar to Fig. 9 of Ref.\cite{Cha72}, shows the ratio
of the total two-neutron L=0 transfer intensity to 0$^+_2$ and 0$^+_3$ states to the ground-state
L=0 transfer intensity. In the drawing we used relative cross sections for (t,p) transfer reaction
from Refs.\cite{Cha72,Bje66} and from the recent review work \cite{Apr25}. In addition to the (t,p)
rates we have drawn in Fig. \ref{A150_Nd_S2n_transfer_excess} (b) relative cross sections for the
(p,t) transfer reaction using data from Ref. \cite{Apr25}.

One notes extraordinary peaks in (p,t) and (t,p) cross sections at N=88 and N=90, respectively,
which correlate with the start of the $\Delta$S$_{2n}$(N,Z) (and deformation) increase. The jump is
observed in Nd, Sm and Gd isotopes for the (t,p) reaction and in Sm and Gd isotopes for the (p,t)
(it is of great interest to find the (p,t) cross section in $^{148}$Nd). The suddenness of the
observed changes suggests that the effect relates to local population of valence orbitals by
neutron pairs at N=88 and N=90.

To understand the effects shown in Figs. \ref{A150_Nd_S2n_transfer_excess} (a) and (b) and their
correlation with the evolution of the 0$^+$ configurations, we will utilize the proposed concept
of neutron-pair transfer from the extruder orbital.

Figure \ref{A150_Nd_transfer} is a schematic representation of the dominating valence-neutron
configurations in the $0^+_1$, $0^+_2$ and $0^+_3$ levels of 86$\leq$N$\leq$92 isotones. Blue
arrows show the relative (t,p) and (p,t) cross sections from Fig\ref{A150_Nd_S2n_transfer_excess}
(b). The arrow thickness is is proportional to cross sections averaged over Nd, Sm and Gd. Dashed
red arrows link more collective (deformed) $0^+$ levels and dashed green arrows link less
collective $0^+$ levels. The crossing of red and green arrows marks the avoided crossing of
0$^+_1$ and 0$^+_2$ levels around N=89, shown by dashed arrows in Fig. \ref{A150_Nd_0plus_2plus}.

Above N=88 the wave function of the $0^+_1$, ground-state acquires subsequent neutron pairs on
the down-slopping, low-$\Omega$ orbitals of the $\nu i_{13/2}$ shell. This drives the nucleus
towards prolate deformation, compatible with the ``prolate geometry'' of the populated,
low-$\Omega$ neutron orbits. Therefore, one expects an increase of the S$_{2n}$ separation energy
as seen in Fig. \ref{A150_Nd_S2n_transfer_excess} (a). The S$_{2n}$ energy saturates at N=92
where the prolate-driving orbitals of the $\nu i_{13/2}$ shell are filled. At higher neutron
number high-$\Omega$, ``oblate'' neutron orbitals of the $\nu i_{13/2}$ shell in the prolate-
deformed potential are populated and, therefore, the S$_{2n}$ separation energy decreases.

The key proposition is that in the $0^+$ levels linked by green dashed arrows the 11/2$^-$[505]
extruder is occupied by a pair of neutrons whereas in  $0^+$ levels linked by red dashed arrows
this neutron pair is passed from the 11/2$^-$[505] extruder to a deformation-driving orbital.
Consequently the $0^+$ levels linked by red dashed arrows are more collective. In particular, at
N=88 the $0^+_2$ level is more collective than the $0^+_1$ ground state. This is indicated by
lower energy of the in-band 2$^+_2 \rightarrow 0^+_2$ transition and the newly obtained, higher
total B(E2) decay strength from the 2$^+_2$ level, compared to analogous values for the
2$^+_1\rightarrow 0^+_1$ transition as seen in Fig. \ref{A150_Nd_BE2} (a) and (b). Furthermore,
the new B(E2) rate for the 2$^+_2 \rightarrow 0^+_2$ transition at N=90 found in this work is
lower than that for the 2$^+_1\rightarrow 0^+_1$ transition. In the avoided crossing between
N=88 and N=90 the $0^+_1$ and $0^+_2$ levels exchange their properties and from N=90 the ground
states are more collective (deformed).

The proposed dominating valence-neutron configurations shown in Fig. \ref{A150_Nd_transfer}
allow to understand sudden local changes of relative transfer cross sections in Fig.
\ref{A150_Nd_S2n_transfer_excess}

- the population of the 0$^+_2$ level at N=88 in the (p,t) reaction on N=90 target is
favoured because removing the pair of neutrons from 3/2$^+[651]$ orbital in the 0$^+_1$,
ground-state configuration leads to the proposed wave function of the 0$^+_2$ level
at N=88. At the same time the (p,t) transfer to the 0$^+_1$ ground state at N=88 is hindered
because its wave function differs from the wave function of the 0$^+_1$ ground state at N=90.
This results in large relative (p,t) cross section to the 0$^+_2$ level at N=88

- the population of the 0$^+_2$ level at N=90 in the (t,p) reaction on the N=88 target is
favoured because adding a pair of neutrons to the 1/2$^+[660]$ orbital in the 0$^+_1$,
ground-state configuration at N=88 leads to the wave function of the 0$^+_2$ level at N=90.
At the same time the wave function of the 0$^+_1$ ground state at N=90 is much different
hindering the (t,p) transfer to this level. This will result in a large relative (t,p)
cross section to the 0$^+_2$ level at N=90

- populations of the 0$^+_1$ and  0$^+_2$ levels at N=92 in the (t,p) reaction on the N=90
target and of the 0$^+_1$ level at N=90 in the (p,t) reaction on the N=92 target are favoured
whereas the population of the 0$^+_2$ at N=90 in the (p,t) is hindered. This results in a small
relative (p,t) cross sections to 0$^+_2$ level at N=90, and a moderate (t,p) cross sections to
the 0$^+_2$ level at N=92.

- the population of the 0$^+_3$ level at N=86 in the (p,t) reaction on N=88 target is
favoured whereas (p,t) transfer to the 0$^+_2$ level is strongly hindered because of its much
different, probably oblate, structure. No such transfer has been observed despite devoted
efforts \cite{Apr25}. One also notes the weak (t,p) strength to the 0$^+_2$ level at N=88.

Transfer rates to levels shown in Fig. \ref{A150_Nd_S2n_transfer_excess}  (b) and to some other
levels need further comments

- transfer rates shown in Fig. \ref{A150_Nd_S2n_transfer_excess} (b) are relative values and,
apparently, the high rate for the 0$^+_{g.s.} \rightarrow 0^+_2$ (t,p) transfer rate on N=88
target is rather due to a significant decrease in the 0$^+_{g.s.}\rightarrow 0^+_{g.s.}$ (t,p)
rate on N=88 target, as shown by the absolute (t,p) transfer strength for Sm isotopes reported
in Ref. \cite{Bje66}. This is supported by calculations in Ref. \cite{Gar20} (see Fig. 9 (a) there).

- the (t,p) strength to both, 0$^+_2$ and 0$^+_3$ states in $^{152}$Sm and$^{154}$Gd is
large and of similar amplitude. This indicates identical neutron but different proton structure
of the two levels, which repeal each other from original positions, as marked in Fig.
\ref{A150_Nd_0plus_Z} by black arrows. The similarity of their neutron structure is the reason
for showing in Figs. \ref{A150_Nd_S2n_transfer_excess} and \ref{A150_Nd_transfer} summed strength
to 0$^+_2$ and 0$^+_3$ states. In contrast, the transfer strength in $^{150}$Nd is concentrated
predominantly in the 0$^+_2$ level. This suggests different proton structure in Nd compared to Sm
and Gd (see also the discussion in Ref. \cite{For19}).

- oblate structures in transitional nuclei just above closed shells are rare phenomena, though
persistently suggested by calculations \cite{Rod10,Pet12,Lun25}. We have proposed that the
0$^+_2$ levels in $^{98}$Sr and $^{100}$Zr \cite{Urb19} and now in $^{146}$Nd are oblate
structures created by a specific action of extruder orbitals. All three levels are reported as
certain in the ENSDF compilation \cite{ENSDF} though the recent review \cite{Apr25} have
questioned the existence of the 0$^+$ level at 915.4 keV in $^{146}$Nd. The present work supports
its presence. It is of interest to verify further this very specific level.

Recent calculation in the A$\approx$150 region, called PNBCS approach \cite{Guo25}, has suggested
that the population of the $f_{7/2}$ proton and $g_{9/2}$ neutron shells, positioned above Z=82
and N=126 shell closures, respectively, is crucial for describing the sudden shape change in
the Nd isotopes between N=88 and N=90. Reproducing in PNBCS the S$_{2n}$ separation energies and
the 2n-transfer cross sections discussed above would be a useful test of this idea.

\subsubsection{Proton 9/2[404] extruder and S$_{2p}$ separation energies }

\begin{figure}[]
\centering
\scalebox{.33}{\includegraphics{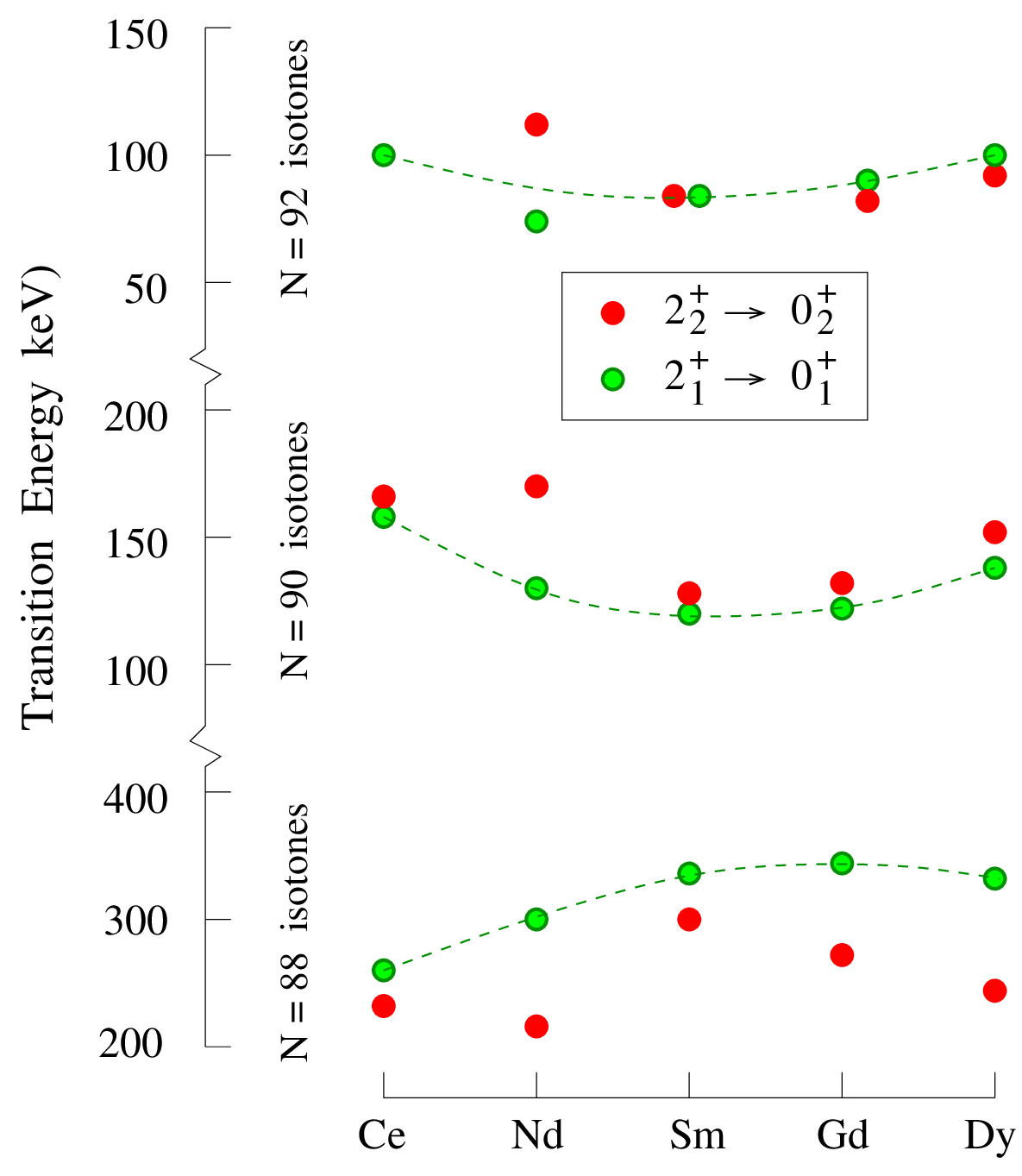}}
\caption{Energies of 2$^+_1\rightarrow 0^+_1$ and 2$^+_2\rightarrow 0^+_2$ transitions in
         Nd isotopes. Dashed lines are drawn to guide the eye. The data are taken from the
         compilations \cite{NDS142,NDS144,NDS146,NDS148,NDS150,NDS152,NDS154}.}
\label{A150_Nd_2plus_N90-N92}
\end{figure}

The study of the 0$^+_2$ level in $^{150}$Nd$_{90}$ \cite{Fog86} reported that its deformation is
significantly lower than the deformation of the ground state, in contrast to other N=90 isotones,
as illustrated in Fig. \ref{A150_Nd_2plus_N90-N92}. Dashed lines mark ``smooth''trends
for ground states along N=88, N=90 and N=92 isotones. For Nd isotopes one sees deviations from
these trends. An interesting conclusion of Ref. \cite{Fog86} was that the anomaly may be
due to a proton orbital.

The effect was studied further in $^{152}$Nd$_{92}$ in Ref. \cite{Hel92}. Again, lower deformation
of the 0$^+_2$ level compared the 0$^+_1$ ground state was reported. It was concluded that the
effect is due to protons and is not present in $^{148}$Nd. Figure \ref{A150_Nd_2plus_N90-N92}
shows that in $^{148}$Nd the deformation of the 0$^+_2$ level is higher than that of the 0$^+_1$
level, suggesting structure exchange between 0$^+_1$ and  0$^+_2$ around N=89. This is also
suggested by a reversal of ``smooth'' trends in the two isotonic chains. The anomaly reported in
Refs. \cite{Fog86,Hel92} is consistent with the scenario shown in Fig. \ref{A150_Nd_transfer}

Figure \ref{A150_Nd_2plus_N90-N92} suggests that at N=92 there is an extra increase of deformation
in the ground state. We proposed recently \cite{Urb20} that in addition to the 11/2$^-$[505]
neutron extruder involvement in creating 0$^+$ excited levels in mass A$\approx$150 region there
may be an analogous action on the proton side employing the 9/2$^+$[404] proton extruder.

\begin{figure}
\centering
\scalebox{.35}{\includegraphics{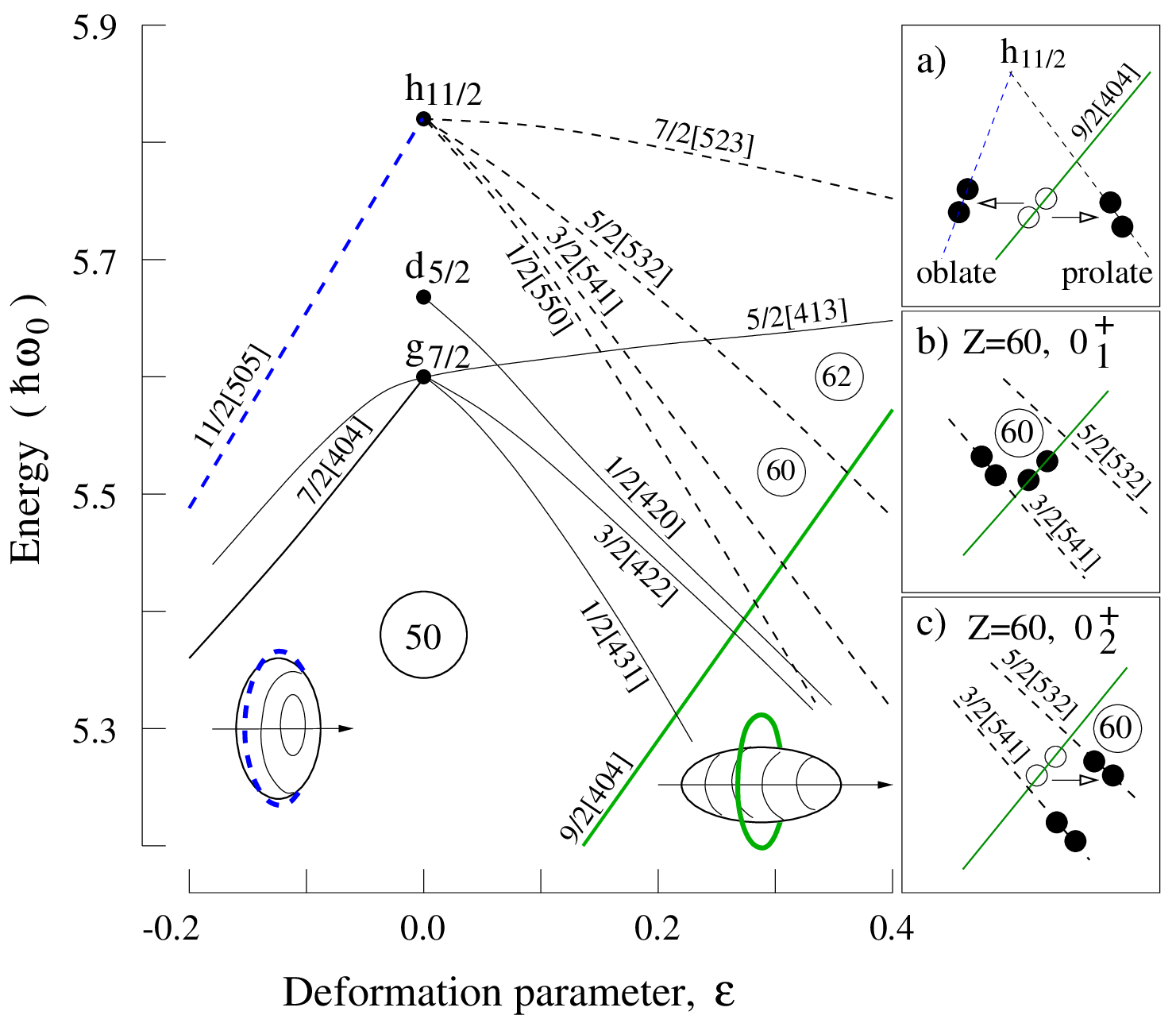}}
\caption{Energies of selected Nilsson levels for protons. The levels are drawn after Ref.
\cite{And76}.}
\label{A150_Nd_Nilsson_protons}
\end{figure}

Figure \ref{A150_Nd_Nilsson_protons} illustrates a possible mechanism creating excited 0$^+$
levels in A$\approx$150 nuclei, which involves the 9/2$^+$[404] proton extruder. It may produce
an oblate as well as a prolate 0$^+$ state, as shown in the inset (a). Furthermore, on the
prolate side one may expect two nearby 0$^+$ states with different deformations, as sketched
in the insets (b) and (c).

\begin{figure}
\centering
\scalebox{.34}{\includegraphics{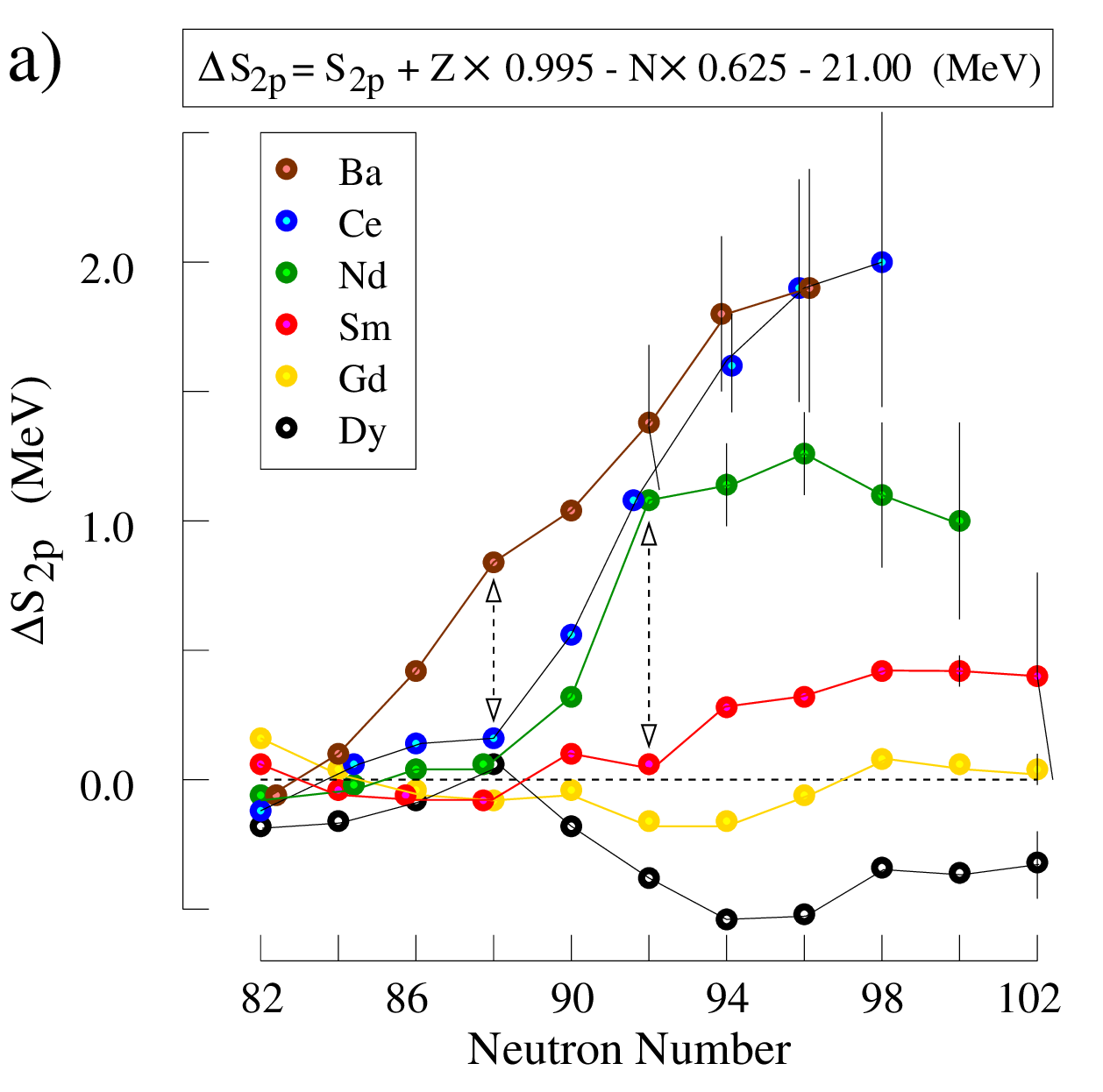}}
\scalebox{.34}{\includegraphics{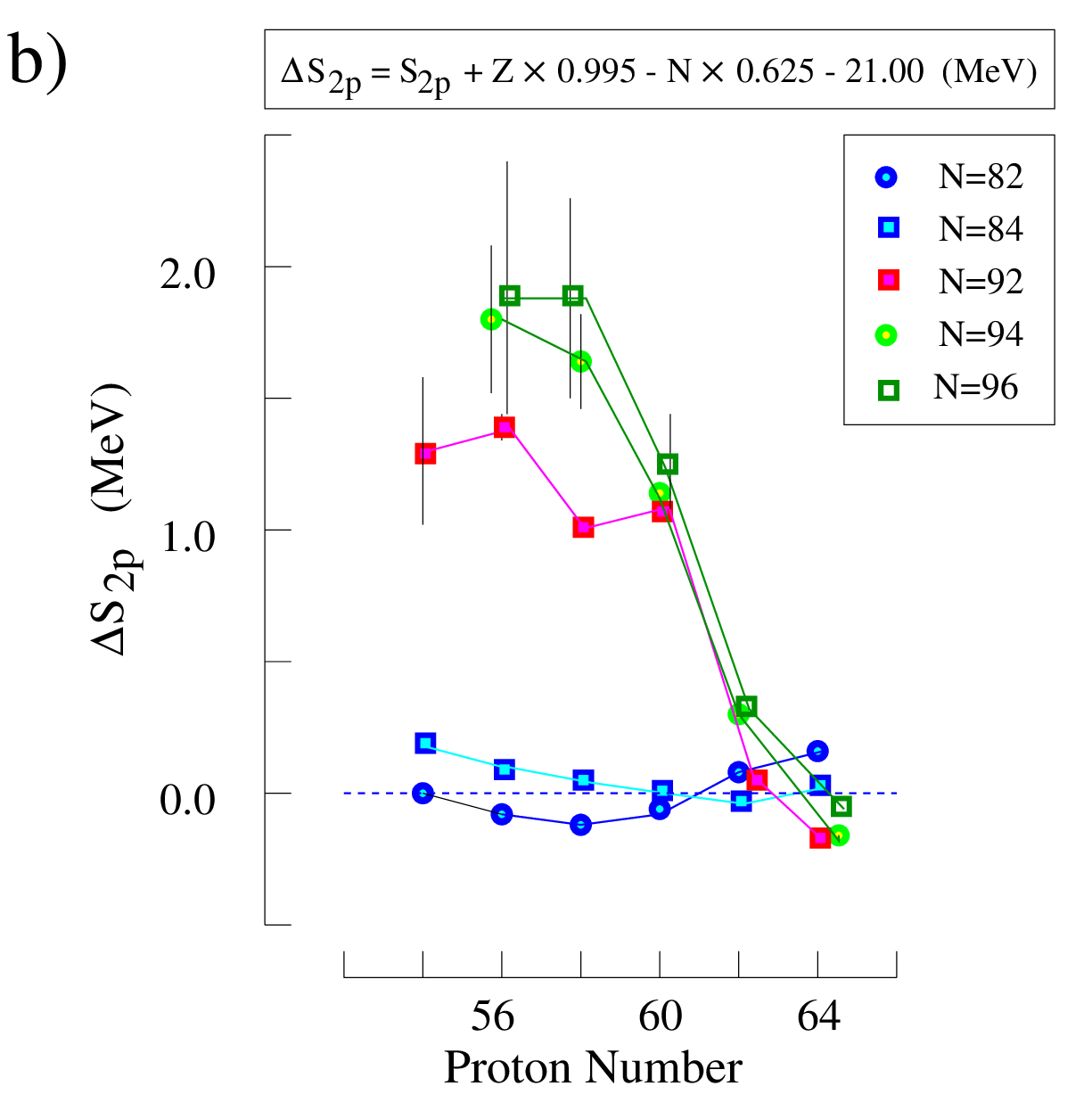}}
\caption{Local excess, $\Delta$S$_{2p}$(N,Z), of S$_{2p}$ separation energy in the A$\approx$150
         region a) in function of N, and b) in function of Z. The data are taken from the NuDat
         data base \cite{NUDAT}. See text for more comments and the definition of
         $\Delta$S$_{2p}$(N,Z).}
\label{A150_Nd_S2p_excess}
\end{figure}

It requires further studies to understand the role of the $\pi$9/2$^+$[404] extruder in Nd
nuclei as well as in Sm, Gd and Dy isotopes, where it may help explaining the 0$^+_2$-0$^+_3$
``doublets'' marked by black arrows in Fig. \ref{A150_Nd_0plus_Z}. However, some clues are
already seen in Fig. \ref{A150_Nd_S2p_excess}, displaying an excess of two-proton separation
energy, $\Delta$S$_{2p}$(N,Z), defined similarly to the $\Delta$S$_{2n}$(N,Z) excess discussed
above, as $\Delta$S$_{2p}$(N,Z) = S$_{2p}$(Z,N) - S$_{2p}^{ref}$(N,Z). Here the coefficients of
the local reference plane S$_{2p}^{ref}$(N,Z) = Z$\times$A$_p$ - N$\times$B$_p$ + C$_p$ have
been obtained from fitting to twelve Nd, Sm, and Gd nuclei along the N=82, 84, 86 and 88 isotonic
lines to reproduce, on average, the S$_{2n}$ energy, which changes almost linearly in this (N,Z)
fitting range.

As seen in Fig. \ref{A150_Nd_S2p_excess} (a), above N=90 a large gap opens between
$\Delta$S$_{2p}$(N,Z) values in Ba, Ce and Nd isotopes and $\Delta$S$_{2p}$(N,Z) values in
Sm, Gd and Dy isotopes. This is even more evident in Fig. \ref{A150_Nd_S2p_excess} (b), which
reveals an abrupt rise of $\Delta$S$_{2p}$(N,Z) values from Gd and Sm to Nd, Ce and Ba isotopes
along N=92, 94 and 96 isotonic lines.

A possible explanation is that in the strongly deformed, prolate potential developed above N=90
in Sm, Gd and Dy isotopes by neutrons populating the deformation-driving orbitals of the
$\nu i_{13/2}$ shell, the occupied 9/2$^+$[404] proton extruder is elevated by proton-neutron
interactions and passes its pair of protons to down-slopping orbitals of the $\pi h_{11/2}$
shell (see Fig. \ref{A150_Nd_Nilsson_protons}). This mechanism is particularly effective in
$^{152}$Nd, as seen in Fig. 8 of Ref. \cite{Urb20}.

One may conclude, that at N$>$90 in ground states below Z=62 protons occupy prolate orbitals of
the $\nu h_{11/2}$ shell while the $\pi 9/2^+[404]$ oblate orbital is empty. In the prolate
potential this results in large $\Delta$S$_{2p}$(N,Z). However, at higher Z the protons stay
on the oblate $\pi 9/2^+[404]$ extruder, because at Z$>$60 passing proton pair from the
$\pi 9/2^+[404]$ extruder to the $\pi h_{11/2}$ shell is less effective, as seen in Fig.  \ref{A150_Nd_Nilsson_protons}. The low $\Delta$S$_{2p}$(N,Z) in Sm, Gd and Dy isotopes seen in
Fig. \ref{A150_Nd_S2p_excess} reflect the fact that the oblate $\pi 9/2^+[404]$ extruder is
incompatible with the prolate potential. A two-proton pickup reaction on the $^{154}$Sm target
could help verifying this explanation.

\subsubsection{Calculations of 0$^+$ levels}

\begin{figure}
\centering
\scalebox{.21}{\includegraphics{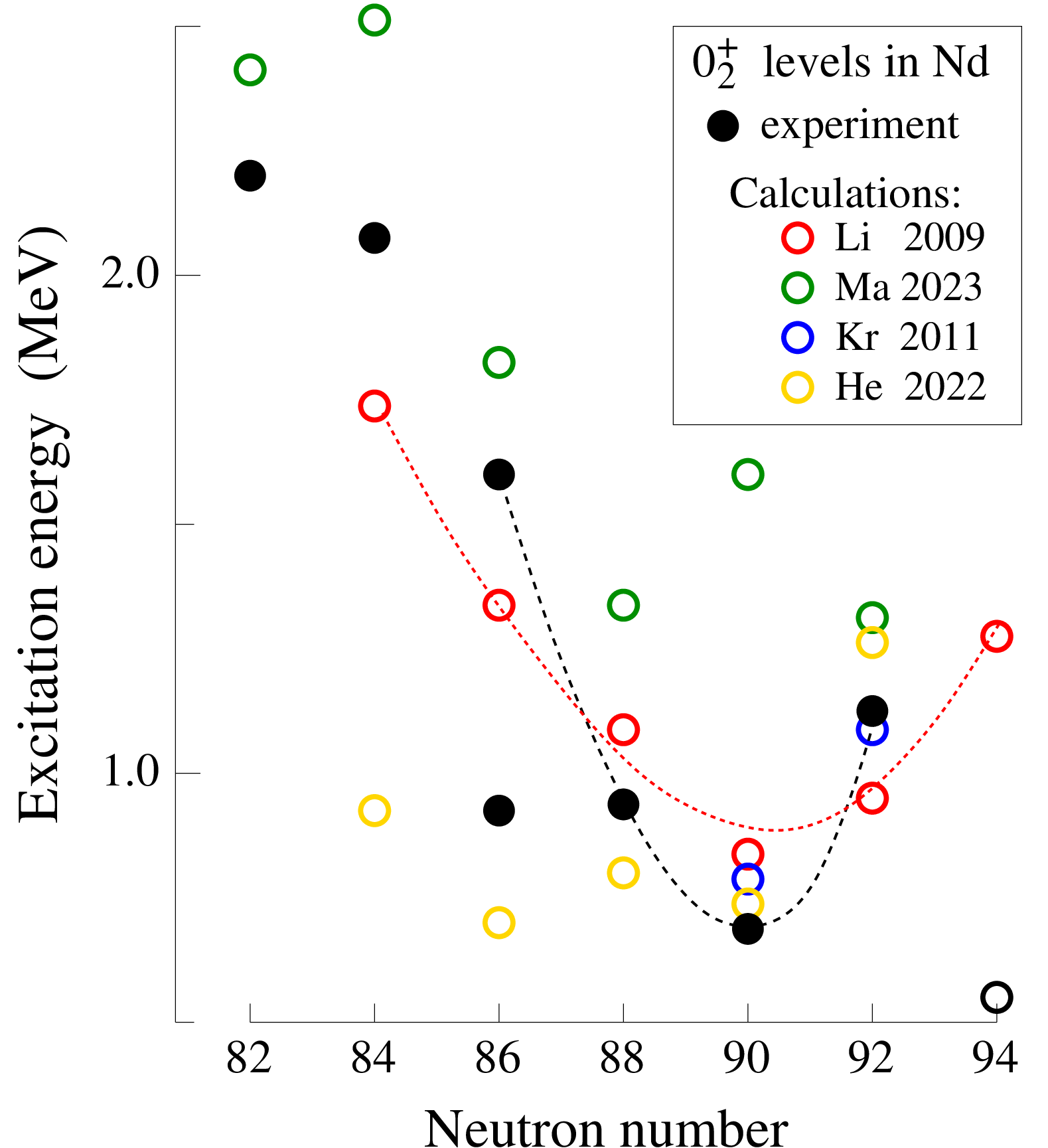}}
\caption{Experimental excitation energies of $0^+_2$ levels in even-even Nd isotopes compared
         to calculations of Li2009 \cite{Li009}, Ma2023 \cite{Ma023}, Kr2011 \cite{Kru11} and
         He2022 \cite{He022}. Dashed lines are drawn to guide the eye. See text for more comments.}
\label{A150_Nd_0plus_N_th}
\end{figure}

In the study of transitional Sr isotopes \cite{Urb21} we used the Large-Scale Shell Model (LSSM)
to interpret excited levels. In contrast to the excellent description of 2$^+$ excitations, the
LSSM  was not able to describe properly $0^+$ excitations. In the A$\approx$150 region the
situation is even more difficult as one should use rather truncated basis of shells in LSSM
calculations to allow calculations in reasonable time. Therefore, other model calculations
available in the literature have been copmared to experimental data,  as shown in Fig. \ref{A150_Nd_0plus_N_th}.

The calculations of Ref. \cite{Li009} (red circles) predict 0$^+_2$ excitation energies, which
are close to experimental values in five Nd isotopes and reproduce properly the position of
the minimum energy in function of N. However, at N=86 the calculations fail to reproduce the
0$^+_2$ level, proposed above to be the special oblate structure generated by the extruder.
They are closer to the 0$^+_3$ level in $^{146}$Nd, interpreted above as collective configuration.
The  0$^+_3$ at N=86 and 0$^+_2$ at N=84 calculated levels are too collective. It seems that
the collective-model calculations exaggerate the collective component in wave functions,
resulting in more gradual changes with N compared to the experimental trend.

Recent nucleon-pair SM-approximation study of Nd isotopes \cite{Ma023} (green circles) reproduces
the general trend but is far off the experiment at N=88 and N=90, where the $\nu 11/2^-[505]$
extruder is particularly active.

The confined $\beta$-soft model describes well properties of $0^+_2$ levels in $^{150}$Nd and
$^{152}$Nd \cite{Kru11} (blue circles). This may be unexpected in view of the systematics presented
in Fig. \ref{A150_Nd_0plus_Z}, which does not suggests $0^+_2$, $\beta$ excitations below 1 MeV,
especially in $^{150}$Nd. We note that at N=90 and 92 other models also do well. For example,
the SD-pair shell model \cite{He022} (yellow circles) reproduces well $0^+_2$ energies at N=88, 90
and 92 but is far off at N=84.

\subsection{Octupole excitations}

\begin{figure}
\centering
\scalebox{.32}{\includegraphics{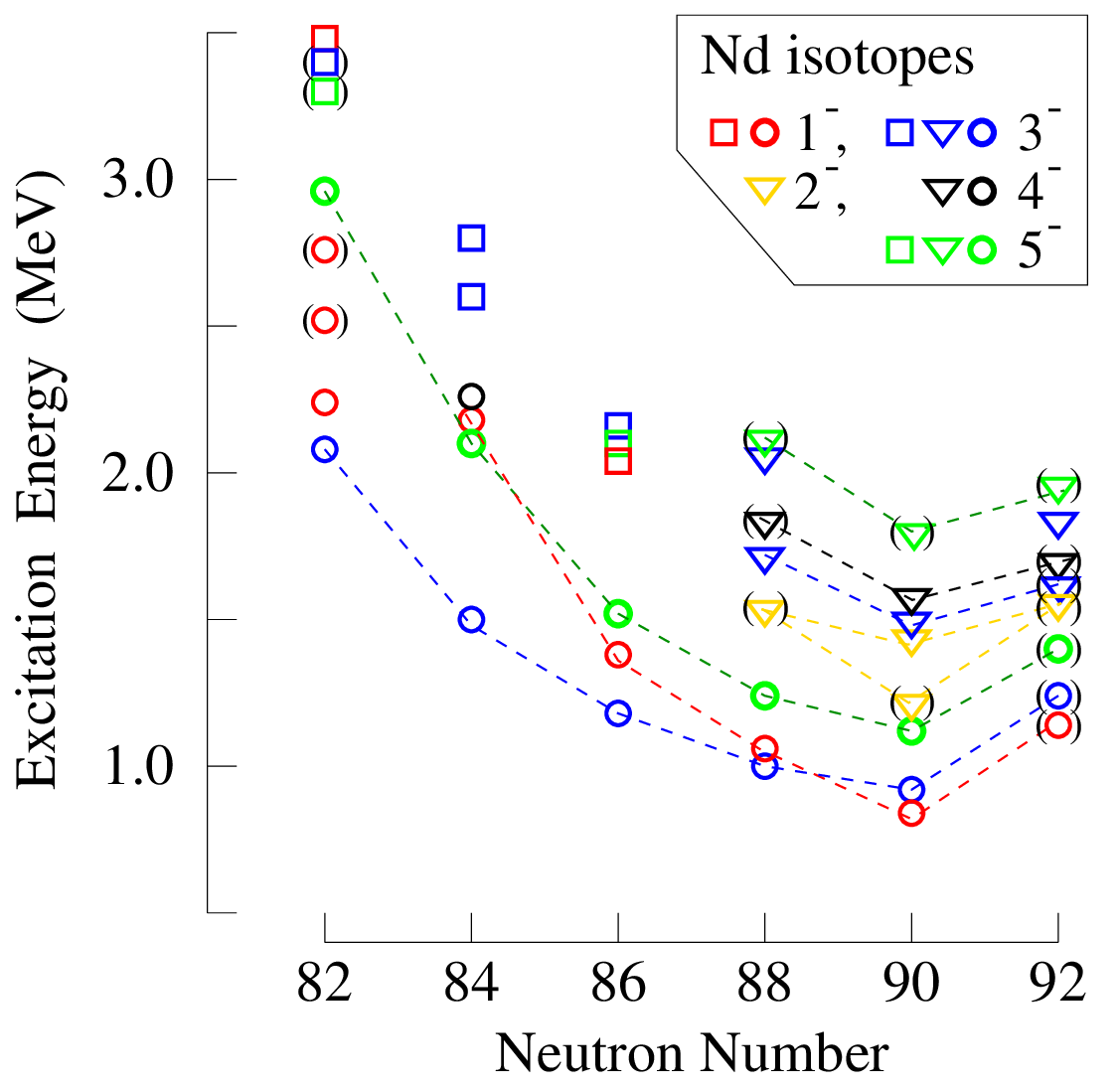}}
\caption{Energies of low-energy negative-parity levels in Nd isotopes. The data are taken from
         the data base \cite{ENSDF} and the present work. Points in parenthesis have tentative
         spin-parity. Dashed lines are drawn to guide the eye.}
\label{A150_Nd_octupole}
\end{figure}

Excitation energies of 3$^-_1$ levels shown in Fig. \ref{A150_Nd_low_exc} indicate maximum octupole
collectivity in Nd isotopes at N=90. Figure \ref{A150_Nd_octupole} shows more low-energy levels
of negative parity in Nd isotopes where three groups of such levels can be distinguished.

Levels shown by open circles, correspond to K$^{\pi}$=0$^-$ octupole excitations \cite{Urb88,Iac96}.
In  the A$\approx$150 region the K$^{\pi}$=0$^-$ collectivity has its maximum in $^{144}$Ba
\cite{Phi86,Urb97}. The gap between $\Delta$S$_{2p}$(N,Z) values for Ba and higher-Z, N=88 isotones
marked in Fig. \ref{A150_Nd_S2p_excess} (a) may be partly caused by an increased binding due to
octupole deformation at the N=88, Z=56 numbers. As discussed in Ref. \cite{Sch25} this
is due to octupole coupling between the {\it{$\Delta$j=$\Delta$l=}}3 shells, here
$\pi (d_{5/2},h_{11/2})$ and $\nu (f_{7/2},i_{13/2})$.

Levels shown by triangles, which appear at N$>$86 were interpreted as due to K$^{\pi}$=2$^-$ and
K$^{\pi}$=3$^-$ configurations with $\nu (3/2^+[532] \otimes1/2^+[660])_{2^-}$ and $\nu (3/2^+[532] \otimes1/2^+[651])_{3^-}$ dominant 2-qp components. In $^{152}$Nd these configurations
correspond to the 1541.8- and 1826.9-keV band heads, respectively \cite{Hel93}.

\begin{figure}
\centering
\scalebox{.35}{\includegraphics{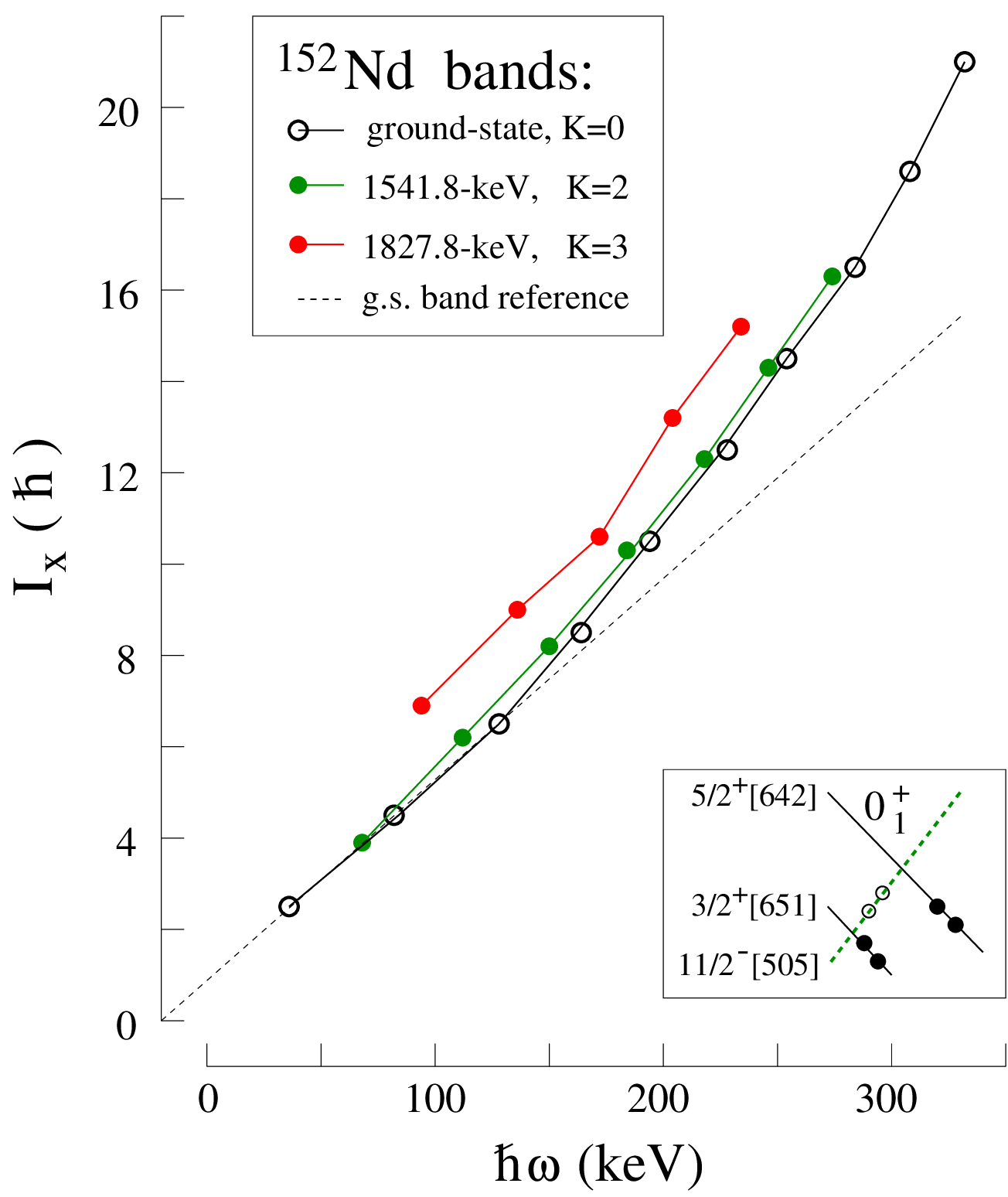}}
\caption{Total aligned angular momentum, I$_x$, for bands in $^{152}$Nd. The data are taken from
         the data base \cite{ENSDF} and the present work. See text for more comments.}
\label{A150_alignment_Nd152_2}
\end{figure}

Figure \ref{A150_alignment_Nd152_2} shows the total aligned angular momentum
$I_x=\sqrt{I_i(I_i+1)-K^2}$, where $I_i$ is the spin of the initial level. The rotational
frequency, $\omega$, is defined as $\hbar\omega=[E(I_i)-E(I_f)]/2$, where $E(I_i)$ and $E(I_f)$ is
the excitation energy of the initial and the final level, respectively. The dashed line is a
rigid-rotor reference drawn through the (0,0) point and the three lowest points of the ground
state band. The inset shows the dominating configuration of the  0$^+_1$, ground
state in $^{152}$Nd.

The ground-state band steadily gains alignment up to 5.5$\hbar$ above the rigid-rotor reference at
spin I=20. The K=2 band on top of the 1541.8-keV level follows the ground state band very closely
with 0.5$\hbar$ more aligned angular momentum, only, whereas the K=3 band on top of the 1826.8-keV
level (shown from the 1951.7-keV level up), shows the alignment higher by 2$\hbar$ relative to the
g.s. band. Therefore, the two bands seems to have configurations different from those proposed in
Ref. \cite{Hel93}. They are not the $\alpha$=0 and $\alpha$=1 signature partners of the 2-qp,
K$^{\pi}$=1$^-$, $\nu (3/2^-[521]\otimes 5/2^+[642])$ configuration discussed in Ref. \cite{Yeo10}
(one notes that with K=1 the difference in alignments is even larger than shown in
Fig. \ref{A150_alignment_Nd152_2}).

The third group of levels shown in Fig. \ref{A150_Nd_octupole} by empty squares, may by due to
mixed-symmetry (isovector) octupole excitations identified recently in $^{144}$Nd
\cite{Thu19,Sch25b}. Their occurrence, as well as the observation of mixed-symmetry quadrupole
excitations discussed above, points to the universal character of proton-neutron excitations in
spherical nuclei of the A$\approx$150 region.

\subsection{Two-quasi-particle configurations in Nd}

The 7$^{(+)}$ spin-parity proposed in the present work for the 2242.70-keV isomer in $^{152}$Nd
and the rearrangement of the band structure above it call for reinterpretation of its 2-qp
structure reported previously \cite{Yeo10}.

In Ref. \cite{Yeo10} the 2242.70-keV isomer was given tentative spin-parity I$^{\pi}$$\geq$7$^-$
and interpreted as a 2-qp, 5/2$^-$[532]$\otimes$9/2$^+$[404] proton configuration, based on
calculations, which predicted this configuration below the 2-qp, 3/2$^+$[651]$\otimes$ 11/2$^-$[505]
neutron configuration. The Authors noted that the calculated 2-qp proton level is 200 keV below
the experiment but did not provide the energy of the 2-qp neutron level. In contrast, the
calculations of Ref. \cite{Gau98} predicted 2-qp neutron configurations below 2-qp proton
configurations, reporting 2-qp neutron levels with spin-parity 7$^+$ around 2 MeV and with
spin-parity 8$^-$ around 2.2 MeV, not providing their configurations.

\begin{figure}
\centering
\scalebox{.35}{\includegraphics{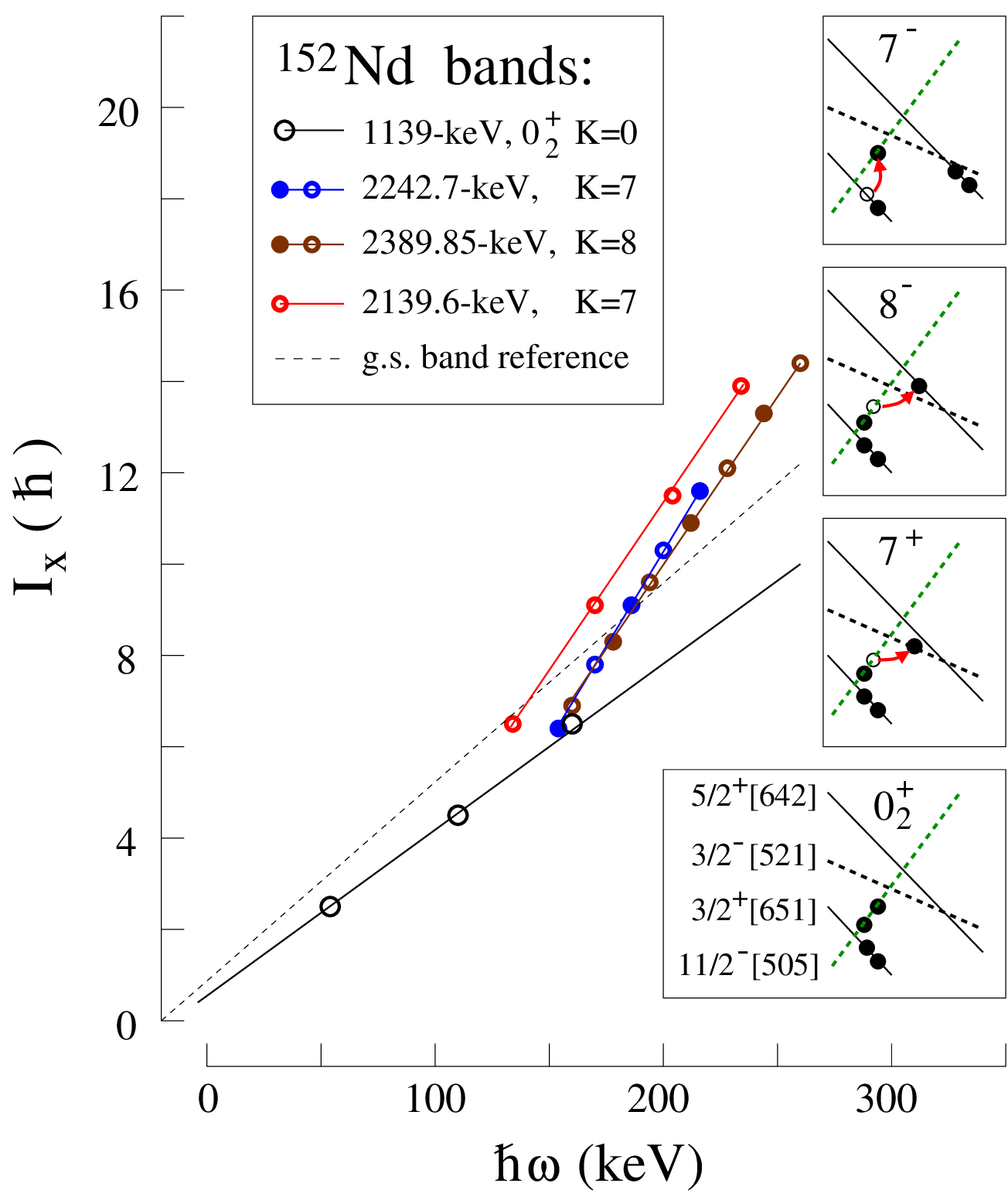}}
\caption{Total aligned angular momentum, I$_x$, for bands in $^{152}$Nd. The data are taken from
         the data base \cite{ENSDF} and the present work.See text for more comments.}
\label{A150_alignment_Nd152_3}
\end{figure}

The results of Ref. \cite{Gau98} fit well 2-qp structures in $^{152}$Nd found in the present
work, which are shown shown in Fig. \ref{A150_Nd_152_scheme}. One notes that 2-qp excitations with
spin I$\geq$7 have to involve the $\nu 11/2^-[505]$ extruder orbital. Another hint is the negative
value of the aligned angular momentum for the band on top of the 1139-keV, 0$^+_2$ level, relative
to the g.s. band, as seen in Fig. \ref{A150_alignment_Nd152_3}. The negative alignment results from
lower deformation of the 0$^+_2$ band, which contains a pair of neutrons in the $\nu 11/2^-[505]$
extruder orbital as shown in Fig. \ref{A150_Nd_transfer}. The lowest-energy 2-qp neutron
configurations based on the 0$^+_2$ level are created by exciting one neutron from the $\nu
11/2^-[505]$ extruder to the nearby, unoccupied $\nu 3/2^-[521]$ or $\nu 5/2^+[642]$ orbital
\cite{Rie71}. Therefore, we propose the $\nu (11/2^-[505] \otimes 3/2^-[521])_{7^+}$ dominating
configuration for the 2242.70-keV isomer and the $\nu (11/2^-[505] \otimes 5/2^+[642])_{8^-}$
dominating configuration for the 2389.85-keV, 2-qp level.

Figure \ref{A150_alignment_Nd152_3} supports these propositions in that the total aligned angular
momenta for bands on top of 2242.70-keV and 2389.85-keV 2-qp levels start with zero alignment
relative to the 1139-keV, 0$^+_2$ reference configuration, which contains a pair of neutrons
in the $\nu 11/2^-[505]$ extruder. As reported in Refs. \cite{Kle74,Kle77} such 2-qp configurations
correspond to highly deformed rigid rotors, the feature clearly seen in Fig.
\ref{A150_alignment_Nd152_3}.

By analogy, for the 2285.4-keV isomer in $^{150}$Nd, with tentative spin-parity (7$^+$) we propose
the $\nu (11/2^-[505] \otimes 3/2^-[521])_{7^+}$ dominating configuration, produced by exciting
one neutron from the $\nu 11/2^-[505]$ extruder to the $\nu 3/2^-[521]$ orbital. Higher excitation
energy of this isomer relative to the 0$^+_2$ level at 676 keV, compared to an analogous
distance in $^{152}$Nd, may be due to lower position of the Fermi level at N=90 than at N=92.

In Fig. \ref{A150_alignment_Nd152_3} we also show the aligned angular momentum for the band on
top of the 2139.6-keV level, calculated with K=7 (in Fig. \ref{A150_alignment_Nd152_2} this
band is assigned K=3 and starts from the 1827.8 keV level). Figure \ref{A150_alignment_Nd152_3}
shows that this band also corresponds to rigid rotation with deformation as large as that of the
$\nu (11/2^-[505] \otimes 5/2^+[642])_{8^-}$ configuration but its reference configuration is the
0$^+_1$ ground state. For this band we propose the  $\nu (11/2^-[505] \otimes 3/2^+[651])_{7^-}$
configuration where one neutron is excited from the $\nu 3/2^+[651]$ orbital to the $\nu 11/2^-[505]$
extruder. This is the remaining of the three lowest configurations of this type expected at N=92
\cite{Kon81}. Insets in Fig. \ref{A150_alignment_Nd152_3} show population of valence orbitals in
the 0$^+_2$, 7$^+$, 8$^-$ and 7$^-$ configurations in $^{152}$Nd.

Recent theoretical work on high-K isomers \cite{Zha18} reports an advanced study of 2-qp
configurations in the neutron-rich Nd and Sm isotopes, albeit only up to spin I=6. It is of
interest to make such type of calculations extended to higher-spin, to verify the proposed 2-qp
configurations involving the $\nu (11/2^-[505])$ extruder.

\subsection{Gamow-Teller transitions in the A$\approx$150 region.}

In nuclei above Z=50 and N=82 one expects enhanced $\beta^-$ decays due to the allowed,
$\nu h_{9/2}\rightarrow \pi h_{11/2}$, $\Delta \pi$=+, $\Delta l$=0, $\Delta I=0,1$ transition.
In spherical nuclei with mass A$\approx$140, where the $\nu h_{9/2}$ and $\pi h_{11/2}$ shells
are not split by deformation, $\beta^-$ decays with $log ft \leq$5 are reported in $^{140}$Ba
\cite{NDS140} and $^{144}$Nd \cite{NDS144}. In deformed nuclei with mass A$\approx$150, these
shells split into orbitals each holding two nucleons, only. Consequently, with the $\nu h_{9/2}$
and $\pi h_{11/2}$ shells split by deformation, Gamow-Teller transitions there are retarded,
with 5$< log ft <$6.

In $^{146}$Nd, $^{148}$Nd and $^{150}$Nd there are groups of levels with (2$^-$) spin-parity
proposed in the present work, which are strongly populated in $\beta^-$ decay of the (2$^-$)
g.s. of $^{146}$Pr, 1$^-$ g.s. of $^{148}$Pr and 1$^-$ g.s. of $^{150}$Pr, respectively:

- the 3316.8-keV level in $^{146}$Nd was not observed in the $^{145}$Nd(n,$\gamma$) reaction
\cite{NDS146}, known to populate collective states. This suggests its single-particle nature.
Analogous 3197.17-keV level in the N=86 isotone $^{144}$Ce is populated in $\beta^-$ decay
with $log ft$=5.7, most likely by a retarded Gamow-Teller transition.

- newly identified states in $^{148}$Nd at 3999.8, 4063.3, 4074.7 and 4099.5 keV may belong to a
multiplet of 2$^-$ levels populated by a retarded Gamow-Teller transition. The four levels decay
predominantly to the 1683.37-keV, 2$^+$ state, supporting their common origin being a 2$^-$ s.p.
state fragmented by the emerging deformation.

- the 1993.67-, 2008.73- and 2068.70-keV levels in $^{150}$Nd are reported with $log ft <$6 in
the compilation \cite{NDS150}.

In the present work $\gamma$ intensities were obtained from coincidence data. Although normalized
to the strongest singles intensities reported before, they sometimes differ from singles
intensities, influencing level intensity balances, especially inside longer cascades. Therefore,
we did not report $log ft$ values. One may expect, however, that for top levels with no $\gamma$
feeding their intensity balances are correct. With these values and the ground-state feeding from
previous $\beta^-$ decay works we estimated approximate $log ft$ values for the proposed (2$^-$)
levels in $^{146}$Nd, $^{148}$Nd and $^{150}$Nd. Taking average excitation energy and summed
population for the 3316.8- and 3335.3-keV levels in $^{146}$Nd we estimated $log ft$=6.2(2) for
this group, $log ft$=5.1(2) for the group of 3957.4-, 3999.8-, 4063.3-, 4074.7- and 4099.5-keV
in $^{148}$Nd and $log ft$=5.5(2) for the group of 1993.67-, 2008.73- and 2068.70-keV levels in
$^{150}$Nd ($log ft$=5.3(1) is estimated for the group using I$_{\beta}$ values from the
compilation \cite{NDS150}). Dedicated $\beta^-$ decay measurements are needed to verify $log ft$
values estimated in this work.

The obtained $log ft$ values help identifying proton configurations in the proposed (2$^-$),
levels and determining energies of proton orbitals in the region. For example, taking the
$(\pi 5/2^+[413] \otimes\nu 3/2^-[521])$ configuration of the 1$^-$ ground state in $^{150}$Pr
reported in Ref. \cite{Koj15}, one may propose the $(\pi 5/2^+[413]\otimes\pi 1/2^-[550])$
dominating configuration for the discussed (2$^-$) levels in $^{150}$Nd.

\section{Summary and outlook}

Low-to-medium spin excitations in $^{146,148,150,152}$Nd isotopes, populated in $\beta^-$ decay
of corresponding Pr isotopes or in prompt-$\gamma$ fission of $^{252}$Cf have been studied using
Gammasphere array of Ge spectrometers. In the four Nd nuclei we added 161 new levels, 305 new
$\gamma$ transitions and 85 new spin-parity assignments as well as two new isomers found at 2285.4
keV in $^{150}$Nd with T$_{1/2}$=41(14) ns and at 2389.85 keV in $^{152}$Nd with T$_{1/2}$=42(8) ns.
The structure of excited levels in the studied Nd isotopes has been discussed using phenomenological
classifications and systematics and calculations reported in other works. Particular attention was
paid to the $0^+$ and $2^+$ excitations related to the emerging quadrupole collectivity related to
the phase transition in the region and to the role of the 11/2$^-$[505] neutron extruder in this
phenomenon.

The present work provides essential new experimental data to be used for testing models of
nuclear excitations in the A$\approx$ 150 region. The most important observations and suggestions
in the present work are the following:

(i) the proposed classification of low-energy 0$^+$ excitations in A$\approx$150 region suggests
that they are created predominantly by nucleon-pair excitations, helped by the ``catalytic'' action
of the 11/2$^-$[505] neutron extruder orbital. Analogous action of the 9/2$^+$[404] proton extruder
in the region is also likely. It is of interest to trace the presence of the $\pi 9/2^+[404]$
orbital in Nd isotopes using transfer reactions, for example, $^{A+2}$Sm($^{14}$C,$^{16}$O)$^{A}$Nd.

The exceptional 0$^+_2$ level at 915.4 keV in $^{146}$Nd is proposed to be an oblate configuration
produced by neutron-pair excitation from the 11/2$^-$[505] extruder to the 13/2$^+$[606] oblate
orbital.

Several 0$^+$ levels at around 1.3 MeV, which are observed in deformed nuclei of the region, may
be due to K=0 vibrations. We suggest searching for analogous 0$^+$ levels around this energy in
$^{150}$Nd and $^{154}$Nd isotopes.

The configurations of 0$^+_1$, 0$^+_2$ and 0$^+_3$ levels, proposed in the present work, allow
consistent explanation of the abrupt changes of the (p,t) and (t,p) {\it relative} cross sections
in Nd, Sm and Gd isotopes around N=90, where sudden onset of deformation occurs. Systematic
measurements of {\it absolute} transfer cross sections in the region may help better understanding
of these changes.

Recent PNBCS-model calculations of S$_{2n}$ separation energies in rare-earth nuclei \cite{Guo25b}
reproduce properly these values in well deformed nuclei of the region but differ from experiment
in transitional Nd-Dy isotopes (see Fig. 9 in Ref. \cite{Guo25b}). The newly defined,
$\Delta$S$_{2n}$(N,Z) excess of two-neutron separation energy, which precisely correlates with the deformation change in the region, can be explained using neutron configurations and pair-excitation mechanism proposed in the present work.

(ii) Low-energy 2$^+$ levels in Nd isotopes and in other isotopes of the A$\approx$150 region can
be grouped into three categories:

- two kinds of K=0 vibrations with dominating $\nu f_{7/2}$ contribution and different proton
contributions. They are based on top of 0$^+_1$, with characteristic ``U'' shaped energy
systematics in function of neutron number and are similar to analogous excitations observed in
Sr isotopes of the A$\approx$100 region \cite{Urb21,Wis23}.

- K=0 rotational 2$^+$ excitations based on 0$^+_1$ and 0$^+_2$ levels. They show similar energy
decrease in function of an increasing neutron number.

- various K=2, 2$^+$ levels. Below N=90 one observes excitations characteristic of vibrations in
$\gamma$-soft potential. At higher N they evolve into rotational bands in a triaxial potential.
Energy systematics of $\gamma$ bands in Xe-Nd isotopes differ from those in higher-Z isotopes.
These observations are not consistent with recent calculations in the region.
Further experimental work is needed, among others the identification of $\gamma$-type
excitations in $^{152}$Nd and unique spin-parity assignments to new levels shown in Fig.
\ref{A150_N92_gamma}.

(iii) Three isomeric states in $^{150}$Nd and $^{152}$Nd are proposed to be 2-qp configurations
involving 11/2$^-$[505] neutron extruder. This supports the presence of the extruder at the Fermi
surface in these nuclei and its vital role in creating excited configurations in the region.
Further studies of the proposed 7$^+$, 7$^-$ and 8$^-$ configurations are needed to
firmly determine their spin-parity assignments.

(iv) low-energy excitations below the pairing-gap energy in transitional nuclei are dominated by
single-particle excitations, with collective modes admixtures incrising when adding valence
nucleons. Therefore, descriptions using collective modes, only, overestimate collective nature
of such excitations. The shell model approach is also not satisfactory, as it does not describe
properly collective excitations due to quantum effects restoring broken symmetries of a
spherical nuclear potential. For the description of transitional nuclei a new approach is needed,
perhaps along the path sketched by Matsuyanagi and co-workers \cite{Mat10,Mat16}. It should
combine the proposed nucleon-pair excitations at crossings between extruders and down-slopping
orbitals (``pair hopping'' \cite{Bar90,Bro94}) with the emerging collectivity ``dressing'' them
up \cite{Kur75a,Kur75b,Kur75c}.

\medskip

This work was supported in part by the US DOE under grant no. DE-FG02-91ER-40609.

\end{document}